\documentclass[11pt,a4paper]{article}
\pdfoutput=1
\usepackage[centertags]{amsmath}
\usepackage[square, comma, sort&compress,numbers]{natbib}
\usepackage{array,multirow}
\numberwithin{equation}{section}
\usepackage{amssymb}
\usepackage{graphicx}
\usepackage{color}
\usepackage{setspace}

\usepackage{epsfig}
\def\half{{1 \over 2}}

\def\Or[#1]{{\text{O}}\left({#1}\right)}
\def\dotl[#1,#2]{\left\langle #1, #2 \right\rangle}
\def\dotlb[#1,#2]{[ #1, #2 ]}
\def\dotp[#1,#2]{(#1) \cdot (#2)}
\def\aff[#1,#2]{\hat{#1}(#2)}
\def\n4sym{{\cal N}=4 SYM}
\def\>{\rangle}
\def\<{\langle}
\def\weight[#1,#2,#3]{\{(#1),#2,#3\}}
\def\ads[#1]{$\text{AdS}_{#1}$}

\newcommand{\ba}{\begin{eqnarray}}
\newcommand{\ea}{\end{eqnarray}}
\newcommand{\be}{\begin{eqnarray}}
\newcommand{\ee}{\end{eqnarray}}
\newcommand{\bq}{\begin{equation}}
\newcommand{\eq}{\end{equation}}

\newcommand{\CF}{{\cal F}}

\newcommand{\CL}{{\cal L}}

\newcommand{\CN}{{\cal N}}
\newcommand{\CO}{{\cal O}}
\newcommand{\CP}{{\cal P}}

\newcommand{\CV}{{\cal V}}
\newcommand{\nn}{\nonumber}

\newcommand\oo\infty
\newcommand\s\sigma
\newcommand\de\delta
\newcommand\De\Delta

\newcommand\f\phi
\newcommand\g\gamma
\newcommand\x\times
\newcommand{\nin}{\noindent}
\newcommand{\ra}{\rightarrow}
\newcommand{\fr}{\frac}
\newcommand{\comm}[2]{[#1,#2]}

\newcommand\G{\Gamma}

\usepackage{jheppub}
\usepackage{comment}

\title{Universality of long-distance AdS physics from the CFT bootstrap}
\author[a,b]{A.\ Liam Fitzpatrick,}
\author[c]{Jared Kaplan,}
\author[c]{and Matthew T.\ Walters}
\affiliation[a]{Stanford Institute for Theoretical Physics, Stanford University, \\
Via Pueblo, Stanford, CA 94305, U.S.A.}
\affiliation[b]{SLAC National Accelerator Laboratory, \\
Sand Hill Road, Menlo Park, CA 94025, U.S.A.}
\affiliation[c]{Department of Physics and Astronomy, Johns Hopkins University, \\
Charles Street, Baltimore, MD 21218, U.S.A.}
\emailAdd{fitzpatr@stanford.edu}
\emailAdd{jaredk@pha.jhu.edu}
\emailAdd{mwalters@pha.jhu.edu}
\abstract{We begin by explicating a recent proof of the cluster decomposition principle in AdS$_{\geq 4}$ from the CFT$_{\geq 3}$ bootstrap.  The CFT argument also computes the leading interactions between distant objects in AdS$_{\geq 4}$, and we confirm the universal agreement between the CFT bootstrap and AdS gravity in the semi-classical limit.  

We proceed to study the generalization to CFT$_2$, which requires knowledge of the Virasoro conformal blocks in a lightcone OPE limit.  We compute these blocks in a semiclassical, large central charge approximation, and use them to prove a suitably modified theorem.  In particular, from the $d=2$ bootstrap we prove the existence of large spin operators with fixed `anomalous dimensions' indicative of the presence of deficit angles in AdS$_3$.    As we approach the threshold for the BTZ black hole, interpreted as a CFT$_2$ scaling dimension, the twist spectrum of large spin operators becomes dense.  

Due to the exchange of the Virasoro identity block, primary states above the BTZ threshold mimic a thermal background for light operators.    We derive the BTZ quasi-normal modes, and we use the bootstrap equation to prove that the twist spectrum is dense. 
Corrections to thermality could be obtained from a more refined computation of the Virasoro conformal blocks.}
\arxivnumber{1403.6829}

\begin{document}

\maketitle
\flushbottom

\section{Introduction and summary}
\label{sec:Introduction}

Spacetime is a set of coordinate labels associated with the states and operators of a quantum mechanical system.  It becomes a  useful concept when the Hamiltonian of the system is approximately local in these coordinate labels.  One need not resort to holography to find examples;  for instance, this line of thinking underlies the reconstruction of extra dimensions from their Kaluza-Klein spectra.  One can produce even more elementary examples by studying the `emergence' of the coordinate label $x$ from an abstract interacting harmonic oscillator defined in terms of creation and annihilation operators. 
 
In this spirit, the conformal bootstrap \cite{FerraraOriginalBootstrap1,PolyakovOriginalBootstrap2,Rattazzi:2008pe} and related techniques \cite{Callan:1973pu, AldayMaldacena, KomargodskiZhiboedov} have recently led to a rigorous, non-perturbative proof \cite{Fitzpatrick:2012yx} of the cluster decomposition principle in AdS$_{d+1}$ for \emph{all} unitary $d \geq 3$ CFTs.  Both AdS cluster decomposition and the leading corrections to it, including long-distance gravitational  and gauge forces, are the AdS spacetime interpretation of a CFT theorem.  The theorem pertains to the operator content of the operator product expansion (OPE) in the large angular momentum limit.

In this paper we will explain the AdS interpretation in more detail, review the theorem and its proof, and then study its generalization to CFT$_2$/AdS$_3$.  We will show that in a certain semi-classical limit of 2d CFTs it is possible to generalize the theorem. 
 In particular, we will derive the existence of deficit angles in AdS$_3$ from the properties of Virasoro conformal blocks.  We will also study the CFT dual of a light object interacting with a BTZ black hole  \cite{BTZ}.  

The goal of the analysis is to use the conformal bootstrap to constrain the dynamics of an emergent AdS theory in a limit where a pair of objects are well-separated in AdS.\footnote{We emphasize that we are not assuming anything about the existence of an actual description in terms of fields, strings, etc. propagating in AdS.  All our claims about AdS will follow as consequences of the CFT spectrum and OPE.  }    The geodesic distance between the AdS objects will be extremely large and in particular, it may be much larger than the radius of curvature of the AdS theory.  One should therefore think of the results as demonstrating super-AdS scale locality.\footnote{This is in contrast to analyses that demonstrate sub-AdS scale locality after making various additional assumptions about the CFT \cite{JP, Hamilton:2005ju, Katz, ElShowk:2011ag, Papadodimas:2012aq, AdSfromCFT}. }    Below, as in \cite{Fitzpatrick:2012yx}, we will formulate a more precise criterion along these lines that we will term `cluster decomposition' in AdS, since it encodes the constraint that physics in one region of AdS should have no effect on physics in another region in the limit that the separation between the two regions approaches infinity.

\begin{figure}[t!]
\begin{center}
\includegraphics[width=0.75\textwidth]{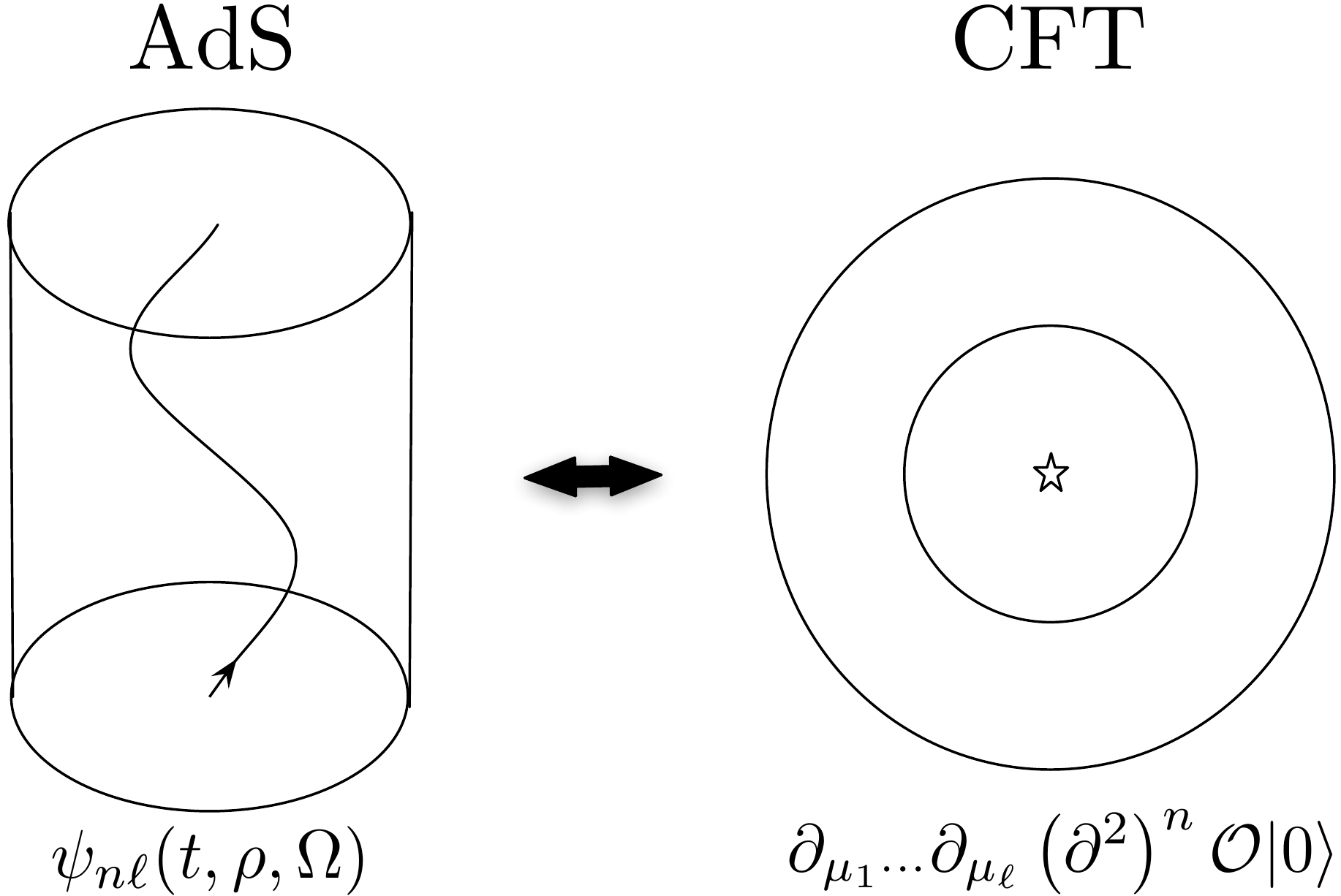}
\caption{ This figure indicates the correspondence between a descendant operator/state in the CFT and a center-of-mass  wavefunction in AdS.  The relationship is entirely kinematical; it follows because the conformal group is the isometry group of AdS. A primary state would have its center of mass at rest near $\rho=0$, the origin of AdS in the metric of equation (\ref{eq:AdSGlobalCoordinates}). }
 \label{fig:SingleDescendantAdSCFT} 
\end{center}
\end{figure}

To motivate our criterion for cluster decomposition, we rely on some basic facts about the kinematics of `objects' in AdS, which we discuss in more detail in section \ref{sec:DefiningLongDistanceAdS}.  The AdS kinematic facts that we will invoke follow almost entirely from the role of the conformal symmetry group as the isometry group of AdS.  We define an `object' in AdS as a state created by any primary operator in the CFT with definite dimension and angular momentum.  
The wavefunction for the center-of-mass of an object can be uniquely determined, and it is mainly supported near the origin of AdS.  All possible center-of-mass motions in AdS arise as linear combinations of conformal descendant states, as pictured in Figure \ref{fig:SingleDescendantAdSCFT}.  In other words, center-of-mass wavefunctions in AdS fill out a single irreducible representation of the conformal group.

\begin{figure}[t!]
\begin{center}
\includegraphics[width=0.3\textwidth]{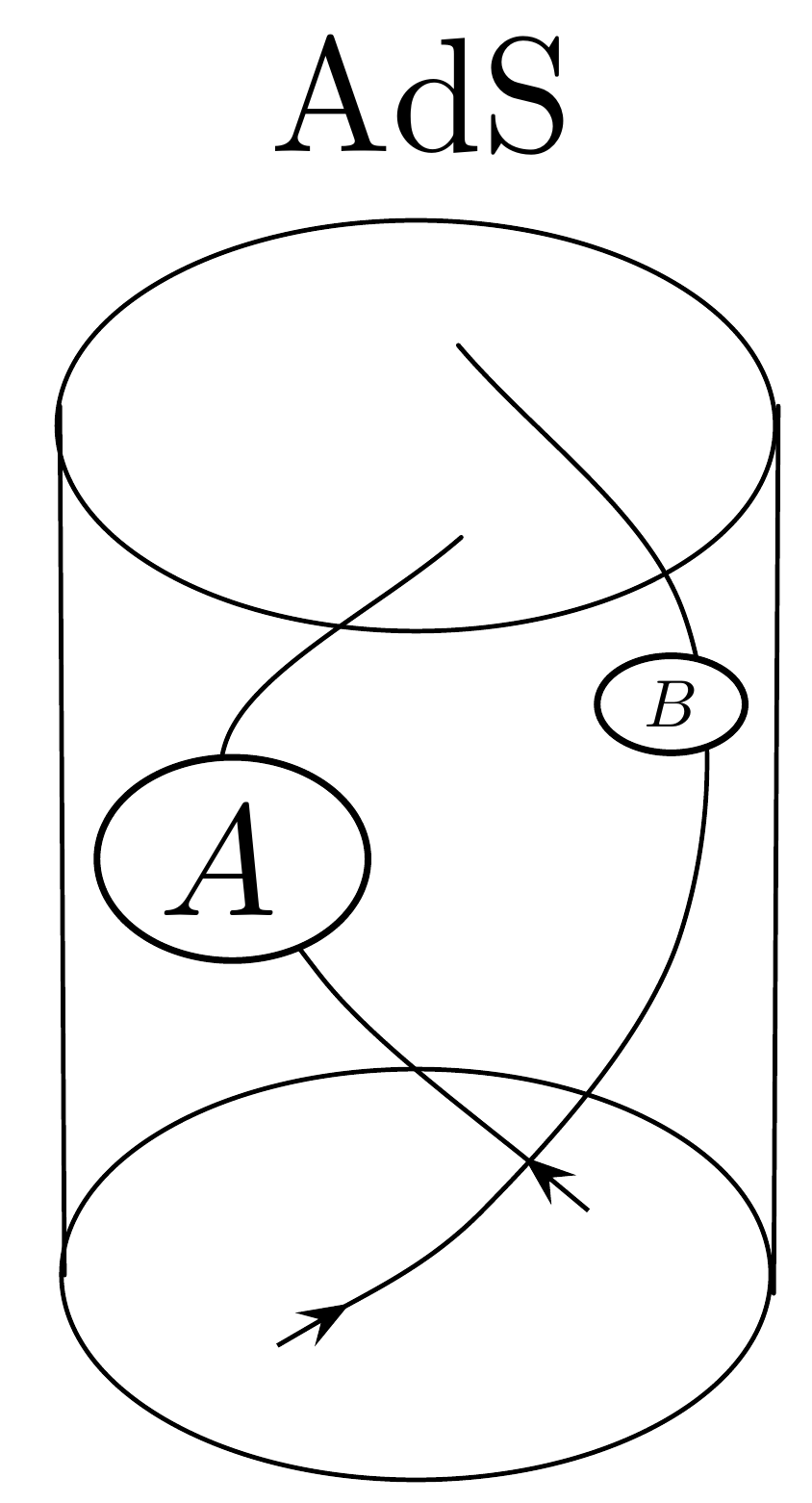}
\caption{ This figure shows two objects created by CFT operators $\CO_A$ and $\CO_B$ orbiting each other at large angular momentum, and therefore at large separation, in AdS.  A major goal will be to show that such states exist and to describe their properties. }
 \label{fig:TwoObjectsOrbitingIntro} 
\end{center}
\end{figure}

Next we would like to understand how to construct a CFT state corresponding to a pair of well-separated objects in AdS.  Naively one might try acting on the vacuum with two primaries, $\CO_A$ and $\CO_B$, but how can we create a large separation between objects $A$ and $B$?  There is no CFT state where the objects are far apart and permanently at rest in AdS, because the AdS potential would cause them to fall towards each other.  However,  if we give the pair of objects a large relative orbital angular momentum, then the centrifugal force will keep them far apart.  A rough definition of cluster decomposition can now be provided: given the existence of primaries $\CO_A$ and $\CO_B$ in a CFT,  there also exist primary operators with large angular momentum $\ell$ that create states with the appearance of objects A and B, spinning around each other at large $\ell$ in AdS, with vanishingly small interactions.  Such a state is pictured in figure \ref{fig:TwoObjectsOrbitingIntro}.

We must clarify what we mean when we say the objects are non-interacting in the limit of wide separation.  If their interactions are negligible, then the interaction or `binding' energy of the two-object state must be negligible as well. The Dilatation operator of the CFT must split up into two pieces that act separately on objects $A$ and $B$.  This translates into the statement that the anomalous dimension of the two-object state should vanish.  In precise terms, given two CFT primary operators, $\CO_A$ and $\CO_B$, their OPE should contain primary operators $[\CO_A \CO_B]_{n, \ell}$ with dimensions
\be 
\Delta_{AB}(n, \ell) = \Delta_A + \Delta_B +  2 n + \ell + \gamma_{AB}(n, \ell), 
\ee
such that $\gamma_{AB}(n, \ell) \rightarrow 0$ as $\ell \rightarrow \infty$.  Here $n$ is an additional quantum number that parameterizes the eccentricity of the orbits in the semi-classical limit, so it allows for relative boosts between the objects.  

This is exactly the spectrum of `double-trace' states in a generalized free theory (GFT).  These are theories whose correlators are entirely determined by two-point Wick contractions, as we discuss in section \ref{sec:BootstrapGFT}.  For our present purposes it is more useful to define GFTs as the dual of free quantum field theories in AdS, since this definition emphasizes that GFTs describe non-interacting objects in AdS.  In the limit $\ell \rightarrow \infty$, not only the anomalous dimensions, but also the OPE coefficients of $[\CO_A \CO_B]_{n, \ell}$ with $\CO_A$ and $\CO_B$ should approach those of a generalized free theory.  In other words, at large angular momentum the CFT should have a spectrum and OPE coefficients that match GFT.  When these criteria are all satisfied, we say that the AdS dual satisfies the cluster decomposition principle.  

Crucially, this implies that at large angular momentum, the Hilbert space of the CFT has the structure of a Fock space.  In other words, associating creation and annihilation operators $a^\dagger_{A,i}, a^\dagger_{B,i}$ and $a_{A,i}, a_{B,i}$ with the $i$-th descendants of $\CO_A$ and $\CO_B$, it is meaningful to write the state $[ \CO_A \CO_B]_{n,\ell}$ as $c_{n, \ell; i,j} a^\dagger_{A,i} a^\dagger_{B,j} | 0\>$, where $c_{n, \ell ;i,j}$ is the appropriate `Clebsch-Gordan coefficient' for irreducible representations of the conformal group.  The Dilatation operator $D$, which is the Hamiltonian for radial evolution, acts at large $\ell$ as 
\be
D= \sum_i (\Delta_{A,i} a^\dagger_{A,i}a_{A,i} +\Delta_{B,i} a^\dagger_{B,i}a_{B,i} ).
\ee
When we study AdS in global coordinates, this is the time translation operator, or in other words, the Hamiltonian.

As shown in \cite{Fitzpatrick:2012yx} and reviewed in section \ref{sec:BootstrapHigherD}, {\it all CFTs in $d\ge 3$ satisfy this cluster decomposition principle}. This result generalizes earlier results found in perturbation theory in large classes of CFTs \cite{Callan:1973pu,Belitsky:2003ys,Korchemsky:1992xv}. It is consistent with, though clearly stronger than, our experience with weakly coupled field theories in AdS$_{\ge 4}$.  Specifically, potentials between particles due to the exchange of massless fields fall off exponentially in proper distance at large separation.  In fact, when the lowest-twist ($\tau = \Delta - \ell$) operator appearing in both the $\CO_A^* \CO_A$ and $\CO_B^* \CO_B$ OPE is a conserved current, such as $T_{\mu \nu}$, the leading anomalous dimension at large angular momentum is  \cite{Fitzpatrick:2012yx, AldayMaldacena,KomargodskiZhiboedov}
\be
\gamma_{AB}(\ell) \propto \frac{1}{\ell^{d-2}}.
\label{eq:leadingcorrection}
\ee
The constant of proportionality is determined by the central charge of the current and the charges of $\CO_A, \CO_B$.  In the case where this conserved current is the energy-momentum tensor, we verify that the numerical value of the coefficient exactly matches the prediction from semi-classical gravity in AdS.  Thus ``Newtonian''  gravity in AdS is a generic long-distance feature for any CFT in $d\ge 3$.  

More generally, if operators with twist $\tau_m < d-2$ are present, the correction behaves like $\gamma_{AB}(\ell) \propto \ell^{-\tau_m}$. By unitarity, the twist cannot be less than $\frac{d-2}{2}$ for scalars, and cannot be less than $d-2$ for operators with spin $\ell\ge 1$.  Violations of the unitarity bound could produce forces that grow at long-distance, so unitarity is intimately connected with AdS locality.

\begin{figure}[t!]
\begin{center}
\includegraphics[width=0.75\textwidth]{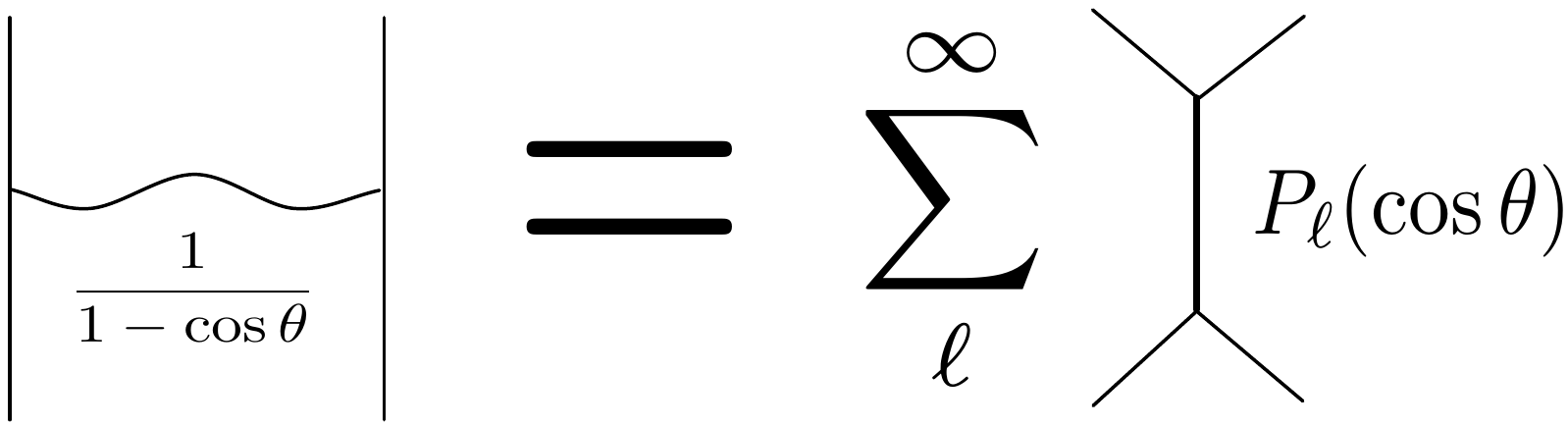}
\caption{ One can only obtain an $s$-channel singularity in a scattering amplitude via an infinite sum of $t$-channel partial waves as $\ell \to \infty$.  The same physical point, adapted to AdS/CFT, underlies the proof of cluster decomposition and the derivation of long-range forces from the CFT bootstrap.  }
 \label{fig:TChannelPartialWave} 
\end{center}
\end{figure}

The key observation that allows us to obtain these constraints is that individual conformal blocks\footnote{  For readers unfamiliar with the conformal bootstrap, we give a brief overview in section \ref{sec:bootstraprecap}.  For a more thorough review, see e.g. \cite{Rattazzi:2008pe}.} in the decomposition of  the four-point CFT correlator
\be
 \< \CO_A^*(x_1) \CO_A(x_2) \CO_B(x_3) \CO_B^*(x_4)\>
 \ee
 predict singularities in the $\CO_A^* \CO_A \rightarrow \CO_B \CO_B^*$, or `s-channel' that cannot be reproduced by any sum over a finite number of spins in the decomposition in the $\CO_A \CO_B \rightarrow \CO_A \CO_B$, or `t-channel'.  An analogous phenomenon in scattering theory is indicated in Figure \ref{fig:TChannelPartialWave}.
These singularities occur in the limit $x_{12}^2 \rightarrow 0$, which is often referred to as a ``light-cone'' limit since the position  $x_2$ is being brought onto the light-cone of the position $x_1$.  In the s-channel, these singularities are controlled by the exchange of operators with minimum twist, which generically includes the identity operator $1$ and conserved currents.

The situation becomes both more difficult and richer in $d=2$, as we discuss in section \ref{sec:VirasoroBootstrap}.  On the one hand, this difficulty can already be seen from the exchange of weakly coupled massless fields in AdS$_3$, where the potential at long distances no longer falls off at wide separation; we discuss AdS$_3$ dynamics in detail in sections \ref{sec:DeficitAngles} and \ref{sec:BTZ}.  This is related to the fact that the minimum twist of operators allowed by unitarity in $d=2$ is zero, so the leading correction from equation (\ref{eq:leadingcorrection}) to the anomalous dimension does not decay at large angular momentum $\ell$. More precisely, in $d=2$, the Virasoro algebra implies that there are infinite towers of zero-twist operators, which are the (anti-)holomorphic descendants of any (anti-)holomorphic primary operator, and these contribute singularities at the same order as the identity operator. At a minimum, the spectrum always contains the holomorphic and anti-holomorphic descendants of the identity operator itself.  

Therefore to make progress in $d=2$ we must take these contributions into account, which means we must determine the Virasoro conformal block for the identity operator.  Fortunately we can use technology that has been specifically developed to exploit the full Virasoro symmetry. 
In particular, by focusing on the case of large central charge $c$, we can use powerful techniques \cite{Monodromy} to calculate various contributions to correlators, and in particular the contribution from the OPE exchange of any number of products of the energy-momentum tensor. The conformal blocks holomorphically factorize, so in such a calculation we can focus on the holomorphic piece. In all cases, we are looking at the conformal block for an operator with weight $h_p$ contributing to the the four-point function $\< \CO_A(0) \CO_A(z) \CO_B(1) \CO_B(\infty) \>$ of operators $\CO_A, \CO_B$ with weight $h_A, h_B$.  In the semi-classical limit $c\rightarrow \infty$ and formally $ \frac{h_A}{c},\frac{h_B}{c} $ fixed, the conformal blocks
 ${\cal F}(z)$  take the form
\be
 {\cal F}(z) &=& \exp \left( - \frac{c}{6} f(z) \right) 
\ee
for a function $f(z)$ that depends on $c$ only through the various ratios $h/c$.  In the limit $h_A \ll c, h_p \ll c$ but keeping $h_B/c$ arbitrary, we find
\begin{equation}
\frac{c}{6}f(z) = (2 h_A - h_p) \log \left(\frac{1-(1-z)^{\alpha _B}}{\alpha _B}\right)+h_A \left(1-\alpha _B\right) \log (1-z) + 2h_p \log \left( \frac{1+(1-z)^{\frac{\alpha_B}{2}}}{2} \right), 
\label{eq:2dblock}
\end{equation}
where  
 $\alpha_B \equiv \sqrt{1 -24h_B/c}$, and we neglect terms of order $\CO(h_A^2/c^2, h_p^2/c^2)$. Further results using these methods for the conformal blocks are presented in appendix \ref{app:ComputingVirasoroViaMonodromy}.

The identity conformal block is the special case of (\ref{eq:2dblock})  with $h_p=0$.  In AdS$_3$, this captures the exchange of arbitrary numbers of gravitons in the semi-classical (large $m_{\rm pl}$) limit. 
 By taking appropriate limits of the positions $x_i$, one can reinterpret the four-point function equivalently as 
 the two-point function of $\CO_A$, not in the vacuum state, but in the state created by a heavy operator. 
A remarkable fact is that in this semi-classical limit, we find that {\it the identity conformal block exactly reproduces the two-point function for the light operator $\CO_A$ in a CFT at finite temperature}  \cite{Cardy:1984rp, Cardy:1986ie}
\be \label{eq:BlockAsThermalCorrelator}
\< \CO_B| \CO_A(i t) \CO_A(0) | \CO_B \> &=& \frac{(\pi T_B)^{2h_A}}{\sinh^{2h_A}(\pi T_B t)},
\ee
 set by the conformal weight of the heavy operator $\CO_B$
 \be
T_B =  \frac{\sqrt{24 h_B/c-1}}{2\pi},
\ee
where we have conformally mapped (\ref{eq:BlockAsThermalCorrelator}) to radial time coordinates $t=-\log(z)$.  
An identical formula with $h_A,T_B, z \rightarrow\bar{h}_A, \bar{T}_B, \bar{z}$ holds for the anti-holomorphic piece $\bar{\cal F}(\bar{z})$ of the identity conformal block, so for spinning operators $\CO_B$ one finds distinct left- and right-moving temperatures.  The effective temperatures $T_B,\bar{T}_B$ obtained here from the bootstrap match the semi-classical temperature of a black hole in AdS$_3$ with mass and spin given by the conformal weights of $\CO_B$.  
Consequently, the effect of multi-$T_{\mu\nu}$ exchange ({\it i.e.}, multi-graviton exchange in AdS$_3$) between a light ``test mass'' and a heavy operator has exactly the same effect that the BTZ black hole geometry has on light fields in AdS$_3$.  This provides a derivation of a version of the Eigenstate Thermalization Hypothesis \cite{ETH2, ETH} for CFT$_2$ at large central charge.

Because we take the large $c$ limit, the results we obtain in 2d have a more limited range of applicability than in $d\ge 3$, where we made no assumptions whatsoever about the CFT other than unitarity and the OPE.  However, in the large $c$ limit we have a transparent physical interpretation in AdS$_3$, and we can prove striking results about the dual dynamics, including the presence of deficit angles from particles in AdS$_3$, as well as the 
modes in a BTZ black hole background.  
A summary of the results from our bootstrap analyses follows.

\subsubsection*{Summary: CFT$_{d }$ with $d \geq 3$ }

It is convenient to state the results \cite{KomargodskiZhiboedov, Fitzpatrick:2012yx} in terms of the anomalous dimension $\gamma_{AB}(n,\ell) \equiv \Delta_{AB} - (\Delta_A + \Delta_B + 2n+\ell) $ and the OPE coefficients $c_{AB}(n,\ell)$ for the operator $[ \CO_A \CO_B]_{n,\ell}$.  These operators are implicitly defined by the proof that in the limit of large $\ell$, there exists a sequence of operators with the stated properties for every integer $n$.  We begin with the result for the general case, which assumes only unitarity and the OPE:
\begin{center}
\begin{tabular}{c}
\hline
\\
\vspace{0.5cm} General: $\gamma_{AB}(n,\ell) \sim \begin{aligned}[t]  \frac{\gamma_n}{\ell^{\tau_m}} \end{aligned}$  \quad \ $P_{AB}(n,\ell) \sim P_{\rm GFT}(n,\ell) \Big(1 + \CO(\gamma_{AB}(\ell,n))\Big)$ \\
\hline
\end{tabular}
\end{center}
In the above expression, the symbol $\sim$ denotes the behavior in the limit of large $\ell$. The function $P_{\rm GFT}(n,\ell)$ is the OPE coefficient-squared in generalized free theories; the explicit expression can be found in \cite{Unitarity}.    $\tau_m$ is defined as the smallest twist of any operator that appears in both the $\CO_A^* \CO_A$ and $\CO_B^* \CO_B$ OPE, and by unitarity this cannot be less than $\frac{d-2}{2}$.   

Using the results of \cite{Maldacena:2011jn}, it is convenient to separate out the case of CFTs whose correlators are exactly those of free fields, and all other CFTs.  The reason is that only the former case can have conserved currents with spin $\ell \ge 3$, so eliminating this one essentially trivial case allows us to restrict the minimal twist $\tau=d-2$ operators to spin-1 currents and the energy-tensor.  The result in this large class of CFTs is:
\begin{center}
\begin{tabular}{lc}
\hline \\
\vspace{0.1cm}  non-free CFT, &  \multirow{2}{*}{: $\gamma_{AB}(n, \ell) \sim \begin{aligned}[t] \frac{\gamma_{\rm grav}+\gamma_{\rm gauge}}{\ell^{d-2}} \end{aligned} $ \quad  $\gamma_{\rm grav} \approx \begin{aligned}[t]- \frac{2^{\frac{d+2}{2} } \pi  G_N (\Delta_A \Delta_B)^{\frac{d}{2}}}{{\rm vol}(S^{d-1}) (d - 1)} \end{aligned}$ \quad  $\gamma_{\rm gauge} \propto q_A q_B $ 
}\\
\vspace{0.5cm}  
$\tau^{\rm (scalar)} > d-2$&   \\
\hline
\end{tabular}
\end{center}
The coefficients $\gamma_{\rm grav}$ and $\gamma_{\rm gauge}$ can be calculated in the CFT by using the Ward identities to constrain the coefficients of conserved currents in the $\CO_A^* \CO_A$ OPE in terms of the charge of $\CO_A$, which for a spin-1 current is defined above as $q_A$, and for $T_{\mu\nu}$ is the dimension $\Delta_A$.  For simplicity we have approximated $\gamma_{grav}$ in the limit of large $\Delta_A$ and $\Delta_B$.  
The conserved current contributions can be interpreted in terms of AdS parameters by using their relation  to the CFT central charges at weak coupling; in section \ref{sec:NewtonianGravityHigherD}, we perform this matching in $d=4$ for the gravitational term and find complete agreement.

\subsubsection*{Summary:  CFT$_2$}

In the limit where $h_A h_B / c$ is fixed while $h_A / c$ and $ h_B / c \to 0$ as $c \to \infty$, the Virasoro conformal block for the identity is particularly simple. Assuming the identity is the only zero-twist primary being exchanged, the bootstrap leads to: 
\begin{center}
\begin{tabular}{c}
\hline
\\
\vspace{0.5cm} $h_A, h_B \ll c$:  $\gamma_{AB}(n, \ell) = \begin{aligned}[t]  -24 \frac{h_A h_B }{c} = -4G_N E_A E_B \end{aligned}$   \\
\hline
\end{tabular}
\end{center}
The above anomalous dimension gets corrections at  order $ \CO(\frac{h_i^3}{c^2}, \frac{n h_i}{c})$.  As indicated in the final equality above, this agrees exactly with the binding energy for two test masses in linearized gravity in AdS$_3$.  

We can also go beyond this  ``test mass'' limit, and analyze the bootstrap constraints in the limit that $h_B/c$ is fixed but $h_A/c$ is small.  It is well known that AdS$_3$ has a gap in energy of $\frac{1}{8G_N}$ between the vacuum and the lightest BTZ black hole. Below this threshold, masses in AdS$_3$ just create local conical ``deficit angle'' singularities.  Using the relation $c= \frac{3}{2 G_N}$, this energy gap translates to a threshold in the weight of a scalar operator at $h = \bar{h} = \frac{c}{24}$.  It is convenient to  
separate our results into $h_B>\frac{c}{24}$ and $h_B<\frac{c}{24}$, {\it i.e.} into weights that correspond to AdS geometries above and below threshold for a BTZ black hole.    As we review in section \ref{sec:DeficitAngles}, the deficit angle created by a particle with mass $2h_B$ in AdS$_3$ is just $\Delta \phi =2 \pi( 1- \sqrt{1- 24h_B/c})$.   
In this more general limit, we find: 
\begin{center}
\begin{tabular}{c|ll}
\hline \\
  $ \begin{aligned}[t] \frac{h_B}{c} 
   \end{aligned}  $  fixed   & 
 $h_B < \frac{c}{24}$  & : $\tau_{AB}(\ell,n) \sim \begin{aligned}[t] 2 \left( h_B + \sqrt{1 - 24h_B/c} (h_A + n)\right) = E_B + \left( 1 - \frac{\Delta \phi}{2 \pi} \right) E_A  \end{aligned} $ 
 \\
and $\begin{aligned}[t] \frac{h_A}{c} \end{aligned} \ll 1 $& $h_B > \frac{c}{24}$ & :   $\tau_{AB}(\ell,n) = $ dense   $ \sim 2h_B+ 4 \pi i  T_{\rm BTZ}(h_A + n) $ \\ \\
\hline
\end{tabular}
\end{center}
where we have listed the case of scalar $\CO_A$ and $\CO_B$, for simplicity.  

The energy spectrum below the BTZ black hole threshold exactly matches the semi-classical result from AdS$_3$ with a deficit angle $\Delta \phi$, as we discuss in more detail in section \ref{sec:DeficitAngles}.  The spacing between modes becomes vanishingly small as one approaches the BTZ threshold at $h_B =c/24$.  Above the BTZ threshold we derive a dense discretum of twists in the large $\ell$ spectrum of the $\CO_A \CO_B$ OPE. One can also identify the spectrum of BTZ quasi-normal modes. 
For this, one should use a basis not of primary operators (which must have real and positive dimensions by unitarity), but rather of in and out states, obtained in practice by adopting an appropriate $i \epsilon$ prescription.
As shown in equation (\ref{eq:BlockAsThermalCorrelator}), the semi-classical identity conformal block matches the two-point function evaluated in a thermal background, so the full spectrum\footnote{Our methods are  generally only reliable for the large angular momentum modes. } of BTZ quasinormal modes can be reproduced \cite{CFTquasinormal}.

\section{Defining long-distance AdS physics in CFT terms}
\label{sec:DefiningLongDistanceAdS}

\begin{figure}[t!]
\begin{center}
\includegraphics[width=0.75\textwidth]{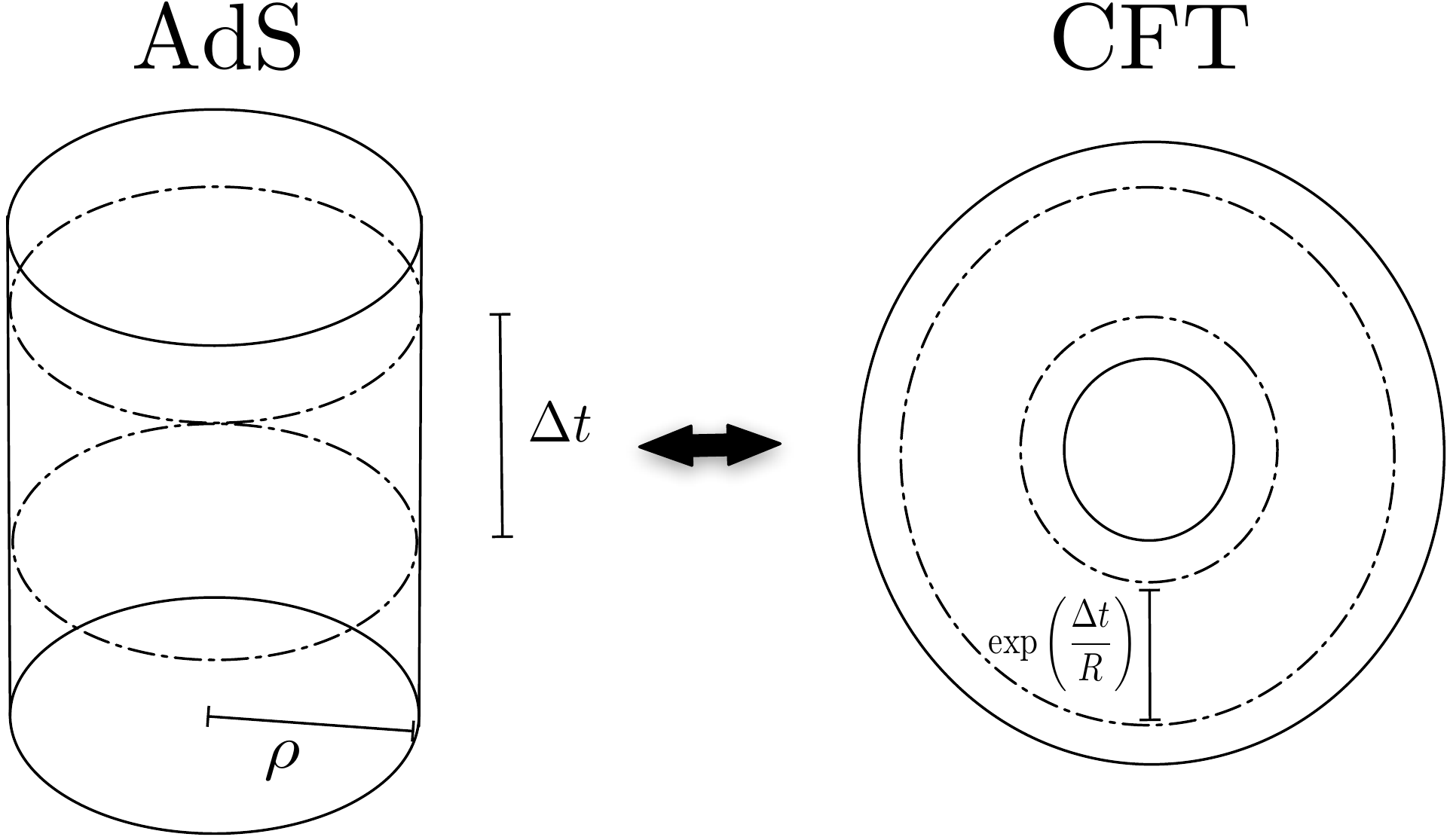}
\caption{ This figure depicts the AdS/CFT correspondence in global coordinates, emphasizing that AdS time translations are generated by the Dilatation operator, so that bulk energies correspond to operator/state dimensions in the CFT. }
 \label{fig:AdSCFTSetup} 
\end{center}
\end{figure}

In this section we will formulate a version of the AdS cluster decomposition principle and translate it into a statement about the spectrum and OPE of a CFT.  Brief in situ reviews of some necessary aspects of AdS/CFT \cite{Maldacena:1997re,GKP,Witten} will be given where required.

We will be considering CFTs in radial quantization, taking the Dilatation operator $D$ as the Hamiltonian.  Since the angular momentum generators commute with $D$, we label CFT states according to their scaling dimension $\Delta$, which is their $D$ eigenvalue, and their angular momentum quantum numbers, which we denote by $\ell$.  In this basis the momentum generators $P_\mu = -i \partial_\mu$ act as raising operators of the dimension $\Delta$, while the special conformal generators $K_\mu$ act as lowering operators.  Irreducible representations of the conformal group are labeled by the quantum numbers of a primary state, which is a state annhilated by all the $K_\mu$.  Descendant states are created by acting with $P_\mu$  on a primary.  In radial quantization, local operators can be identified with the states they create on a tiny circumscribing ball (see e.g. Chapter 2 of \cite{JPStringTheory}).

We will study AdS$_{d+1}$ in global coordinates, with metric 
\be \label{eq:AdSGlobalCoordinates}
ds^2 = \frac{R_{\rm AdS}^2}{\cos^2 \rho} \left(dt^2 - d \rho^2 - \sin^2 \rho \, d \Omega^2 \right).
\ee
 This coordinate system has a natural correspondence with a CFT in radial quantization, as pictured in Figure \ref{fig:AdSCFTSetup}.  We identify the unit $d$-vector $\hat \Omega$ with coordinates on a sphere about the origin in the CFT, and $e^t$ with the radius of the sphere.  The Dilatation operator generates $t$-translations, so that bulk energies correspond to CFT dimensions via 
\be
\Delta_{\textrm{CFT}} = E_{\textrm{AdS}} R_{\rm AdS} . 
\ee
The other global conformal generators also correspond to AdS isometries.    For the most part we will work in units with $R_{\rm AdS} = 1$, although we will occasionally reintroduce the AdS length for clarity and emphasis.

Conformal invariance uniquely determines an AdS$_{d+1}$ wavefunction for the center of mass coordinate of any primary or descendant state, as pictured in Figure \ref{fig:SingleDescendantAdSCFT}.  This is a general result; it follows because the conformal symmetries form the isometry group of AdS, so there is a one-to-one map between conformal representations and AdS coordinates.   A primary wavefunction must be annihilated by all the special conformal generators $K_\mu$, and this provides $d$ distinct first order differential equations that must be satisfied by a primary wavefunction in AdS$_{d+1}$.  In the scalar case primary wavefunctions necessarily take the form
\be \label{eq:ScalarPrimaryWavefunction}
\psi_{prim}(t, \rho, \Omega) = e^{i \Delta t} \cos^\Delta \rho .
\ee
Since the Dilatation operator $D = -i \partial_t$ we see that the undetermined parameter $\Delta$ is the scaling dimension of the state.

Equation (\ref{eq:ScalarPrimaryWavefunction}) describes a wavefunction centered at $\rho = 0$, falling off quickly at large distances, with a characteristic rate set by $\Delta$.  In the large $\Delta$ limit this can be approximated by a Gaussian wavepacket at the center of AdS, with a width $\sim 1 / \sqrt \Delta$.  It is held in place by the effect of the AdS curvature.  Descendant state wavefunctions filling out a full irreducible representation of the conformal group can be computed by acting on the primary wavefunction with the raising operator $P_\mu$, the CFT momentum generator.  A typical descendant state is portrayed in Figure \ref{fig:SingleDescendantAdSCFT}.

Let us be a bit more precise about the kinematics of the descendant states.  The AdS wavefunction for the center of mass of a state descending from a scalar primary is (see e.g. \cite{BF,Katz})
\be \label{eq:AdSScalarWavefunction}
\psi_{n, \ell J}(t, \rho, \Omega) = \frac{1}{N_{\Delta n \ell}} e^{-i E_{n,\ell} t} Y_{\ell J} (\Omega) \left[ \sin^\ell \rho \cos^\Delta \rho \, {}_2 F_1 \left(-n, \Delta+\ell+n, \ell+\frac{d}{2}, \sin^2 \rho \right) \right] \nn \\
\ee
with normalizations
\be
\label{eqn:PhiAdSNorm}
N_{\Delta n \ell} = 
(-1)^n \sqrt{\frac{n! \Gamma^2(\ell + \frac{d}{2}) \Gamma(\Delta + n - \frac{d-2}{2})} {\Gamma(n+\ell+\frac{d}{2}) \Gamma(\Delta +n+\ell)} } , 
\ee
where $E_{n, \ell} = \Delta + 2n + \ell$.  The two quantum numbers $n$ and $\ell$ index changes in the twist and angular momentum, respectively, where the twist $\tau \equiv \Delta - \ell$.  If we consider the simple case of $n=0$ and $\ell \gg \Delta \gg 1$, corresponding to minimal twist and large angular momentum, then we find that the norm of the wavefunction has a maximum at a geodesic distance\footnote{The geodesic distance $\kappa$ from the center of AdS is related to the $\rho$ coordinate by $\sinh \kappa = \tan \rho$.  }
\be
\label{eq:GeodesicDistanceOrbit}
\langle \kappa \rangle \approx \frac{R_{\textrm{AdS}}}{2} \log \left( \frac{2 \ell} {\Delta} \right)
\ee
from the center of AdS, with a width of order $R_{\textrm{AdS}} / \sqrt{\Delta}$ in $\langle \kappa \rangle$.  In this limit the wavefunction represents an object in a circular orbit about the center of AdS.

The preceding discussion of CFT states and AdS center-of-mass wavefunctions was completely general.  Now let us specialize for a moment and consider CFTs with AdS duals whose spectra include weakly coupled particles.  The 2-particle primary states in such an AdS theory are dual to operators that we will represent as $[\CO_1 \CO_2]_{n, \ell}$ in the CFT, where $\CO_1$ and $\CO_2$ are primaries that create single-particle states.  

\begin{figure}[t!]
\begin{center}
\includegraphics[width=0.85\textwidth]{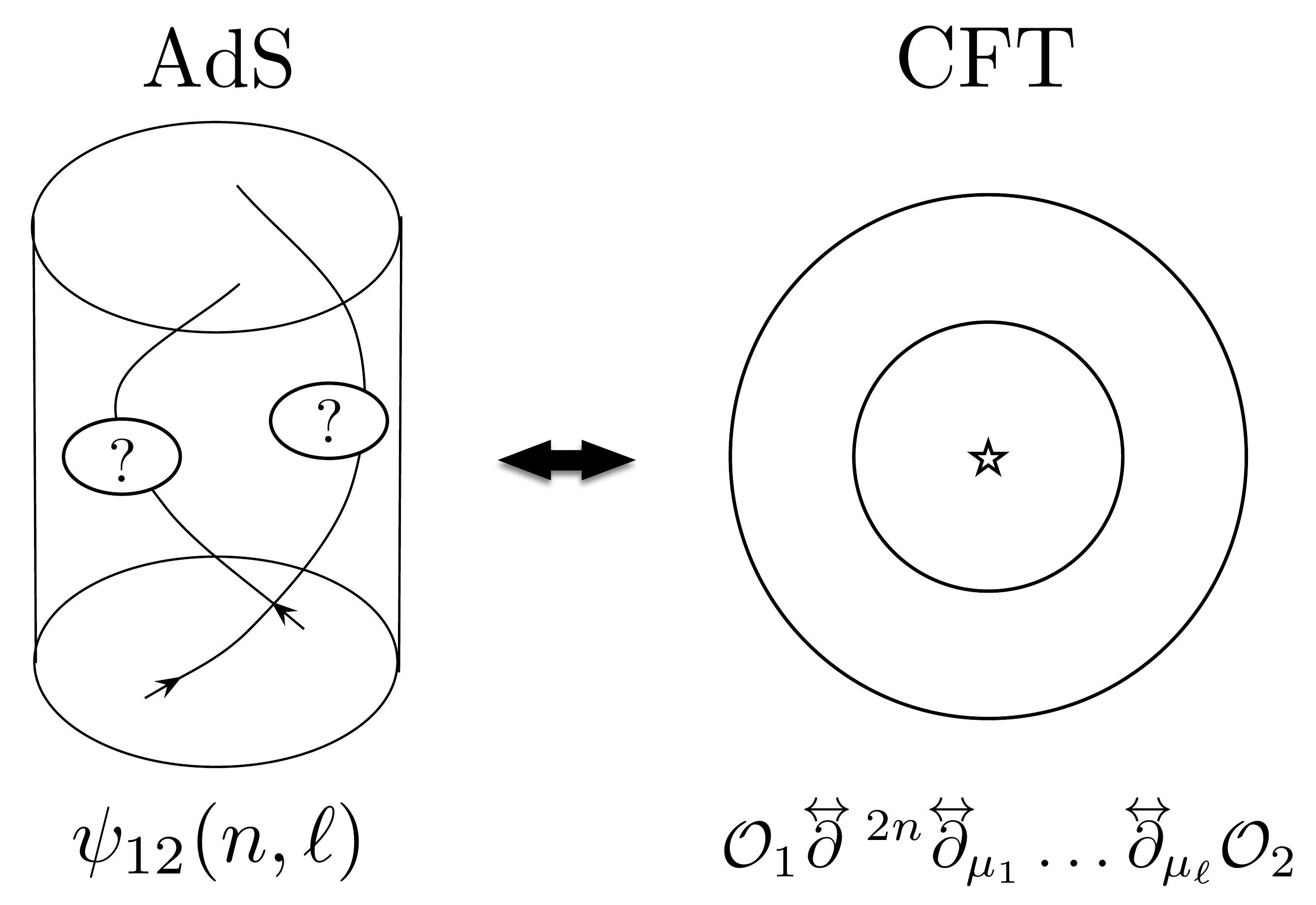}
\caption{ This figure is suggestive of the relationship between certain $\ell \gg 1$ operators in the OPE of $\CO_1$ and $\CO_2$ and a `2-blob' state in AdS, corresponding to the two states created by the CFT primaries $\CO_1(0)$ and $\CO_2(0)$ in an orbit about each other at large separation $\kappa \sim  \log \ell$.  The existence and asymptotic dimension of these 2-blob operators at large $\ell$ in the CFT defines a cluster decomposition principle in AdS.}
 \label{fig:TwoBlobsAdSCFT} 
\end{center}
\end{figure}

The primary operators $[\CO_1 \CO_2]_{n, \ell}$ create 2-particle states whose center of mass is supported near our chosen origin at $\rho=0$ in AdS, but the pair of particles themselves can have a large relative motion.  In particular, we can study the state where the particles both orbit the center of AdS precisely out of phase, so that they are opposite each other across the center of AdS.  This configuration is pictured in Figure \ref{fig:TwoBlobsAdSCFT}.   The particles are very well-separated at large $\ell$, because they are balanced across the center of AdS.  In the case of free particles the primary operators $[\CO_1 \CO_2]_{n, \ell}$ have dimension
\be
\Delta_1 + \Delta_2 + 2n + \ell .
\ee
This CFT scaling dimension corresponds to the rest mass of the two AdS particles plus a contribution from the kinetic energy of their relative motion.  

In the case of a pair of non-interacting AdS objects, including the case of free particles, we can work out the kinematics exactly.  In the appendices of \cite{JoaoMellin, Unitarity} it was shown how to decompose a primary operator $[\CO_1 \CO_2]_{n, \ell}$ in a generalized free theory\footnote{A generalized free theory is the conformal theory dual to a free field theory in AdS.  It can also be described as a CFT whose correlators can all be obtained by Wick contractions into 2-point correlators. } into the descendants of $\CO_1$ and $\CO_2$.  This is identical to decomposing 2-particle primary wavefunctions into sums of products of one-particle descendant wavefunctions in AdS.  In the case of $n=0$ one finds
\be \label{eq:DoubleTraceFromSingleTrace}
[\CO_1 \CO_2 ]_\ell = \sum_{\ell_1 + \ell_2 = \ell} s_{\ell_1, \ell_2} \left( \partial_{\mu_1} \cdots \partial_{\mu_{\ell_1}} \CO_1 \right) \left( \partial_{\nu_1} \cdots \partial_{\nu_{\ell_2}} \CO_2 \right)
\ee
with coefficients
\be
s_{\ell_1, \ell_2} = \frac{(-1)^{\ell_1}}{\ell_1! \ell_2 ! \Gamma(\Delta_1 + \ell_1) \Gamma(\Delta_2 + \ell_2)} .
\ee
This means that at large $\ell$, the CFT primary $[\CO_1 \CO_2]_\ell$ is dominated by contributions from descendants with
\be
\label{eq:AngularMomentumSharing}
\ell_1 \approx \frac{\ell}{2} \left( 1 + \frac{\Delta_2 - \Delta_1}{2 \ell - \Delta_1 - \Delta_2} \right) \approx \frac{\ell}{2} + \frac{\Delta_2 - \Delta_1}{4} .
\ee
We see that at large angular momentum, such operators are composed of pairs of descendants of $\CO_1$ and $\CO_2$ with nearly equal angular momenta.  The  relation (\ref{eq:AngularMomentumSharing}) will be useful for the semi-classical gravity calculations that follow in section \ref{sec:NewtonianGravityHigherD}.

The operators $[\CO_1 \CO_2]_{n, \ell}$ always appear in the OPE of $\CO_1$ and $\CO_2$ if the conformal theory is a generalized free theory.  If the theory is perturbative in either an AdS coupling (e.g. $1/N$) or some weak coupling in the CFT, then these operators are also guaranteed to exist \cite{Callan:1973pu} and to make an appearance in the $\CO_1(x) \CO_2(0)$ OPE.  But away from free theory they will acquire an anomalous dimension $\gamma(n, \ell)$.  

From the AdS viewpoint, this anomalous dimension arises due to the interaction energy between the two objects. This means that at large $\ell$ we can use the relationship between $\langle \kappa \rangle$ and $\ell$ from equation (\ref{eq:GeodesicDistanceOrbit}) to write the total dimension of $[\CO_1 \CO_2]_{n, \ell}$ as
\be
 \Delta_1 + \Delta_2 + 2n + \ell + \gamma(n, \ell(\kappa)) ,
\ee
where $\kappa$ is the geodesic distance between the objects in AdS.  Since $\ell$ grows exponentially with $\kappa$, the strength of the AdS interaction at large distances is determined by the magnitude of the anomalous dimensions $\gamma(n, \ell)$ at very large $\ell$.  In perturbative examples \cite{CostaEikonal, CostaCPW, CostaSW} the anomalous dimension $\gamma(n, \ell)$ falls off as a power-law in $\ell$ as $\ell \to \infty$ in the case $d \geq 3$.

Do operators like $[\CO_1 \CO_2]_{n, \ell}$ always exist in the OPE of $\CO_1$ and $\CO_2$ in any CFT?  If so, then every CFT has a Hilbert space that can be interpreted in terms of states moving in AdS.  The anomalous dimensions $\gamma(n, \ell)$ would give information about the properties of AdS interactions, with the large $\ell$ behavior corresponding to the effects of long-range forces in AdS.

We are finally ready to formulate our version of the AdS cluster decomposition principle as a statement about the OPE and the CFT spectrum: \emph{In the OPE of any two primary operators $\CO_1$ and $\CO_2$,  for each non-negative integer $n$, there exists an infinite tower of operators $[ \CO_1 \CO_2]_{n, \ell}$ in the limit that $\ell \to \infty$, with dimension $\Delta_1 + \Delta_2 + 2n + \ell + \gamma(n, \ell)$ where $\gamma(n, \ell) \to 0$ as $\ell \to \infty$. Furthermore, one can show that 
\be \label{eq:LeadingAnomalousDimension}
\gamma(n, \ell) = \frac{\gamma_n}{\ell^{\tau_m}},
\ee
where $\tau_m$ is the twist of the minimal twist operator appearing in the OPE of both $\CO_1$ with $\CO_1^\dag$ and $\CO_2$ with $\CO_2^\dag$.}  Generically $\tau_m \leq d-2$, since the energy momentum tensor $T_{\mu \nu}$ always appears in both of these OPEs, and in fact it is straightforward to go beyond equation (\ref{eq:LeadingAnomalousDimension}) to derive the anomalous dimension at subleading order in $1/\ell$.   In section \ref{sec:NewtonianGravityHigherD} we will give an explicit computation of the long-distance gravitational effects for $d \geq 3$, which match the universal contribution from $T_{\mu \nu}$ that we will obtain from the CFT bootstrap in section \ref{sec:CoulombPotential}.  

This theorem has been proven \cite{Fitzpatrick:2012yx, KomargodskiZhiboedov} for \emph{all} CFT$_{\geq 3}$, without any assumptions beyond unitarity.   However, our formulation of the cluster decomposition principle is false in the case of AdS$_3$/CFT$_2$.  In fact, the 2d Ising model provides an explicit counter-example \cite{Fitzpatrick:2012yx}.  

We will see what goes wrong in section \ref{sec:DeficitAngles}, but the intuition from AdS$_3$ is simple.  Gravitational effects in $2+1$ dimensions lead to deficit angles surrounding massive sub-Planckian objects, and these deficit angles can be detected from arbitrarily large distances.  This means that they make finite corrections to the spectrum of operator dimensions, so that $\gamma(n, \ell)$ approaches a finite constant $\gamma(n)$ as $\ell \to \infty$.  The CFT$_2$ interpretation is that the presence of zero twist operators, such as the Virasoro descendants of the identity, imply that in equation (\ref{eq:LeadingAnomalousDimension}) we have $\tau_m = 0$.  However, with proper caveats we will show that a modified theorem holds, and that we can compute the finite anomalous dimensions $\gamma(n)$ directly from the CFT bootstrap in two dimensions.  We study the AdS$_3$ expectations for deficit angles in section \ref{sec:DeficitAngles}.  Then in section \ref{sec:BTZ} we will obtain even more interesting expectations when we consider BTZ black holes.   We will review the fact that there are no stable orbits about these objects, so we do not expect that cluster decomposition can hold above the BTZ threshold.  However, what we can expect is a thermal spectrum of quasi-normal modes.  In the remainder of this work we will then provide a universal CFT proof of these results without making further reference to AdS expectations.

\subsection{AdS$_{\geq 4}$: the Newtonian gravitational potential }
\label{sec:NewtonianGravityHigherD}

In this section we will compute the shift in energy due to the gravitational interactions between very distant, uncharged, scalar masses in AdS$_{\geq 4}$.   This corresponds to the CFT computation of the anomalous dimension of the primary operator $[\CO_1 \CO_2]_{n, \ell}$ in the OPE of primaries $\CO_1$ and $\CO_2$, in the large $\ell$ limit.  We will derive this anomalous dimension directly from the CFT bootstrap in section \ref{sec:CoulombPotential} and find that the results match.

The idea of the calculation is to do perturbation theory in the inverse distance between the objects, resulting in a `Newtonian' approximation in AdS.  This approximation is good only when $d \geq 3$, because gravitational interactions do not fall off with distance in $2+1$ bulk dimensions. 
In section \ref{sec:DeficitAngles}, we use a different method to derive the interaction energy in $2+1$ dimensions assuming that $G_N$ is sufficiently small.

We will obtain the first order energy shift by computing the expectation value of the gravitational interaction Hamiltonian using the unperturbed wavefunction for the orbiting object.  First we will compute the interaction Hamiltonian (gravitational potential) at large distances due to the presence of a point mass, and then we will evaluate the expectation value.  

In AdS$_{\geq 4}$, the AdS-Schwarzschild metric \cite{Hawking:1982dh} is the solution to Einstein's equations in the presence of a spherically symmetric, uncharged mass.   In $d+1$ dimensions it is
\be \label{eq:AdSSchwarzschildMetric}
ds^2 = U(r) dt^2 - \frac{1}{U(r)} dr^2 - r^2 d \Omega^2,
\ee
where
\be
U(r) = 1 - \frac{\mu}{r^{d-2}} + \frac{r^2}{R_{\textrm{AdS}}^2}
\ee
and the mass of the black hole is
\be
M = \frac{ (d-1) \Omega_{d-1} \mu }{16 \pi G_N},
\ee
where $\Omega_{d-1} = {\rm vol}(S^{d-1})$.  
This coordinate system is useful because $\sqrt{-g}$ is independent of $M$, so only $g^{00}$ and $g^{rr}$ are affected by the mass $M$.  We need compute only to first order in $M$, since this is equivalent to expanding in the inverse distance. 

The energy shift to first order in $M$ is then
\be
\delta E_{orb} &=& \langle n, \ell_{orb} | \delta H | n, \ell_{orb} \rangle
\nn \\
&=& - \frac{\mu}{4} \int dr \, r^{d-1} d^{d-1} \Omega \< n,\ell_{orb}| \left( \frac{r^{2-d}}{(1+r^2)^2} (\partial_t \phi)^2 + r^{2-d} (\partial_r \phi)^2 \right) |n,\ell_{orb}\>.
\ee
The two pre-factors of $\frac{1}{2}$ in the above equation come from the normalization of the action for a scalar field in AdS and the inclusion of both the scalar and gravitational energy shifts (see e.g. \cite{Fitzpatrick:2011hh}).  We have attached an `orb' label to emphasize that we are currently studying one mass, described by the scalar field $\phi$, orbiting a second mass $M$ at the origin of AdS.  This is not a primary state in the CFT, since its center of mass is not at rest, and so we will need to translate this result to obtain the anomalous dimension of a primary operator $[\CO_1 \CO_2]_{n, \ell}$.  

Using the wavefunctions from equation (\ref{eq:AdSScalarWavefunction}) transformed to $r = \tan \rho$ coordinates, we find
\be
\delta E_{orb}(n, \ell_{orb}) = - \frac{\mu}{2}  \int \frac{r dr }{N_{\Delta n \ell_{orb}}^2} \left( \frac{1}{(1+r^2)^2} E_{\Delta n \ell_{orb}}^2 |\psi_{n \ell_{orb}}(r) |^2 + (\partial_r \psi_{n \ell_{orb}}(r))^2 \right),
\ee
where $\psi_{n\ell_{orb}}(r)$ just includes the $r$ dependence of the wavefunctions.  Taking the $n=0$ case as an example and expanding the result as $\ell \to \infty$, we find the two terms
\be
\label{eq:EnergyShiftOrb}
\delta E_{orb}(0, \ell_{orb}) = - \frac{8 \pi G_N M \Delta}{(d - 1)\Omega_{d-1}}
\left( \frac{ \Gamma (\Delta )}{2 \Gamma \left(-\frac{d}{2}+\Delta +1\right)}  \right)
  \left(
\left(\frac{1}{\ell_{orb}}\right)^{\frac{d-2}{2}}
+ \left(\frac{1}{\ell_{orb}}\right)^{\frac{d}{2}}
\right),
\ee
and clearly the first term is dominant at large $\ell_{orb}$.  This follows from the familiar fact that the Newtonian approximation requires us to keep track only of shifts in the metric component $g_{tt}$.  In fact, we could have obtained this energy shift to leading order at large $\Delta$ via a computation in classical gravitational perturbation theory.

Equation (\ref{eq:EnergyShiftOrb}) is not yet the formula of interest, since it is the energy shift associated with a configuration where one mass is at rest at the center of AdS, while the other orbits.  To get the energy shift or anomalous dimension of the primary operator $[\CO_1 \CO_2]_{n, \ell}$, we need to use equation (\ref{eq:AngularMomentumSharing}) to relate the double-trace primary to this `orbit' state.  In the semi-classical limit the orbit state has the same energy shift as a primary with equal geodesic separation between the two objects, so that $\kappa_{orb} = \kappa_1 + \kappa_2$ with
\be
\kappa_{1} = \frac{1}{2} \log \left( \frac{\ell_{prim}}{\Delta_1} \right)  \ \ \ \mathrm{and} \ \ \ \kappa_2 =  \frac{1}{2} \log \left( \frac{\ell_{prim}}{\Delta_2} \right) .
\ee
Using equation (\ref{eq:GeodesicDistanceOrbit}) for the geodesic radius of an orbit, the angular momentum of orbit can be related to that of the primary by
\be
\ell_{orb} = \frac{\ell_{prim}^2}{2 \Delta_1} .
\ee
Taking $\ell_{prim} \to \ell$, $M \approx \Delta_1$, and $\Delta = \Delta_2$, we find a semi-classical energy shift
\be
\delta E(0, \ell) \approx - \frac{2^{\frac{d+2}{2}} \pi  G_N (\Delta_1 \Delta_2)^{\frac{d}{2}}}{\Omega_{d-1} (d - 1)} 
\left(\frac{1}{\ell}\right)^{d-2}
\ee
in the approximation that $\ell \gg \Delta_1, \Delta_2 \gg 1$.  In the case of $d=4$, using the relation $c = \frac{\pi}{8 G_N}$, this gives
\be \label{eq:GravityPredictionAnomalousDimension}
\gamma(0, \ell) \approx - \frac{1}{6} \frac { (\Delta_1 \Delta_2)^{2}}{c} 
\left(\frac{1}{\ell}\right)^{2} ,
\ee
which matches the result we will derive from the CFT bootstrap in section \ref{sec:CoulombPotential}.

As a final consideration, one might ask if these AdS$_{\geq 4}$ configurations are unstable due to the emission of gravitational and other radiation.\footnote{We thank Gary Horowitz for discussions of this point.}  For a variety of reasons we expect that radiation will be an extremely small effect at large $\ell$.  First, it is worth emphasizing that unlike binary star systems in our own universe, the pair of objects we consider here are held in their orbit by the AdS curvature.  The gravitational binding energy between the objects vanishes at large $\ell$ even though the orbital period remains constant.  Each object in the orbiting pair closely resembles a conformal descendant, as indicated in equation (\ref{eq:DoubleTraceFromSingleTrace}), and such states are exactly stable.  This means that an emitted graviton would have to `know' about both objects, despite their very large separation, and so we would expect emission to be exponentially suppressed.  Furthermore, since the gravitational binding energies vanish at large $\ell$, while gravitons in AdS have an energy or dilatation gap $d/R_{\textrm{AdS}}$,  considerations of energy and angular momentum conservation also suggest that graviton emission should be an exponentially suppressed process.  Thus we expect that our orbiting pairs will have a highly suppressed radiation rate at very large $\ell$.

\subsection{Deficit angles in AdS$_3$ from sub-Planckian objects}
\label{sec:DeficitAngles}

Although there are no propagating gravitons in $2+1$ dimensional gravity, Einstein's equations have well-known, non-trivial solutions \cite{Deser:1983tn, Deser:1983nh} in the presence of sources.  In particular, a point particle of sub-Planckian mass placed in AdS$_3$ will produce a deficit angle at its location, while the spacetime remains locally AdS$_3$ everywhere else.  This explicit solution for a particle at the origin can be written as
\be
ds^2 = \frac{(1 - 8 G_N M)}{\cos^2( \rho )}  \left( dt^2 - \frac{d \rho^2}{1 - 8 G_N M} -  \sin^2(\rho) d \theta^2 \right),
\label{eq:deficitmetric}
\ee
where $M$ is the mass of the particle and $\theta \in [0, 2 \pi)$.  This looks exactly like the usual AdS$_3$ metric except for the presence of an angular deficit of $2 \pi (1-\sqrt{1- 8 G_N M})$, which is $\approx 8 \pi G_N M$ in the limit $G_N M \ll 1$.  We have made our choice for the normalization of $t$ and $\theta$ so that these coordinates have the usual relationship with CFT coordinates in radial quantization.  In particular, the Dilatation operator $D = i \partial_t$.

Now let us compute the energy shift of a particle in AdS$_3$ due to the presence of the deficit angle.  In fact, there is no computation to do.  The usual bulk wavefunctions $\psi_{n \ell}(t, \rho, \theta)$ in AdS$_3$ from equation (\ref{eq:AdSScalarWavefunction}) are also the wavefunctions in our AdS-deficit spacetime if we send   
\be
\Delta \to \Delta \sqrt{ 1 - 8 G_N M} , \ \ \ \ n \to n \sqrt{ 1 - 8 G_N M} , \ \ \ \ \ell \to \ell .
\ee
In particular, this means that the eigenspectrum for a scalar field in this spacetime is
\be \label{eq:ExactSpectrumDeficit}
E_{n, \ell} = (\Delta + 2n)\sqrt{1 - 8 G_N M } + \ell   . 
\ee
An interesting feature of this equation is that as $8G_NM \to 1$ the spectrum of twists, labeled by $n$, becomes more and more closely spaced, until we obtain a dense spectrum  at $8G_NM = 1$, the BTZ black hole  threshold.  In section \ref{sec:deficitanglespectrum} we will derive this result 
in CFT$_2$ with large central charge in the large $\ell$ limit, without making reference to AdS$_3$.

It is also useful to consider an expansion in the limit that $G_NM \ll 1$.
Using this result, we thus have a prediction that in the limit $1,n \ll \Delta_1, \Delta _2  \ll c$, we should find an anomalous dimension
\be
\gamma(n, \ell) \approx - \frac{6}{c} \Delta_1 \Delta_2 
\ee
for the shift in dimension of large $\ell$ operators that dominate the OPE of $\CO_1$ and $\CO_2$, where we have identified $c = \frac{3}{2G_N}$ for the case \cite{Brown:1986nw} of AdS$_3$/CFT$_2$.

\subsection{Quasi-normal mode spectrum from super-Planckian objects}
\label{sec:BTZ}

In AdS$_3$ there exist the well-known BTZ black hole \cite{BTZ} solutions.  As our last example we will be interested in the quantum mechanical spectrum associated with a sub-Planckian object moving with large angular momentum around a $2+1$ dimensional black hole.  We can approach this question by studying the scalar wave equation in the BTZ background.  

The BTZ metric for a spinless, uncharged black hole is 
\be
ds^2 = (r^2 - r_+^2) dt^2 - \frac{dr^2}{r^2 - r_+^2} - r^2 d \phi^2.
\ee
  The black hole has a horizon at the coordinate $r = r_+$.   Unlike in the case of higher dimensional AdS black holes, there are no timelike geodesics \cite{Cruz:1994ir} in this spacetime that avoid entering the black hole horizon.  This is easy to see from the metric of equation (\ref{eq:AdSSchwarzschildMetric}), which naturally accords with the BTZ metric when $d=2$.  Timelike geodesics in this metric can be characterized by a radial equation
\be
\dot r^2 = E^2 - V(r) \ \ \ \mathrm{where} \ \ \ V(r) = \left( 1 - \frac{\mu}{r^{d-2}} + r^2 \right) \left( 1 + \frac{\ell^2}{r^2} \right)
\ee
We see that when $d = 2$, the effective potential $V(r)$ is always monotonic in the presence of a BTZ black hole (when $\mu > 1$), whereas $V(r)$ can have a stable minimum in $d > 2$.  So there are no classical orbits about the BTZ black hole, sharply differentiating the behavior in AdS$_3$ from AdS$_{\geq 4}$.

The wave equation for a scalar with squared mass $m^2 = \Delta (\Delta - 2)$ in the spinless BTZ background has solutions of the form
\be
\phi(t, r, \phi) = e^{-i \omega t + i \ell \phi} \,U_{\omega \ell}(r),
\ee
where the normalizable radial wavefunction is
\be
U_{\omega \ell}(r) = \left( r^2 - r_+^2 \right)^{\frac{i \omega}{2 r_+}} \left( r \right)^{\frac{i \omega}{r_+} - \Delta} 
{}_2 F_1 \left(\frac{i \ell + i \omega}{2 r_+} + \frac{1}{2}\Delta  , \frac{i \ell -i \omega }{2 r_+} + \frac{1}{2}\Delta ,\Delta, \frac{r_+^2}{r^2} \right) .
\ee
One can check that these solutions analytically continue to the pure AdS$_3$ solutions of equation (\ref{eq:AdSScalarWavefunction}) if one takes $r_+ \to i$.  

The BTZ-background solutions differ in an important qualitative way from those for a scalar in empty AdS$_3$.  The BTZ solutions are oscillatory in $\log r$, whereas the AdS$_3$ solutions are exponentially suppressed as $\log r \to - \infty$.   As a consequence, even at very large $\ell$, the BTZ solutions are not suppressed in the vicinity of the black hole horizon.  This is the quantum mechanical reflection of the absence of stable orbits.
This behavior sharply distinguishes the BTZ solutions from those in $d \geq 3$, as in the latter we can make the orbital lifetime as large as desired by taking the limit of large angular momentum.

This feature of the solutions has an immediate consequence for the quasi-normal mode frequencies $\omega$.  To determine these frequencies we need to impose some sort of boundary condition on the radial wavefunction, and quasi-normal modes are taken to be purely ingoing solutions at $r_+$, the black hole horizon.  Imposing this boundary condition leads to
\be
\omega_{n, \ell} = \ell +  i r_+ (\Delta + 2 n),
\label{eq:qnmodes}
\ee
where $\ell$ is the angular momentum, and $n$ is another quantum number analogous to that which labels the twist in the pure AdS$_3$ case.  In fact this is just the analytic continuation of equation (\ref{eq:ExactSpectrumDeficit}).  
We see that for all values of the angular momentum $\ell$ the AdS$_3$ energies have a constant, finite imaginary part.  We therefore expect that after diagonalizing the CFT$_2$ Dilatation operator we will obtain a dense spectrum of twists $\tau \equiv \Delta - |\ell |$.  This matches expectations from equation (\ref{eq:ExactSpectrumDeficit}), which showed that as the mass of a deficit angle approaches the minimal BTZ mass, the spectrum of twists becomes more and more closely spaced. We also expect to reproduce the quasi-normal mode spectrum (\ref{eq:qnmodes}) after analytic continuation in radial time of the CFT correlators.  We will prove both of these predictions using the CFT$_2$ bootstrap in section \ref{sec:BTZspectrum}.

The fact that there are no stable orbits around a BTZ black hole has the surprising consequence that one cannot make stable configurations of multiple deficit angle singularities orbiting each other, if their total mass is above the BTZ mass threshold.   This is all the more surprising since above $d=2$, we know how to make such states by spreading high-energy particles diffusely throughout space and giving them large angular momentum.  To understand this phenonemon better, let us see qualitatively why such a state forms a black hole in AdS$_3$.  Consider the case of $k$ identical particles each with large angular momentum $\ell$, so that they are well-separated.  The Schwarzschild radius for a state with dimension $E$  is
\be
r_+ &=& \sqrt{8G_NE -1},
\ee
 On the other hand, each of the $k$ particles is localized at a radial coordinate
$r \approx \sqrt{\ell/\Delta}.$
The $k$-particle state has total dimension $E \approx k(\Delta+ \ell)$, and so the condition for them to be outside their Schwarzschild radius is
\be
r > r_+ \Rightarrow \frac{\Delta+\ell}{\Delta} > 8G_N k(\Delta + \ell),
\ee
or equivalently, since $\Delta$ and $\ell$ are positive,
\be
 k \Delta < \frac{1}{8 G_N}. 
\ee
The important point is that increasing $\ell$ does not help in satisfying this condition.  Once the total rest mass $k\Delta$ of the particles increases beyond the BTZ mass threshold $1/8G_N$ black hole formation cannot be avoided by increasing the angular momentum, as was possible in higher dimensions.

\section{Review of the bootstrap derivation for $d \geq 3$}
\label{sec:BootstrapHigherD}

Despite the various technical details we will discuss along the way, the argument presented here is conceptually quite simple. In any CFT, the correlation function of four scalar operators can be expressed in terms of a series of basis functions, known as conformal partial waves or conformal blocks, in a calculation directly analagous to the standard partial wave expansion of scattering amplitudes. This expansion can be performed in any of three channels, yielding different expressions which must be identical.  The equality of these expressions is referred to as the conformal bootstrap equation, which is a powerful tool used to constrain the structure of any CFT.   The bootstrap was originally developed in the case of CFT$_2$ \cite{FerraraOriginalBootstrap1,PolyakovOriginalBootstrap2}, and has recently seen extensive analytic \cite{JP,  AdSfromCFT, Fitzpatrick:2012yx, Komargodski:2012ek,Alday:2013cwa, Alday:2013opa, Beem:2013sza} and numerical \cite{Rattazzi:2008pe,Rychkov:2009ij,Caracciolo:2009bx,Poland:2010wg,Rattazzi:2010gj,Rattazzi:2010yc,Vichi:2011ux,Poland:2011ey,Rychkov:2011et,ElShowk:2012ht,Liendo:2012hy,ElShowk:2012hu,Fitzpatrick:2012yx, Beem:2013qxa,Kos:2013tga,Gliozzi:2013ysa,El-Showk:2013nia,Gaiotto:2013nva,Bashkirov:2013vya, El-Showk:2014dwa} application in CFT$_{\geq 3}$.

\begin{figure}[t!]
\begin{center}
\includegraphics[width=0.95\textwidth]{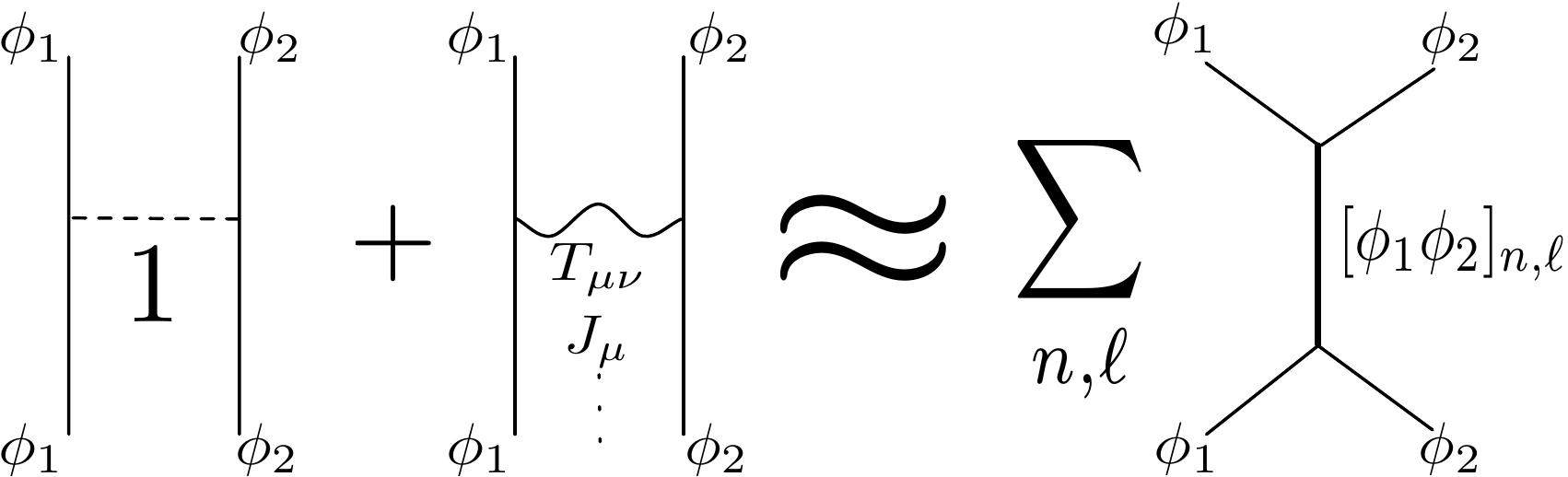}
\caption{ This figure indicates the form that the Bootstrap equation takes in the lightcone OPE limit where the conformal cross-ratio $u \to 0$.  The first and dominant term on the left-hand side comes from the exchange of the $1$ operator, and corresponds to `free propagation' or 2-point Wick contraction.  The other terms indicate the exchange of low-twist operators, such as the energy-momentum tensor. }
 \label{fig:BootstrapPictureLargeSpin} 
\end{center}
\end{figure}

We consider the bootstrap equation in a particular kinematic limit, the lightcone OPE limit, such that the left-hand side of the equation has a manifest singularity, as pictured in figure \ref{fig:BootstrapPictureLargeSpin}.   This singularity simply arises from the disconnected correlator, and would correspond to free propagation in a scattering amplitude.  The lightcone OPE singularity must be reproduced by the other side of the bootstrap equation, but this can arise only from an infinite sum of conformal blocks, with very specific scaling behavior.  Since conformal blocks correspond to the exchange of definite states in the theory, this analysis has far-reaching implications for the structure of the Hilbert space and the spectrum of the Dilatation operator.

We will provide a brief review of the arguments in \cite{Fitzpatrick:2012yx}, which specifically studied the case correlators involving a single primary operator $\phi$. We will give the argument for the case of two distinct scalar primaries $\phi_1$ and $\phi_2$, but the core of the analysis will remain the same, such that interested readers may consult \cite{Fitzpatrick:2012yx} for details and for a more rigorous proof.

\subsection{Bootstrap recap}
\label{sec:bootstraprecap}

In a CFT, the product of any two local operators can be rewritten using the operator product expansion (OPE), which is a sum over all primary operators in the theory,
\be
\phi_1(x) \phi_2(0) = \sum_\CO \lambda_{12\CO} C_{12\CO} (x,\partial) \CO(0) \textrm{.}
\ee
The function $C_{12\CO}(x,\partial)$ corresponds to the contribution of all operators in the conformal multiplet associated with the primary operator $\CO$, and its structure is completely fixed by conformal invariance. The OPE coefficients $\lambda_{12\CO}$ are theory-dependent and undetermined by conformal invariance. The OPE can be used within a four-point correlation function, rewriting the correlator as a sum over the exchange of irreducible representations of the conformal group. The contribution of each representation, associated with the primary operator $\CO$ of dimension $\Delta$ and spin $\ell$, is referred as the conformal block $g_{\tau,\ell}(u,v)$, where $\tau = \Delta - \ell$ is the twist of $\CO$ and the conformally-invariant cross-ratios $u$ and $v$ are defined as
\be
u = \left( \frac{x_{12} x_{34}}{x_{24} x_{13}} \right)^2 \textrm{ , } v = \left( \frac{x_{14} x_{23}}{x_{24} x_{13}} \right)^2 \textrm{,}
\ee
{\nin}with $x_{ij} = x_i - x_j$. In terms of these conformal blocks, the four-point correlator takes the form
\be
\< \phi_1(x_1) \phi_2(x_2) \phi_3(x_3) \phi_4(x_4) \> = \frac{1}{x_{12}^{\Delta_1+\Delta_2} x_{34}^{\Delta_3+\Delta_4}} \left( \frac{x_{24}}{x_{14}} \right)^{\Delta_{12}} \left( \frac{x_{14}}{x_{13}} \right)^{\Delta_{34}} \sum_{\tau,\ell} P_{\tau,\ell} \, g_{\tau,\ell} (u,v) \textrm{,} \nn \\
\ee 
{\nin}where $\Delta_{ij} = \Delta_i - \Delta_j$ and the conformal block coefficient $P_{\tau,\ell}$ is proportional to the product of OPE coefficients $\lambda_{12\CO} \lambda_{34\CO}$.

In this expansion, we specifically took the OPE of the products $\phi_1 \phi_2$ and $\phi_3 \phi_4$. However, we could have instead taken the OPE of $\phi_1 \phi_4$ and $\phi_2 \phi_3$. The conformal bootstrap equation is simply the statement that these two different expansions, or channels, give the same correlator
\be \label{eq:BasicBootstrapEquation}
\frac{1}{x_{12}^{2 \Delta_1} x_{34}^{2\Delta_2}} \sum_{\tau,\ell} P_{\tau,\ell} \, g_{\tau,\ell} (u,v) = \frac{1}{(x_{14} x_{23})^{\Delta_1 + \Delta_2}}  \left( \frac{x_{24}}{x_{12}} \right)^{\Delta_{12}} \left( \frac{x_{13}}{x_{12}} \right)^{\Delta_{12}} \sum_{\tau,\ell} P_{\tau,\ell} \, g_{\tau,\ell} (v,u) \textrm{,} \nn \\
\ee
where we have taken the first two operators to be $\phi_1$ with dimension $\Delta_1$, and the latter two to be $\phi_2$ with dimension $\Delta_2$, as this will be the case we examine below. The bootstrap equation for $\langle \phi_1 \phi_1 \phi_2 \phi_2 \rangle$ provides a strong constraint on the spectrum and OPE coefficients of the CFT.

\subsection{The bootstrap in generalized free theories}
\label{sec:BootstrapGFT}

As a simple but far-reaching example of our bootstrap argument, we will consider four-point correlation functions in a generalized free theory (GFT), where all correlators are determined by 2-point Wick contractions.  GFTs can also be defined as the dual correlators derived from free quantum field theories in AdS.
In the case where we consider two different operators $\phi_1$ and $\phi_2$ we simply have
\be
\< \phi_1(x_1) \phi_1(x_2) \phi_2(x_3) \phi_2(x_4) \> = \frac{1}{x_{12}^{2\Delta_1} x_{34}^{2\Delta_2}} \textrm{.}
\ee
We can also express this correlator as an expansion in conformal blocks. This calculation is trivial in the `s-channel', as the only contribution in the series is from the identity,
\be
\frac{1}{x_{12}^{2\Delta_1} x_{34}^{2\Delta_2}} \sum_{\tau,\ell} P^{(11,22)}_{\tau,\ell} \, g^{(11,22)}_{\tau,\ell} (u,v) = \frac{1}{x_{12}^{2\Delta_1} x_{34}^{2\Delta_2}} \textrm{,}
\ee
{\nin}where the superscripts for $P_{\tau,\ell}$ and $g_{\tau,\ell}$ simply indicate that this channel corresponds to the OPE of $\phi_1 \phi_1$ and $\phi_2 \phi_2$. However, the expansion in the `t-channel' takes a very different form, setting up the non-trivial equality of equation (\ref{eq:BasicBootstrapEquation}), which we can write as
\be
u^{-\half (\Delta_1+\Delta_2)} = v^{-\half(\Delta_1+\Delta_2)} u^{-\half \Delta_{12}} \sum_{\tau,\ell} P^{(12,12)}_{\tau,\ell} \, g^{(12,12)}_{\tau,\ell} (v,u) \textrm{.}
\label{eq:MFTbootstrap}
\ee
If we consider this expression in the limit $u \ll v \ll 1$, we see that the left side contains a very specific power-law singularity $u^{-\half (\Delta_1+\Delta_2)}$. This singularity must be reproduced by the right side, and our focus will be on precisely \textit{how}  it is reproduced.

Since we are considering a GFT, the only primary operators appearing in this conformal block expansion are the operators $[\phi_1 \phi_2]_{n,\ell}$, which schematically take the form
\be
[\phi_1 \phi_2]_{n,\ell} \sim \phi_1 \partial^{2n} \partial_{\mu_1} \cdots \partial_{\mu_\ell} \phi_2 \textrm{,}
\ee
{\nin}with fixed twist $\tau_n = \Delta_1 + \Delta_2 + 2n$. As discussed in \cite{Fitzpatrick:2012yx}, the corresponding conformal blocks $g_{\tau_n,\ell}(v,u)$ are known exactly and possess at most a logarithmic divergence in the limit $u \ra 0$. For the bootstrap equation to be satisfied, the full sum over the t-channel conformal blocks must not converge uniformly in $u$ and $v$.

In order to understand this series, we need to study the conformal blocks at large $\ell$, in the limit $u \ll v \ll 1$.  In fact, we need the specific limit $\ell \to \infty$ with $\ell \sqrt{u}$ fixed.  As shown in appendix \ref{app:ConformalBlocks}, in this limit the conformal blocks at fixed $\tau$ take the approximate form
\be
g_{\tau,\ell}(v,u) \approx 2^{\tau+2 \ell} v^{\frac{\tau}{2}} u^{\half \Delta_{12}} \sqrt{\frac{\ell}{\pi}} \, K_{\Delta_{12}}(2 \ell \sqrt{u}) \textrm{,}
\ee
{\nin}where $K_x(y)$ is a modified Bessel function.  We see that at small $v$  the lowest twist terms ($n=0$) will dominate. In addition, the universal prefactor of $u^{-\half \Delta_{12}}$ in eq.\ (\ref{eq:MFTbootstrap}) will cancel with a corresponding term from each conformal block, such that the only remaining $u$-dependence arises from the Bessel function.

Let us now consider the conformal block coefficients $P_{\tau_n,\ell}$, specifically for the minimal twist operators. As shown in \cite{Unitarity}, these coefficients can be calculated precisely in a generalized free theory, and for $n=0$ take the form
\be
P_{\tau_0,\ell} = \frac{(\Delta_1)_\ell (\Delta_2)_\ell}{\ell! (\Delta_1 + \Delta_2 + \ell - 1)_\ell} \textrm{,}
\ee
{\nin}where $(q)_x = \frac{\Gamma(q+x)}{\Gamma(q)}$ is the rising Pochhammer symbol. In the large $\ell$ limit, these coefficients take the approximate form
\be
P_{\tau_0,\ell} \approx \frac{4 \sqrt{\pi}}{\Gamma(\Delta_1) \Gamma(\Delta_2) 2^{\tau_0+2\ell}} \ell^{\Delta_1 + \Delta_2 - \frac{3}{2}} \textrm{.}
\ee
Combining these results, the sum of large $\ell$ conformal blocks can be approximated as
\be
v^{-\half(\Delta_1+\Delta_2)} u^{-\half \Delta_{12}} \sum_{\tau_n, \textrm{large } \ell} P_{\tau_n,\ell} \, g_{\tau_n,\ell}(v,u) \approx \frac{4}{\Gamma(\Delta_1) \Gamma(\Delta_2) } \sum_{\textrm{large } \ell} \ell^{\Delta_1 + \Delta_2 - 1} K_{\Delta_{12}}(2 \ell \sqrt{u}) \textrm{.} \nn \\
\ee
This sum over large $\ell$ can be further approximated as an integral, which we can write as
\be
\sum_{\textrm{large } \ell} \ell^{\Delta_1 + \Delta_2 - 1} K_{\Delta_{12}}(2\ell\sqrt{u}) \approx u^{-\half (\Delta_1 + \Delta_2)} \int d \ell \, \ell^{(\Delta_1 + \Delta_2 - 1)} K_{\Delta_{12}}(2 \ell) \textrm{,}
\ee
{\nin}where we are specifically considering the limit of large $\ell$ at fixed $\ell \sqrt{u}$. As we can see, the large $\ell$ conformal blocks perfectly replicate the $u \ra 0$ behavior present on the left side of eq.\ (\ref{eq:MFTbootstrap}). 

The takeaway lesson  from this discussion is that the full sum of large $\ell$ conformal blocks contains a singularity in $u$ that is not present in any individual term.  This singularity was required by the bootstrap equation, and it is simply the result of a 2-point Wick contraction, also known as the exchange of the identity operator, or `free propagation'.

\subsection{Lightcone OPE limit and cluster decomposition}

Let us now study the existence and properties of large $\ell$ operators in any CFT$_{\geq 3}$. Separating the contribution of the identity operator, the bootstrap equation can be written as
\be
u^{-\half (\Delta_1+\Delta_2)} \left( 1 + \sum_{\tau,\ell} P^{(11,22)}_{\tau,\ell} \, u^{\frac{\tau}{2}} f^{(11,22)}_{\tau,\ell} (u,v) \right) = v^{-\half(\Delta_1+\Delta_2)} \sum_{\tau,\ell} P^{(12,12)}_{\tau,\ell} \, v^{\frac{\tau}{2}} f^{(12,12)}_{\tau,\ell} (v,u) \textrm{,} \nn \\
\label{eq:CFTbootstrap}
\ee
{\nin}where we have rewritten the conformal blocks as $g^{(ij,pq)}_{\tau,\ell} (u,v) = u^{\frac{\tau}{2}} v^{\half \Delta_{ij}} f^{(ij,pq)}_{\tau,\ell} (u,v)$ to highlight their behavior at small $u,v$.

For $d \geq 3$, unitarity separates the twist of the identity from that of all other operators, placing the bounds
\be
\tau \geq \left\{ \begin{matrix} \frac{d-2}{2} & (\ell = 0) \textrm{,} \\ d-2 & (\ell \geq 1) \textrm{.} \end{matrix} \right.
\ee
With these bounds in mind, we can see that the identity provides the dominant contribution to the left side of eq.\ (\ref{eq:CFTbootstrap}) in the limit $u \ra 0$. In fact, in this limit the left side of the bootstrap equation for any CFT is approximately the same as in GFT. Our arguments will again hinge on the simple statement that the right side must reproduce this contribution from the identity in the limit $u \ll v \ll 1$. This statement can be written as the approximate constraint
\be
1 \approx \left( \frac{u}{v} \right)^{\half(\Delta_1+\Delta_2)} \sum_{\tau,\ell} P_{\tau,\ell} \, v^{\frac{\tau}{2}} f_{\tau,\ell} (v,u) \qquad (u \ra 0) \textrm{,}
\label{eq:approxBootstrap}
\ee
{\nin}where we have suppressed the superscripts on $P_{\tau,\ell}$ and $f_{\tau,\ell}(v,u)$, as we will only consider conformal blocks in the t-channel for the remainder of this discussion.

We can clearly see that the $u,v$-dependence of the right side of eq.\ (\ref{eq:approxBootstrap}) must vanish in the appropriate limit. Just as in the case of GFT, the $u$-dependence cannot be reproduced by any individual conformal block, so it must come from the full infinite sum.  We again need to consider the large $\ell$ portion of this expression, as demonstrated in \cite{Fitzpatrick:2012yx}.

As discussed in appendix \ref{app:ConformalBlocks}, the large $\ell$ conformal blocks in the limit $u \ll 1$ on the right-hand side of equation (\ref{eq:approxBootstrap}) can be approximated as
\be
g_{\tau,\ell}(v,u) \approx v^{\frac{\tau}{2}} k'_{2\ell} (1-z) \, F^{(d)}(\tau,v) \textrm{,}
\ee
{\nin}where $k'_{2\beta}(x) = x^\beta \phantom{}_2 F_1 (\beta-\half \Delta_{12},\beta-\half \Delta_{12};2\beta;x)$, $z$ is defined by $u = z \bar{z}, v = (1-z)(1-\bar{z})$, and the $d$-dependent function $F^{(d)}(\tau,v)$ is positive and analytic near $v=0$, though its exact form will be unimportant for our discussion. Note that the limit $z \ra 0$ at fixed $\bar{z}$ is equivalent to $u \ra 0$ at fixed $v$. For the remainder of this section, we will be using $z$ rather than $u$, as this greatly simplifies the discussion.

Note that in this limit the $z,\ell$-dependence of the conformal blocks factorizes from the $v,\tau$-dependence, such that we may consider the cancellation of each piece separately. Since the conformal blocks are completely theory-independent, the function $k'_{2\ell} (1-z)$ takes the same approximate form as in GFT. The total sum over $\ell$ must then produce the divergence of $z^{-\half(\Delta_1 + \Delta_2)}$ necessary to cancel the prefactor in eq.\ (\ref{eq:approxBootstrap}).

What about the $v$-dependence? As mentioned above, the function $F^{(d)}(\tau,v)$ approaches a finite positive value as $v \ra 0$. In this limit, the $v$-dependence of each large $\ell$ conformal block is approximately $v^{\frac{\tau}{2}}$. Since this dependence must cancel the prefactor of $v^{-\half(\Delta_1 + \Delta_2)}$, we can obtain a bound on the possible twists that dominate in the large $\ell$ sum. 
While this is already a powerful restriction, we can make a much stronger statement. In order to reproduce the left side of eq.\ (\ref{eq:approxBootstrap}), there must be a contribution from an infinite number of operators of increasing spin with $\tau \ra \Delta_1 + \Delta_2$ as $\ell \ra\infty$. The constraint on the twists comes from the need to cancel the $v$-dependence, while the requirement for an infinite tower of these operators comes from the need to cancel the $z$-dependence.

We can actually take this argument one step further. Consider the conformal block associated with any primary operator in this infinite tower of operators with $\tau \approx \Delta_1 + \Delta_2$. We can then expand this conformal block as a series in $v$,
\be
g_{\tau,\ell}(v,u) \approx v^{\frac{\tau}{2}} k'_{2\ell} (1-z) F^{(d)}(\tau,0) + O(v^{\frac{\tau}{2}+1}) \textrm{.}
\ee
However, to obtain eq.\ (\ref{eq:approxBootstrap}) we only had to take the $z \ra 0$ limit, which means that this equality must hold to all orders in $v$. There must then be \textit{another} conformal block which cancels the $O(v^{\frac{\tau}{2}+1})$ term. More specifically, this additional conformal block must correspond to an operator with twist $\tau' = \tau + 2 \approx \Delta_1 + \Delta_2 + 2$. We can continue this process at every level in this power series, each time requiring the existence of a new operator with twist $\tau_n \approx \Delta_1 + \Delta_2 + 2n$ to cancel the other $O(v^{\frac{\tau}{2}+n})$ terms.

This argument applies to \textit{every} operator in our infinite tower at $\tau \approx \Delta_1 + \Delta_2$. This tells us that, for each non-negative integer $n$, the large $\ell$ spectrum of any CFT must include an infinite tower of operators with twists approaching $\tau_n = \Delta_1 + \Delta_2 + 2n$.  We refer the reader to \cite{Fitzpatrick:2012yx} for a mathematically rigorous version of these arguments.

\subsection{Anomalous dimensions and long-range forces in AdS}
\label{sec:CoulombPotential}

In this section we will explain how subleading corrections to the $u \to 0$ lightcone OPE limit of the bootstrap equation make it possible to constrain the anomalous dimensions and OPE coefficients of the operators $[\phi_1 \phi_2]_{n,\ell}$.  This means that we can use the bootstrap to derive the effects of long-range forces in AdS$_{\geq 4}$.  The universal exchange of $T_{\mu \nu}$ in the bootstrap leads to the universal long-range gravitational potential in AdS.

So far we have  considered only the dominant s-channel behavior due to the identity. However, we can extend our argument by considering the subleading contributions of conformal blocks associated with the CFT's minimal nonzero twist $\tau_m$. In the limit of small $u$, these minimal twist operators provide a correction to the left side of the bootstrap equation,
\be
1 + \sum_{\ell_m = 0}^2 P_m^{(11,22)} u^{\fr{\tau_m}{2}} f^{(11,22)}_{\tau_m,\ell_m}(u,v) \approx \left( \frac{u}{v} \right)^{\half(\Delta_1+\Delta_2)} \sum_{\tau,\ell} P^{(12,12)}_{\tau,\ell} \, v^{\frac{\tau}{2}} f^{(12,12)}_{\tau,\ell} (v,u) \qquad (u \ra 0) \textrm{.} \nn \\
\label{eq:subleadBootstrap}
\ee
We have limited this sum to $\ell_m \leq 2$, because higher spin operators either possess twist greater than that of the energy-momentum tensor or couple \cite{Maldacena:2011jn,Maldacena:2012sf} as in a free field theory.   

However, it is worth emphasizing that the $u \to 0$ contribution of all operators on the left-hand side with $\tau < \Delta_1 + \Delta_2$ 
 must be matched by the large $\ell$ sum on the right-hand side.  This follows because these operators on the left-hand side of the bootstrap equation create a power-law singularity in $u$, while any finite sum of conformal blocks on the right-hand side can only produce a logarithmic singularity in $u$.  This means that  one could use the bootstrap to compute the OPE coefficients and  dimensions of the $[\phi_1 \phi_2]_{n,\ell}$ operators contributing on the right-hand side to  $\CO \! \left(\frac{1}{\ell^{\Delta_1 + \Delta_2}} \right)$ in the large $\ell$ limit.  

In the limit $u \ll 1$, the minimal twist conformal blocks can be written as \cite{Dolan:2011dv}
\be
g_{\tau_m,\ell}(u,v) \approx u^{\fr{\tau_m}{2}} (1-v)^{\ell_m} \phantom{}_2 F_1 \left( \fr{\tau_m}{2}+\ell_m, \fr{\tau_m}{2}+\ell_m;\tau_m+2\ell_m;1-v \right) \textrm{.}
\ee
As we can see, these blocks factorize into a $u$-dependent piece, with simple scaling behavior, and a $v$-dependent piece, which can be expanded in a power series at small $v$ by using the relation
\be
\phantom{}_2 F_1 (\beta,\beta;2\beta;1-v) = \fr{\G(2\beta)}{\G^2(\beta)} \sum_{n=0}^\oo \left( \fr{(\beta)_n}{n!} \right)^2 v^n \Big( 2 \left( \psi(n+1) - \psi(\beta) \right) - \ln v \Big) \textrm{,}
\ee
{\nin}where $\psi(x) = \fr{\G'(x)}{\G(x)}$ is the digamma function. The precise form of this expansion is largely irrelevant to our discussion. All that matters to us is the presence of logarithmic terms of the form $v^n \ln v$. Since eq.\ (\ref{eq:subleadBootstrap}) is true to all orders in $v$, these terms must again be replicated by the t-channel conformal blocks.

To see how these logarithmic terms are reproduced by the right side of equation (\ref{eq:subleadBootstrap}), we shall consider the situation where one of the special twist values $\tau_n = \Delta_1 + \Delta_2 + 2n$ is approached by a single tower of operators $\CO_{\tau_n,\ell}$ which at large $\ell$ are separated by a twist gap from all other operators in the spectrum. For simplicity, we will specifically consider the case where there is one operator accumulating near $\tau_n$ for each $\ell$, and the corresponding conformal block coefficients approach those of GFT. However, this approach can be generalized to more complicated scenarios \cite{Fitzpatrick:2012yx}.

Generically, the twists $\tau(n,\ell)$ for this tower of operators will not be precisely $\tau_n$. Instead, they will be shifted by some anomalous dimension $\g(n,\ell) = \tau(n,\ell) - (\Delta_1 + \Delta_2 + 2n)$. For sufficiently large $\ell$, we can expand the associated conformal blocks in terms of the anomalous dimension to obtain the approximate form
\be
g^{(12,12)}_{\tau_n + \g,\ell}(v,u) \approx v^{\frac{\tau_n}{2}} \left( 1 + \fr{\g(n,\ell)}{2} \ln v \right) k'_{2\ell} (1-z) \, F^{(d)}(\tau_n,v) \textrm{.}
\ee
We see that the logarithmic terms due to minimal twist operators in the s-channel are replicated by the anomalous dimensions of large $\ell$ operators in the t-channel. By matching both sides of the bootstrap equation, we can then constrain the form of $\g(n,\ell)$.

While it is clear that we can match the $v$-dependence of both sides, we still need to consider the $z$-dependence. As we can see in eq.\ (\ref{eq:subleadBootstrap}), the right side must not only cancel the original factor of $z^{\half(\Delta_1+\Delta_2)}$, it must produce an additional factor of $z^{\fr{\tau_m}{2}}$. Just like in GFT, we simply need to consider the contribution of conformal blocks at large $\ell$. Focusing on only the relevant terms, we need the approximate relationship
\be
z^{\fr{\tau_m}{2} - \half(\Delta_1+\Delta_2)} \sim \sum_{\textrm{large } \ell} \g(n,\ell) \ell^{\Delta_1 + \Delta_2 - 1} K_{\Delta_{12}}(2\ell \sqrt{z}) \textrm{.}
\ee

Since we are considering the $\ell \ra \oo$ limit, we can approximate the anomalous dimension with its leading $\ell$ dependence $\g(n,\ell) \approx \gamma_n \ell^a$, such that we obtain

\be
\sum_{\textrm{large } \ell} \gamma_n \ell^{a + \Delta_1 + \Delta_2 - 1} K_{\Delta_{12}}(2\ell \sqrt{z}) \approx \gamma_n \, z^{-\half( a + \Delta_1 + \Delta_2)} \int d\ell \, \ell^{a + \Delta_1 + \Delta_2 - 1} K_{\Delta_{12}}(2\ell) \textrm{.}
\ee
Matching this to the left side of the bootstrap, we see that $a = -\tau_m$, such that the anomalous dimension takes the asymptotic form
\be
\gamma(n,\ell) \approx \fr{\gamma_n}{\ell^{\tau_m}} \qquad (\ell \ra \oo) \textrm{,}
\ee
{\nin}where the $\ell$-independent coefficient $\gamma_n$ can be determined by carefully matching the $v^n \ln v$ terms on both sides.

As a simple example, consider the case of stress-energy tensor exchange in $d=4$, which has $\tau_m = \ell_m = 2$ and $P_m = \fr{\De_1\De_2}{360c}$. Matching all terms proportional to $\ln v$, we then obtain the approximate relation
\be
- \fr{u\De_1\De_2}{6c} \left( \fr{1+4v+v^2}{(1-v)^3} \right) \approx \left( \fr{u}{v} \right)^{\half(\De_1+\De_2)} u^{-\half \De_{12}} \sum_{n,\ell} P_{\tau_n,\ell} \, v^{\frac{\tau_n}{2}} \fr{\g_n}{\ell^2} k'_{2\ell} (1-z) \, F^{(d=4)}(\tau_n,v) \textrm{.} \nn \\
\ee
Note that the conformal block coefficients $P_{\tau_n,\ell}$ are approximately those of GFT, which in the limit $\De_1,n \ll \De_2 \ll \ell$ take the form
\be
P_{\tau_n,\ell} \approx \fr{(\De_1)_n}{n!2^{2n}} P_{\tau_0,\ell} \textrm{.}
\ee
As every term is proportional to $P_{\tau_0,\ell}$, we can evaluate the sum over $\ell$ to cancel the $z$-dependence of both sides, yielding the relation
\be
- \fr{\De_1(\De_1-1)\De_2^2}{6c} \left( \fr{1+4v+v^2}{(1-v)^2} \right) \approx (1-v)^{\De_2} \sum_n \fr{(\De_1)_n}{n!2^{\tau_n}} \g_n v^n F^{(d=4)}(\tau_n,v) \textrm{.}
\ee
In this particular limit, we can also apply the results of appendix \ref{app:ConformalBlocks} to the $d=4$ conformal blocks derived in \cite{Dolan:2003hv} to obtain the approximation
\be
F^{(d=4)}(\tau_n,v) \approx 2^{\tau_n} (1-v)^{\De_{12}-1} \textrm{.}
\ee
Using this result, we then have the simplified expression
\be
- \fr{\De_1(\De_1-1)\De_2^2}{6c} \left( \fr{1+4v+v^2}{(1-v)^{\De_1+1}} \right) \approx \sum_n \fr{(\De_1)_n}{n!} \g_n v^n \textrm{.}
\ee
If we expand the left side as a series in $v$, we can then match corresponding terms from the two series to determine the anomalous dimension coefficients $\gamma_n$. For the terms with $n \ll \De_1$, this takes the simple form $\gamma_n \approx -\fr{(\De_1\De_2)^2}{6c}$, which matches precisely with the AdS gravity computation that produced equation (\ref{eq:GravityPredictionAnomalousDimension}) in the $\Delta_1, \Delta_2 \gg 1$ limit.  This has a nice physical interpretation in terms of the picture of section \ref{sec:NewtonianGravityHigherD}:  when $n \ll \Delta_1, \Delta_2 \ll \ell$ the variation of $n$ does not significantly alter the distance between objects $1$ and $2$ in AdS, and so $\gamma_n$, which corresponds to the gravitational binding energy, is independent of $n$.

This same approach can be applied to theories with an arbitrary number of minimal twist primary operators. For example, the general $n=0$ anomalous dimension coefficient is
\be
\g_0 \approx -\fr{2\G(\De_1)\G(\De_2)}{\G(\De_1-\fr{\tau_m}{2})\G(\De_2-\fr{\tau_m}{2})} \sum_{\ell_m} P_m \fr{\G(\tau_m+2 \ell_m)}{\G^2 (\fr{\tau_m}{2} + \ell_m) } \textrm{.}
\ee

Furthermore, as noted above, we could in principle use the existence of the singularity $u^{\half(\tau - \Delta_1 - \Delta_2)}$ on the left-hand side of equation (\ref{eq:subleadBootstrap}) to match the large $\ell$ anomalous dimensions and OPE coefficients to $\CO \! \left(\frac{1}{\ell^{\Delta_1 + \Delta_2}} \right)$ on the right-hand side.  For large values of $\Delta_1$ or $\Delta_2$ this could be extremely powerful.

\section{Virasoro blocks and the lightcone OPE limit}
\label{sec:VirasoroBootstrap}

We would like to generalize the bootstrap arguments from the previous section to the case of CFTs in $d=2$, which possess an infinite-dimensional Virasoro symmetry. For $d \geq 3,$ our argument relied on the fact that once a CFT correlator in the OPE limit is decomposed into conformal blocks, it can then be expanded in increasing powers of $u$, beginning with the identity contribution,
\be
\langle \phi_1 \phi_1 \phi_2 \phi_2 \rangle = u^{-\half(\De_1+\De_2)} + \sum_{\tau, \ell} P_{\tau, \ell} \, u^{\half(\tau - \De_1 - \De_2)} f_{\tau, \ell}(u, v) \textrm{.}
\ee
Two features were crucial for the analysis -- firstly that $\tau \geq \frac{d}{2} -1 > 0$, so that the identity was clearly separated from the contributions of other operators, and secondly, that there were only a finite number of conformal block contributions at the minimum twist $\tau_m > 0$.  Neither of these properties holds in the case of 2d CFTs.  So it is not surprising that many 2d CFTs, including the 2d Ising model \cite{Fitzpatrick:2012yx}, violate the conclusions of the theorem we proved for $d \geq 3$.  In fact we saw in sections \ref{sec:DeficitAngles} and \ref{sec:BTZ} that explicit AdS$_3$ calculations provide different expectations for the large spin spectrum in CFT$_2$. 

We will overcome the aforementioned hurdles by computing the Virasoro conformal blocks in various semi-classical limits and then using them in a more general lightcone OPE bootstrap analysis.  Due to the technical nature of the computation of the blocks themselves, we have confined these calculations to the appendices, with the general method described in appendix \ref{app:MonodromyMethod} and the specific computations in appendix \ref{app:ComputingVirasoroViaMonodromy} and \ref{app:TChannelBlocks}.  We also provide a more straightforward brute force computation in a more restricted limit in appendix \ref{app:Direct}.  With the blocks in hand, the bootstrap analysis proceeds along the same line of reasoning that we saw in section \ref{sec:BootstrapHigherD}, although with qualitatively different conclusions.

\begin{figure}[t!]
\begin{center}
\includegraphics[width=0.95\textwidth]{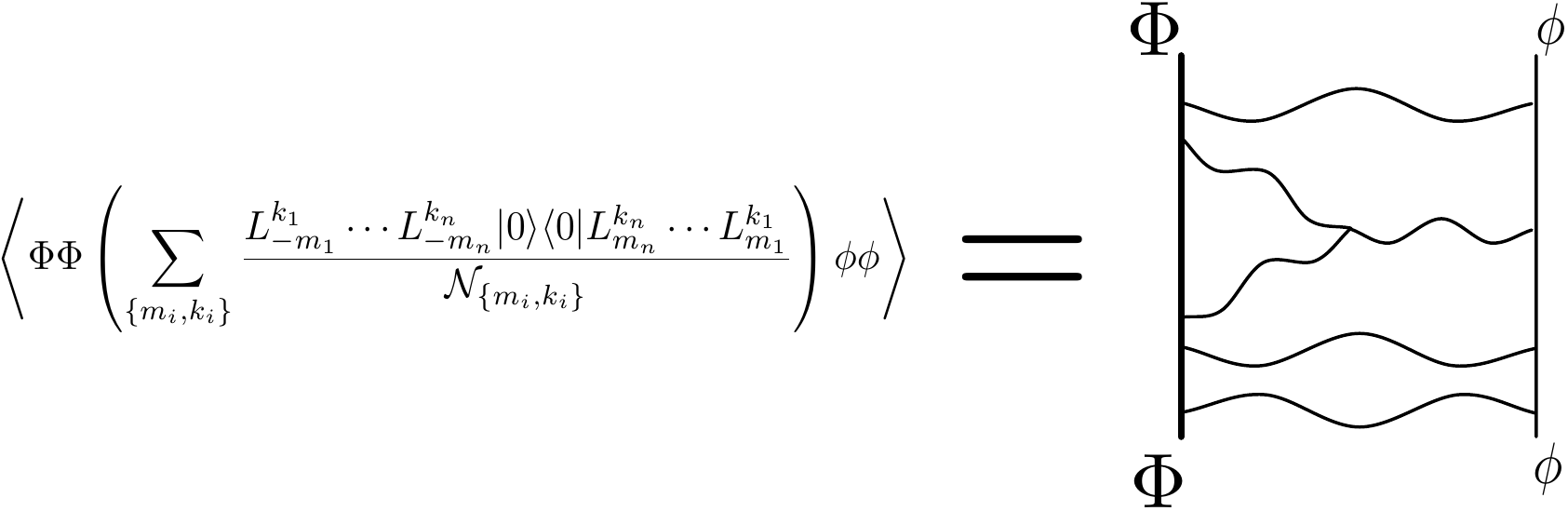}
\caption{ This figure suggests how the exchange of all descendants of the identity operator in the Virasoro algebra  corresponds to the exchange of all multi-graviton states in AdS$_3$.  This is sufficient to build the full, non-perturbative AdS$_3$ gravitational field entirely from the CFT$_2$.
}
 \label{fig:VirasoroDescendantsAsGravitons} 
\end{center}
\end{figure}

Let us now briefly discuss the bootstrap equation in CFT$_2$.  In $d=2$ we can make use of holomorphic factorization to discuss operators of general spin; nevertheless we will mostly discuss scalar external operators for simplicity and uniformity with section \ref{sec:BootstrapHigherD}.   Correlators of local operators in CFT$_2$ can be expanded in Virasoro conformal blocks corresponding to the exchange of irreducible representations of the Virasoro group.  Each of these Virasoro blocks (or Virasoro partial waves) is associated with a primary operator $\CO_{h,\bar{h}}$ of scaling dimension $\Delta = h + \bar{h}$ and spin $\ell = |h - \bar{h}|$.  The Virasoro block decomposition of a four-point correlation function takes a very similar form to the conformal block expansion in $d \geq 3$, namely
\be
\< \phi_1(x_1) \phi_2(x_2) \phi_3(x_3) \phi_4(x_4) \> = \frac{1}{x_{12}^{\Delta_1+\Delta_2} x_{34}^{\Delta_3+\Delta_4}} \left( \frac{x_{24}}{x_{14}} \right)^{\Delta_{12}} \left( \frac{x_{14}}{x_{13}} \right)^{\Delta_{34}} \sum_{h,\bar{h}} P_{h,\bar{h}} \, \CV_{h,\bar{h}} (u,v) \textrm{,} \nn \\
\ee 
where $P_{h,\bar{h}}$ is the set of theory-dependent Virasoro block coefficients and $\CV_{h,\bar{h}}$ are the Virasoro blocks. The bootstrap equation can then be written in terms of Virasoro blocks as
\be
\frac{1}{x_{12}^{2 \Delta_1} x_{34}^{2\Delta_2}} \sum_{h,\bar{h}} P_{h,\bar{h}} \, \CV_{h,\bar{h}} (u,v) = \frac{1}{(x_{14} x_{23})^{\Delta_1 + \Delta_2}}  \left( \frac{x_{24}}{x_{12}} \right)^{\Delta_{12}} \left( \frac{x_{13}}{x_{12}} \right)^{\Delta_{12}} \sum_{h,\bar{h}} P_{h,\bar{h}} \, \CV_{h,\bar{h}} (v,u) \textrm{,} \nn \\
\ee
where we are specifically considering the Virasoro block decomposition of a correlator with only two independent scalar operators $\phi_1,\phi_2$.  We have written the bootstrap equation in terms of $x_i$ and the cross-ratios $u$ and $v$ to make contact with section \ref{sec:BootstrapHigherD}, but it is often more natural to use variables $z$ and $\bar z$, with $u = z \bar z$ and $v = (1-z)(1- \bar z)$, since the full two-dimensional conformal group breaks up into holomorphic and anti-holomorphic Virasoro algebras.

\subsection{AdS$_3$ deficit angles from semi-classical Virasoro blocks}
\label{sec:deficitanglespectrum}

Factoring out the contribution due to the identity operator, we can rewrite the CFT$_2$ bootstrap equation as
\be
\CV_{0,0}(u,v) + \sum_{h,\bar{h}} P^{(11,22)}_{h,\bar{h}} \, \CV^{(11,22)}_{h,\bar{h}}(u,v) = \left( \fr{u}{v} \right)^{\half(\Delta_1 + \Delta_2)} u^{-\half \Delta_{12}} \sum_{h,\bar{h}} P^{(12,12)}_{h,\bar{h}} \, \CV^{(12,12)}_{h,\bar{h}}(v,u) \textrm{.} \nn \\
\label{eq:ViraBootstrap}
\ee
We can clearly see the first difference between 2d CFTs and those in higher dimensions. In our previous discussion, the contribution of the identity operator was simple, with no additional $u,v$-dependence. More concretely, there was no extended `conformal block' associated with the identity, but only a single, trivial operator. This is not the case in 2d, as we now have contributions from all of the descendants of the vacuum.  In terms of the global conformal symmetry, these descendants are simply the states that we obtain by acting with the stress-energy tensor on the CFT$_2$ vacuum.

As in section \ref{sec:BootstrapHigherD}, we are specifically interested in studying eq.\ (\ref{eq:ViraBootstrap}) in the lightcone OPE limit $u \ra 0$. In order to make their small $u$ behavior manifest, the Virasoro blocks can be rewritten as
\be
\CV_{h,\bar{h}}(u,v) = u^{\fr{\tau}{2}} \CF_{h,\bar{h}}(u,v) \textrm{.}
\ee
where $\mathcal{F}$ is analytic at small $u$.
The left side of the bootstrap equation will  be dominated by operators with zero twist. However, unitarity no longer forbids additional operators with $\tau = 0$.  The stress-energy tensor is an example, but it has already been included in the Virasoro identity block.
Any other local primary operator with  $h=0$ or $\bar{h}=0$, and therefore of zero twist, will be a conserved current. We will limit our discussion to theories with no additional continuous global symmetries, such that the only zero twist operators are contained within the identity Virasoro block. In the small $u$ limit, eq.\ (\ref{eq:ViraBootstrap}) can be written as
\be
\CV_{0,0}(u,v) \approx \left( \fr{u}{v} \right)^{\half(\Delta_1 + \Delta_2)} u^{-\half \Delta_{12}} \sum_{\tau,\ell} P_{\tau,\ell} \, g_{\tau,\ell}(v,u) \qquad (u \ra 0) \textrm{.}
\label{eq:approxViraBootstrap}
\ee
We have chosen to explicitly write the t-channel or right-hand side in terms of global conformal blocks.  We discuss the limitations of this approximation below, when it becomes relevant, with most calculations confined to appendix \ref{app:TChannelBlocks}.  

We want to study the behavior of $\CV_{0,0}$ in the $u \to 0$ limit. Unlike global conformal blocks, there is no general closed-form  expression for Virasoro blocks. However, as discussed in appendix \ref{app:ComputingVirasoroViaMonodromy}, the approximate structure of these blocks can be determined in the semi-classical limit where the CFT central charge $c \ra \oo$ with
\be
1 \ll \Delta_1 \ll c  \ \ \ \mathrm{and}  \ \ \ \ \fr{\Delta_2}{c} \ \ \mathrm{fixed.}
\ee
In the semi-classical limit, the $u \ra 0$ form of the identity block is approximately
\be
\CV_{0,0}(u,v) \approx \alpha^{\Delta_1} v^{-\half \Delta_1 (1-\alpha)} \left( \fr{1-v}{1-v^\alpha} \right)^{\Delta_1} \textrm{,}
\ee
where we have defined $\alpha \equiv \sqrt{1 - 12 \fr{\Delta_2}{c}}$. We have assumed that $\phi_i$ are scalar operators with $\Delta_i = h_i + \bar h_i = 2 h_i$, although it is easy to generalize to the case with $h_i \neq \bar h_i$ using the results of appendix \ref{app:ComputingVirasoroViaMonodromy} and holomorphic factorization.  The identity Virasoro block contains new $v$-dependence that arises due to the Virasoro descendants of the vacuum. However, we can also see that in this limit the left side of eq.\ (\ref{eq:approxViraBootstrap}) is completely independent of $u$, which tells us that the right side must also have no $u$-dependence. 

As in higher dimensions, it is impossible for any one conformal block to cancel the $u$-dependent prefactor.  One might wonder if this remains true in CFT$_2$, where the Virasoro blocks replace the simpler global blocks.  We argue in appendix \ref{app:TChannelBlocks} that it does. Specifically, at both $\ell \gg  \Delta_2, c$ and for $\ell \ll \Delta_2, c$ we find that the individual t-channel Virasoro blocks $\CV(v,u) $ do not  contain a sufficiently strong singularity as $u \to 0$ to reproduce the identity block in the t-channel.  Furthermore, there is a natural interpolation between the large and small $\ell$ behavior.   So although our approximations do not allow for a rigorous proof, we expect that there must be an infinite sum of large $\ell$ Virasoro blocks on the right-hand side of equation (\ref{eq:ViraBootstrap}) to reproduce the singularity as $u \to 0$.

Now let us study the bootstrap equation in the limit $u \ll v \ll 1$, using the global blocks on the right-hand side so that we can write
\be
1 \approx \alpha^{-\Delta_1} u^{\half(\Delta_1 + \Delta_2)} v^{-\half (\alpha \Delta_1 + \Delta_2)} u^{-\half \Delta_{12}} \sum_{\tau,\ell} P_{\tau,\ell} \, g_{\tau,\ell}(v,u) \textrm{.}
\ee
We are specifically interested in the large $\ell$ conformal blocks, which in this limit take the same approximate form as in higher dimensions,
\be
g_{\tau,\ell}(v,u) \approx 2^{\tau+2\ell} v^{\fr{\tau}{2}} u^{\half \Delta_{12}} \sqrt{\fr{\ell}{\pi}} K_{\Delta_{12}}(2\ell\sqrt{u}) \textrm{.}
\ee
We can easily see that this discussion will be very similar to our arguments from section \ref{sec:BootstrapHigherD}. The overall prefactor of $u^{-\half \Delta_{12}}$ will be cancelled by each individual block, but the necessary power of $u^{-\half(\Delta_1+\Delta_2)}$ can only be produced by an infinite tower of large $\ell$ conformal blocks. In \textit{every} 2d CFT with large $c$, there must then exist an infinite spectrum of large $\ell$ global conformal blocks, just as in higher dimensions.

However, things become much more interesting if we look at the $v$-dependence. In the small $v$ limit, the conformal blocks approximately scale as $v^{\fr{\tau}{2}}$. Since this $v$-dependence must cancel with the overall prefactor, we again obtain bounds on the possible twists which can dominate in the large $\ell$ sum.  More importantly, there must be an infinite tower of large $\ell$ operators with twist $\tau \ra \alpha \Delta_1 + \Delta_2$ as $\ell \ra \infty$.

This behavior is very different from that of CFTs in higher dimensions. In the large $\ell$ limit, we would naively expect the spectrum to approach that of GFT, with $\tau \approx \Delta_1 + \Delta_2$. Phrased in terms of AdS, we would expect the binding energy of two particles to vanish in the long-distance limit. Instead, we see the presence of a universal `anomalous dimension', or binding energy, which does not vanish in the large $\ell$ limit. As discussed in section \ref{sec:DeficitAngles}, this is precisely the behavior we would associate with a deficit angle in AdS, with the corresponding energy shift
\be
\Delta_1 \ra \Delta_1 \sqrt{1-12\fr{\Delta_2}{c}} = \Delta_1 \sqrt{1-8G_N M} \textrm{,}
\ee
where we have identified $\Delta_2 = M$ and $c = \fr{3}{2G_N}$.

Let us now extend our argument by considering the bootstrap at arbitrary $v$. We will find it convenient to use $v$ and $z$ as variables, instead of $v$ and $u$ as above.  As discussed in appendix \ref{app:ConformalBlocks}, conformal blocks with twist $\tau \sim \Delta_2 \gg \Delta_1$ can be approximated as
\be
g_{\tau,\ell}(v,u) \approx 2^{\tau+2\ell} v^{\fr{\tau}{2}} (1-v)^{\half \Delta_{12}} u^{\half \Delta_{12}} \sqrt{\fr{\ell}{\pi}} K_{\Delta_{12}}(2\ell\sqrt{z}) \textrm{,}
\ee
where we have made no assumptions about the size of $v$. Inserting this into the bootstrap equation and expanding the identity Virasoro block as a series in $v^\alpha$, we can obtain the relation
\be
v^{\half(\alpha \Delta_1 + \Delta_2)} \sum_{n=0}^\oo \fr{(\Delta_1)_n}{n!} v^{n\alpha} \approx \alpha^{-\Delta_1} z^{\half(\Delta_1+\Delta_2)} \sum_{\tau,\ell} P_{\tau,\ell} \, 2^{\tau+2\ell} v^{\fr{\tau}{2}} \sqrt{\fr{\ell}{\pi}} K_{\Delta_{12}}(2\ell\sqrt{z}) \textrm{.}
\ee
In order for the $v$-dependence of both sides to match, there must be at least one primary operator with approximate twist
\be
\tau_n \approx \alpha( \Delta_1 + 2n) + \Delta_2 \textrm{,}
\label{eq:BTZtwistspectrum}
\ee
for every non-negative integer $n$. In order for the $z$-dependence of both sides to also match, there must actually be an infinite tower of primary operators with increasing spin for \textit{every} twist $\tau_n$.  Note that since these twists $\tau_n$ have non-integer spacings, they must correspond to distinct Virasoro primaries.

We find that the large $\ell$ spectrum of any CFT with large central charge matches that of the operators $[\phi_1 \phi_2]_{n,\ell}$ in a generalized free theory, but with the rescalings
\be
\Delta \to \Delta \sqrt{ 1 - 8 G_N M} , \ \ \ \ n \to n \sqrt{ 1 - 8 G_N M} , \ \ \ \ \ell \to \ell .
\ee
As discussed in section \ref{sec:DeficitAngles}, this is precisely the spectrum associated with a probe orbiting a deficit angle in AdS$_3$. Using only the bootstrap equation for a 2d CFT, we have rediscovered the universal long-distance effect of gravity in AdS$_3$.

\subsection{BTZ quasi-normal modes from semi-classical Virasoro blocks}
\label{sec:BTZspectrum}

Now we will consider the spectrum of twists $\tau_n$ in the case where one of the external operators is above the BTZ mass threshold, {\it i.e.} $\De_2 > \frac{c}{12}$. In this case $\alpha$ is imaginary, so we will define $\beta \equiv \sqrt{ 12 \frac{\De_2}{c}-1} = - i \alpha$. For small $z$ with fixed $v$, the bootstrap equation now takes the form
\be
 \left( v^{-i \beta/2}- v^{i \beta/2} \right)^{-\Delta_1} \approx (i\beta)^{-\Delta_1} z^{\half(\Delta_1+\Delta_2)} \sum_{\tau,\ell} P_{\tau,\ell} \, 2^{\tau+2\ell} v^{\half(\tau-\Delta_2)} \sqrt{\fr{\ell}{\pi}} K_{\Delta_{12}}(2\ell\sqrt{z}) \textrm{.}
\ee
At large spin, an infinite sum over spins is still necessary in order to cancel the prefactor of $z^{\frac{1}{2}(\Delta_1 + \Delta_2)}$, and this completely constrains the large $\ell$ behavior of $P_{\tau,\ell}$. Thus, taking the $z \rightarrow 0 $ limit, we can simplify to
\be
 \left( \frac{v^{-i \beta/2}- v^{i \beta/2} }{2i} \right)^{-\Delta_1} \approx 
 \int d\tau \, \tilde{P}_{\tau} \,  v^{\half(\tau-\Delta_2)}  
 \label{eq:BTZdecomp},
\ee
where $\tilde{P}_\tau$ is the remaining $\ell$-independent piece of $P_{\tau,\ell}$, and we have replaced the sum on twists with an integral, without loss of generality.
To constrain the spectrum of twists, we can take $v = e^{-s}$ and perform an inverse Laplace transform of each side, obtaining
\be
 \tilde{P}_{\Delta_2 + 2\delta} &=& \int_{\gamma- i \infty}^{\gamma+i \infty} \frac{ds}{2\pi i} e^{s \delta} \frac{1}{\sin^{\Delta_1}(s \beta/2)}.
\ee
where we have rewritten the twists as $\tau = \De_2 + 2 \delta$. Taking $\gamma= \frac{\pi}{\beta}$ to avoid the poles in the denominator, this integral can be evaluated and one finds
\be
\tilde{P}_{\Delta_2 + 2\delta} &=& \frac{\Gamma(\frac{\Delta_1}{2} + i \frac{\delta}{\beta}) \Gamma(\frac{\Delta_1}{2} - i \frac{\delta}{\beta})}{2 \beta \Gamma(\Delta_1)}.
\label{eq:BTZspectraldecomp}
\ee
This is regular for all real $\delta$, and thus indicates that there is a dense spectrum of twists. This result is consistent with the fact that the separation between twists in the deficit angle spectrum from the previous section approaches zero as $\De_2 \ra \fr{c}{12}$, the BTZ threshold.

 To connect to the spectrum of quasi-normal modes for the BTZ black hole, we want to look not for eigenstates of the Dilatation operator, but rather for asymptotic `in' and `out' states. We thus need to Wick rotate $v = e^{-s} \rightarrow e^{-i s (1+ i \epsilon)}$. One can then read off the spectrum from the poles of (\ref{eq:BTZspectraldecomp}). In this case $v^{- i \beta} \ll v^{i \beta}$ at large $s$, so it is already manifest from a small $v^{-i \beta}$ expansion of (\ref{eq:BTZdecomp})  that the spectrum of twists is
 \be
 \tau_n \approx  i \beta (\Delta_1 + 2n) +  \Delta_2 =  2 \pi iT_{\rm BTZ} (\Delta_1 + 2n) + \Delta_2,
 \ee
 which is just the natural analytic continuation of (\ref{eq:BTZtwistspectrum}) to imaginary $\alpha = i \beta$.  This reproduces the spectrum of BTZ quasi-normal modes, as in \cite{CFTquasinormal}.  We emphasize that these are universal results for  the large $\ell$ spectrum of any CFT$_2$ with large $c$ and no twist zero Virasoro primaries aside from the identity.

\section{Discussion}

What does it mean for a bulk gravitational spacetime to emerge holographically from a CFT?  

One approach views the AdS geometry as an ever-present feature of a CFT state.  The Ryu-Takayanagi formula \cite{RT1,RT2} exemplifies this viewpoint beautifully, as it associates bulk geometry with entanglement entropy in the CFT, even in the unperturbed vacuum.  The disadvantage of this philosophy is its static nature, for it does not readily yield information about bulk dynamics, especially the locality of interactions in AdS.   The concept of a geometric distance between physical objects is important only because local interactions fall off with distance; without locality geometry loses much of its meaning.

In this paper we have taken a complementary approach, interpreting bulk geometry as a derivative idea, defining it purely in terms of the dynamics of localized objects.  From this point of view spacetime coordinates are simply a set of approximate, a posteriori labels that can be consistently applied to operators or states as they evolve with time.  The S-Matrix program in flat spacetime and the reconstruction of AdS effective field theory from CFT correlators exemplify this philosophy, and in both cases we have a host of information about the necessary and sufficient conditions on amplitudes for a local bulk theory. In this approach, one attempts to ``hear the shape'' of the geometry by looking at the spectrum of its excitations --
for example, in this paper we worked with energy and angular momentum eigenstates. These can be translated and combined to form local wavepackets in AdS, which in turn can then be used to probe the geometry in a more direct way.
In many cases these states and their local interpretation are already familiar, and our approach has the advantage of connecting geometry to AdS locality in an essential way.

Applying the CFT bootstrap to 2d large central charge theories has allowed us to derive general, non-perturbative results that are ripe for interpretation in terms of AdS$_3$ dynamics.    We saw that the exchange of the identity and its Virasoro descendants, which can be interpreted in AdS$_3$ as multi-graviton states, creates an effect identical to the presence of either a deficit angle or BTZ black hole background.  

Virasoro primaries with dimension $h_\Phi > \frac{c}{24}$ create a universal background in which ``light'' primaries with dimension $h_\phi \ll c$  have thermal correlators, as shown in equation (\ref{eq:BlockAsThermalCorrelator}).  This can be viewed as a derivation of the Eigenstate Thermalization Hypothesis \cite{ETH2, ETH} for CFT$_2$ at large central charge, although it is important to keep in mind that it will receive corrections from $1/c$ effects and, away from the light-cone limit, from other conformal blocks. 
The corrections from other operators could cancel,  since $\phi_1 \phi_1 \to \phi_2 \phi_2$ conformal block coefficients can have either sign, or alternatively the OPE coefficients of these operators might simply be small.  The suppression of these corrections may be related to both eigenstate thermalization and `no hair' theorems for black holes.

  It would be interesting to investigate this approximate thermality for more general correlators of light primaries in future work, by studying Virasoro conformal blocks for $n$-point correlators \cite{HartmanLargeC}. There should also be a nice confluence of the methods employed here with entanglement entropy methods: by taking the light operators to be ``twist'' operators at the edge of an interval, one might compute the entanglement entropy in the background of a ``heavy'' $h_\Phi> \frac{c}{24}$ operator and reconstruct the corresponding bulk geometry by using the Ryu-Takayanagi formula. It is also interesting to note the connection to results on the general instability of excitations of AdS$_3$ with energy above the BTZ threshold \cite{Bizon}. We have seen that the universal background created by any ``heavy'' operator with $h_\Phi > \frac{c}{24}$ produces a spectrum of modes with an instability even at large distances.  Our results are limited to the case where the excitations are in the ``test mass'' limit, meaning they do not back react on the geometry, but further results on the identity operator conformal block would allow one to generalize beyond this case. 

  Formally, the lightcone OPE limit gives reliable information only about the primaries of large angular momentum $\ell$, but in the case where there is a large gap in twists between the identity operator and the remaining primary operators in the theory, we have obtained results that accord with the BTZ geometry for all values of $\ell$.  This indicates that in AdS$_3$, all states above the BTZ threshold look like black holes, up to corrections embodied by the exchange of higher twist operators.  It suggests that up to the horizon, all black hole states look nearly identical, arguing against any proposal for quantum gravity that would lead to large non-local modifications of the dynamics outside the horizon.  

It would be interesting to sharpen these claims for 2d CFTs with a small number of low twist $\tau \sim \CO(1)$ primaries, and more generally to explore the corrections from primaries with twist $\tau >0$ in the bootstrap equation.  These contributions will have a sub-dominant impact on the large spin operators that we have identified, but they would make it possible to estimate the range of interaction length scales in AdS$_3$.  In particular, one might try to control the behavior in the vicinity of a black hole.  It is also possible to study CFT$_2$ with $\tau=0$ operators besides the identity, namely conserved currents.  In that case one would need to include the contributions of the entire zero twist sector at once, including operators of higher spin, perhaps via a generalization of the monodromy method.  One could also study the identity block in theories with a more general $\mathcal{W}_N$ algebra structure \cite{Zamolodchikov:1985wn,Bouwknegt:1992wg}.
  It would be particularly interesting to see if the $\mathcal{W}_N$ blocks can be interpreted as thermal correlators, since it might shed light on whether the AdS duals of these theories have black hole-like states \cite{Gutperle:2011kf,deBoer:2013gz}.

The semi-classical approximation to the Virasoro identity block contains all the information we need to reconstruct a dynamical AdS$_3$ geometry.  However, it would be fascinating to explore the corrections to thermality embedded in the (unknown) exact formula for the Virasoro blocks, by going beyond the semi-classical approximation of the monodromy method.  Formulas for the blocks based on other approximation methods 
should be able to shed light on this question.   In particular, the recursion relations \cite{Zamolodchikov:1987,ZamolodchikovRecursion,Zamolodchikov:1995aa} for the OPE limit might be used directly, perhaps even numerically, or else they might be transformed \cite{Litvinov:2013sxa} to the lightcone OPE limit. 
  It seems reasonable to expect that the AdS$_3$ interpretation we have uncovered will persist in irrational CFTs with finite $c > 1$, so it would be interesting to examine finite central charge Virasoro blocks in general, or simply in the lightcone OPE limit.  As we have argued, this limit by itself provides a great deal of information about the spectrum of the CFT.

\section*{Acknowledgments}    

We are grateful to Chris Beem, Christopher Brust, John Cardy, Liang Dai, Daniel Harlow, Jaume Gomis, Gary Horowitz, Shamit Kachru,  Ami Katz, Zohar Komargodski, Jo\~ao Penedones, David Poland, Jorge Santos, Eva Silverstein, David Simmons-Duffin, Andreas Stergiou, and Alexander Zhiboedov for discussions.  We are especially grateful to Thomas Hartman for discussions and correspondence concerning Virasoro conformal blocks, and to Thomas, Shamit, and Ami for comments on the draft.  ALF was partially supported by ERC grant BSMOXFORD no. 228169.   This work was supported in part by the National Science Foundation grants PHY-1316665, PHY-0756174, PHY-1214000, and PHY11-25915.

\appendix

\section{Properties of global conformal blocks}
\label{app:ConformalBlocks}

Our arguments rely on some key properties of t-channel global conformal blocks, specifically in the small $u$, large $\ell$ limit. The majority of these properties were discussed quite thoroughly in \cite{Fitzpatrick:2012yx}, but were restricted to the case where all four scalar operators in the correlation function are identical. In this appendix, we generalize this discussion to consider two distinct scalar primary operators $\phi_1,\phi_2$. We will specifically focus on the most relevant case of $d=2$, though this discussion can easily be extended to general spacetime dimensions by following \cite{Fitzpatrick:2012yx}.

\subsection{Factorization at large $\ell$ and small $u$}

In general, the t-channel conformal block expansion of a four-point correlator can be written as
\be
\< \phi_1(x_1) \phi_1(x_2) \phi_2(x_3) \phi_2(x_4) \> = \frac{1}{(x_{14} x_{23})^{\Delta_1+\Delta_2}} \left( \frac{x_{13} x_{24}}{x_{12}^2} \right)^{\Delta_{12}} \sum_{\tau,\ell} P_{\tau,\ell} \, g_{\tau,\ell} (v,u) \textrm{.}
\ee
For the specific case of $d=2$, the global conformal blocks in this expansion take the form
\be
g_{\tau,\ell}(v,u) = k'_{\tau+2\ell}(1-z) k'_{\tau}(1-\bar{z}) + k'_{\tau+2\ell}(1-\bar{z}) k'_{\tau}(1-z) \textrm{.}
\ee
where $k'_{2\beta}(x) \equiv x^\beta \phantom{}_2 F_1 (\beta-\half \Delta_{12},\beta-\half \Delta_{12};2\beta;x)$.
Because we are specifically considering the regime with $(1-\bar{z}) < 1$, the second term will be exponentially suppressed at large $\ell$, such that we may ignore it.

We can use the integral representation of hypergeometric functions to rewrite the general function $k'_{2\beta}(1-z)$ as
\be
k'_{2\beta}(1-z) = \fr{1}{B(\beta \pm \half \De_{12})} \int_0^1 \fr{dt}{t(1-t)} \left( \fr{(1-z)t(1-t)}{1-t(1-z)} \right)^{\beta} \left( \fr{(1-t)(1-t(1-z))}{t} \right)^{\half \Delta_{12}} \textrm{,} \nn \\
\ee
where the prefactor is the beta function $B(x \pm y) = \fr{\G(x+y)\G(x-y)}{\G(2x)}$. For the case where $\beta = \fr{\tau}{2} + \ell$, we can see that the integrand of this expression factorizes into a $\tau$-dependent piece and an $\ell$-dependent piece. When $\ell$ is large, this integrand will be sharply peaked at $t_* = \fr{1-\sqrt{z}}{1-z}$. As the $\tau$-dependent piece of the integrand will vary slowly over this region, we can safely approximate that part with its value at $t = t_*$,
\be
\left( \fr{(1-z)t_*(1-t_*)}{1-t_*(1-z)} \right)^{\fr{\tau}{2}} \sim \fr{(1-z)^{\fr{\tau}{2}}}{(1+\sqrt{z})^{\tau}} + O(1/\sqrt{\ell}) \textrm{.}
\ee
If we use Stirling's approximation for the beta function prefactor and take the small $z$ limit, which is equivalent to small $u$, we find
\be
k'_{\tau + 2\ell}(1-z) = 2^\tau k'_{2\ell}(1-z) \times \Big( 1 + O(\sqrt{z},1/\sqrt{\ell}) \Big) \textrm{.}
\ee
In this limit, we can therefore see that the $\tau$-dependence of 2d conformal blocks factorizes from the $\ell$-dependence,
\be
g_{\tau,\ell}(v,u) = k'_{2\ell}(1-z) 2^\tau k'_{\tau}(v) \times \Big( 1 + O(\sqrt{z},1/\sqrt{\ell}) \Big) \textrm{,}
\ee
where we have used the fact that $1 - \bar{z} = v + O(z)$ in the small $z$ limit .

As discussed in \cite{Fitzpatrick:2012yx}, this factorization behavior can be generalized to higher spacetime dimensions, such that we obtain
\be
g^{(d)}_{\tau,\ell}(v,u) = k'_{2\ell}(1-z) v^{\fr{\tau}{2}} F^{(d)}(\tau,v) \times \Big( 1 + O(\sqrt{z},1/\sqrt{\ell}) \Big) \textrm{,}
\ee
where $F^{(d)}(\tau,v)$ is a $d$-dependent analytic function which is regular and positive at $v = 0$.

\subsection{Further approximations at small $u$}

The function $k'_{2\ell}(1-z)$ can be approximated further if we consider the limit $\ell \ra \infty$ with the product $y \equiv z \ell^2$ fixed such that $y \lesssim O(1)$,
\be
k'_{2\ell}(1-z) = \fr{\Gamma(2\ell)}{\Gamma^2(\ell)} \int_0^1 \fr{dt}{t(1-t)} t^{\ell-\half \Delta_{12}} (1-t)^{\Delta_{12}} e^{-\fr{ty}{\ell(1-t)}} \times \Big( 1 + O(1/\ell) \Big) \textrm{,}
\ee
where we have again used Stirling's approximation to simplify the $\G$-functions. The evaluation of this integral can be greatly simplified by defining the new variable $s \equiv \fr{ty}{\ell(1-t)}$,
\bq
\begin{split}
\fr{\Gamma^2(\ell)}{\Gamma(2\ell)} k'_{2\ell}(1-z) &= \left( \fr{y}{\ell} \right)^{\Delta_{12}} \int_0^\infty \fr{ds}{s^{\Delta_{12}+1}} e^{-s - \fr{y}{s}} \times \Big( 1 + O(1/\ell) \Big) \\
&= 2 \, z^{\half \Delta_{12}} K_{\Delta_{12}}(2\ell \sqrt{z}) \times \Big( 1 + O(1/\ell) \Big) \textrm{,}
\end{split}
\label{eq:besselappx}
\eq
where $K_x(y)$ is a modified Bessel function of the second kind. We stress that this approximation breaks down when $y \gg 1$, but provides a valid description in the regime with $y \lesssim O(1)$.

\subsection{Global conformal blocks in the heavy/light probe limit}

So far, we have made no assumptions about the twists or external scaling dimensions associated with these global conformal blocks. However, in this work we are especially interested in pairs of scalar primaries $\phi_1,\phi_2$ in the limit $\Delta_2 \gg \Delta_1$, such that the relevant conformal block twists are $\tau \gtrsim \Delta_2$. To make this manifest, we can rewrite the twists as $\tau = \Delta_2 + \delta$. With this change of variables, the function $k'_\tau(v)$ takes the form
\be
k'_\tau(v) = v^{\half(\Delta_2 + \delta)} \phantom{}_{2} F_1 \left( \Delta_2 + \half (\delta - \Delta_1), \Delta_2 + \half (\delta - \Delta_1); \Delta_2 + \delta; v \right) \textrm{.}
\ee
Using a Pfaff transformation, this can be rewritten as
\be
k'_\tau(v) = v^{\half(\Delta_2 + \delta)} (1-v)^{\half (\Delta_1 - \delta) - \Delta_2} \phantom{}_{2} F_1 \left( \Delta_2 + \half (\delta - \Delta_1), \half (\delta + \Delta_1); \Delta_2 + \delta; \fr{v}{v-1} \right) \textrm{.} \nn \\
\ee
In the limit $\Delta_1, \delta \ll \Delta_2$, the hypergeometric function greatly simplifies, such that this function is approximately
\be
k'_\tau(v) = v^{\fr{\tau}{2}} (1-v)^{\Delta_{12}} \times \Big( 1 + O(\delta/\Delta_2,\Delta_1/\Delta_2) \Big) \textrm{.}
\ee
This extremely simple result is explained, in the general Virasoro context, in appendix \ref{app:TChannelBlocks}.  It arises because the exchange of the primary dominates over all descendant exchanges.

\section{Direct approach to Virasoro conformal blocks}
\label{app:Direct}

In this appendix, we present one method for determining the structure of the identity Virasoro block, specifically in the semi-classical limit $c \ra \oo$. This `direct' approach relies solely on the Virasoro algebra to construct the identity block as a sum over all possible intermediate graviton states in AdS$_3$. While the reach of this approach is rather limited in comparison to the monodromy method discussed in appendix \ref{app:MonodromyMethod}, it serves as a useful and very elementary test of those more general results.  Also, we use these methods in appendix \ref{app:TChannelBlocks} to show that the Virasoro conformal blocks greatly simplify in a certain semi-classical limit relevant for the right-hand side of the bootstrap equation (\ref{eq:ViraBootstrap}). 

\subsection{Virasoro blocks and projection operators}

For any correlation function, we can always insert the identity operator as a sum over all possible intermediate states $|\alpha\>$ of the theory,
\be
\< \phi_1 (x_1) \phi_1(x_2) \phi_2(x_3) \phi_2(x_4) \> = \sum_{\alpha} \< \phi_1 (x_1) \phi_1(x_2) |\alpha\> \<\alpha| \phi_2(x_3) \phi_2(x_4) \> \textrm{.}
\ee
This statement is of course true in any theory, and does not rely on the presence of any conformal symmetry. However, for the case of a 2d CFT, the states $|\alpha\>$ can be organized into irreducible representations of the Virasoro group, each of which is associated with a Virasoro primary operator $\CO_{h,\bar{h}}$,
\be
\< \phi_1 (x_1) \phi_1(x_2) \phi_2(x_3) \phi_2(x_4) \> = \sum_{h,\bar{h}} \sum_{\alpha_{h,\bar{h}}} \< \phi_1 (x_1) \phi_1(x_2) |\alpha_{h,\bar{h}}\> \<\alpha_{h,\bar{h}}| \phi_2(x_3) \phi_2(x_4) \> \textrm{,}
\ee
where the states $|\alpha_{h,\bar{h}}\>$ are those states created by $\CO_{h,\bar{h}}$ and its Virasoro descendants.

This separation of states into representations of the Virasoro group is precisely the Virasoro conformal block decomposition of a correlation function,
\be
\sum_{h,\bar{h}} \sum_{\alpha_{h,\bar{h}}} \< \phi_1 (x_1) \phi_1(x_2) |\alpha_{h,\bar{h}}\> \<\alpha_{h,\bar{h}}| \phi_2(x_3) \phi_2(x_4) \> = \fr{1}{x_{12}^{2\Delta_1} x_{34}^{2\Delta_2}} \sum_{h,\bar{h}} P_{h,\bar{h}} \CV_{h,\bar{h}}(u,v) \textrm{,}
\ee
such that we can associate each Virasoro block with a particular projection operator
\be
\CP_{h,\bar{h}} = \sum_{\alpha_{h,\bar{h}}} |\alpha_{h,\bar{h}}\> \<\alpha_{h,\bar{h}}| \textrm{.}
\ee

The descendant states $|\alpha_{h,\bar{h}}\>$ are created by acting with various linear combinations of the Virasoro generators $L_m, \bar{L}_n$ on the state $|h,\bar{h}\> = \CO_{h,\bar{h}} |0\>$, where these generators obey the algebra
\bq
\begin{split}
\comm{L_m}{\bar{L}_n} &= 0 \textrm{,} \\
\comm{L_m}{L_n} &= (m-n)L_{m+n} + \fr{c}{12} m(m^2-1) \delta_{m,-n} \textrm{,} \\
\comm{\bar{L}_m}{\bar{L}_n} &= (m-n)\bar{L}_{m+n} + \fr{c}{12} m(m^2-1) \delta_{m,-n} \textrm{.}
\end{split}
\eq
Note that $L_{-1}, L_0,$ and $ L_{1}$ form the holomorphic global conformal subalgebra, and $c$ drops out of their commutation relations.  Because the holomorphic generators $L_m$ commute with all of the antiholomorphic $\bar{L}_n$, we can simultaneously diagonalize one generator from each set, which we choose to be the operators $L_0,\bar{L}_0$. Our basis states $|\alpha_{h,\bar{h}}\>$ can then be expressed as a tensor product of eigenstates of $L_0$ with eigenstates of $\bar{L}_0$,
\be
|\alpha_{h,\bar{h}}\> = |\alpha_h\> \otimes |\bar{\alpha}_{\bar{h}}\> \textrm{.}
\ee
Similarly, the projection operator $\CP_{h,\bar{h}}$ can be written as the tensor product
\be
\CP_{h,\bar{h}} = \sum_{\alpha_h} |\alpha_h\>\<\alpha_h| \otimes \sum_{\bar{\alpha}_{\bar{h}}} |\bar{\alpha}_{\bar{h}}\>\<\bar{\alpha}_{\bar{h}}| = \CP_h \otimes \bar{\CP}_{\bar{h}} \textrm{,}
\ee
which tells us that the Virasoro block can be written as the product
\be
\CV_{h,\bar{h}}(u,v) = \CV_h(z) \bar{\CV}_{\bar{h}}(\bar{z}) \textrm{,}
\ee
where $u = z\bar{z}$ and $v = (1-z)(1-\bar{z})$.

As these functions are invariant under any global conformal transformation, we can simplify their calculation by choosing coordinates such that we obtain the relation
\be
\< \phi_1 (\oo) \phi_1(1) \CP_h \phi_2(z,\bar{z}) \phi_2(0) \> = \< \phi_1 (\infty) \phi_1(1) \> \< \phi_2(z,\bar{z}) \phi_2(0) \> P_h \CV_h(z) \textrm{,}
\ee
with a similar relation for $\bar{\CV}_{\bar{h}}(\bar{z})$.

\subsection{Semi-classical graviton basis}

Everything we discussed in the previous section is exact, with no assumptions about the 2d CFT or the primary operator associated with the Virasoro block. Theoretically, any Virasoro block could be constructed in this fashion, by finding the associated projection operator and acting within a particular correlation function. In practice, though, this process is prohibitively difficult for general operators in a general theory. We will therefore restrict our focus to the identity Virasoro block in theories with large central charge.

The identity operator has $h = \bar{h} = 0$ and its associated  state is  the vacuum $|0\>$. The descendant states which make up the projection operators $\CP_0,\bar{\CP}_0$ are therefore linear combinations of $L_m,\bar{L}_n$ acting on the vacuum. Because the identity is a Virasoro primary, the vacuum is annihilated by all the `lowering' operators $L_m,\bar{L}_m$ with $m>0$. In addition, the vacuum transforms trivially under the global conformal group, so it is also annihilated by all the global operators, such that we have
\be
L_m|0\> = \bar{L}_m|0\> = 0 \qquad (m = -1,0,1)\textrm{.}
\ee
Our projection operators will therefore consist of states created by generators of the form $L_{-m},\bar{L}_{-m}$ with $m \geq 2$. We will restrict our discussion to the holomorphic projector $\CP_0$, but all of our results will also apply to the antiholomorphic $\bar{\CP}_0$.

One obvious basis to use is the `graviton' basis, consisting of the states
\be
\label{eq:GravitonBasis}
|\alpha_0\> = \fr{L_{-m_1}^{k_1} \cdots L_{-m_n}^{k_n} |0\>}{\sqrt{\CN_{\{m_i,k_i\}}}} \textrm{,}
\ee
where $\CN_{\{m_i,k_i\}}$ is simply a normalization factor. To avoid redundancy, we will use the ordering convention $m_1 > \cdots > m_n$. In terms of AdS, these basis states can be loosely interpreted as $k$-graviton states, where $k = \sum_i k_i$, though in AdS$_3$ gravitons are not propagating degrees of freedom in the bulk.

In order to work in this basis, we need to determine an expression for the normalization factors $\CN_{\{m_i,k_i\}}$. For example, let us consider the normalization of a general $k$-graviton state,
\be
\CN_{m_1 \cdots m_k} = \<L_{m_k} \cdots L_{m_1} L_{-m_1} \cdots L_{-m_k}\> \textrm{,}
\ee
where again we have the ordering convention $m_1 \geq \cdots \geq m_k$. To determine the precise form of this factor, we simply need to use the structure of the Virasoro algebra to commute each $L_{m_i}$ term through to the far right, where it then annihilates the vacuum. Starting with $L_{m_1}$, we obtain
\bq
\begin{split}
\CN_{m_1 \cdots m_k} &= \< L_{m_k} \cdots L_{m_2} (L_{-m_1} L_{m_1} + \comm{L_{m_1}}{L_{-m_1}}) L_{-m_2} \cdots L_{-m_k} \> \\
&= \< L_{m_k} \cdots L_{m_2} (L_{-m_1} L_{m_1} + 2m_1 L_0) L_{-m_2} \cdots L_{-m_k} \> + \fr{c}{12} m_1(m_1^2-1) \CN_{m_2 \cdots m_k} \textrm{.}
\end{split}
\eq
The $L_0$ originating from $\comm{L_{m_1}}{L_{-m_1}}$ can easily be commuted through the remaining operators, resulting in
\be 
\< L_{m_k} \cdots L_{m_2} (2m_1 L_0) L_{-m_2} \cdots L_{-m_k} \> = 2m_1 \left( \displaystyle{ \sum_{i=2}^k } m_i \right) \CN_{m_2 \cdots m_k} \textrm{.}
\ee
Since we are considering the limit $c \ra \oo$ at fixed $m_i$, this term will be subdominant, such that we can safely ignore it.

As we continue to commute $L_{m_1}$ through the remaining operators, we can immediately see that the only non-negligible terms are those which arise if $m_i = m_1$. We then obtain the semi-classical recursion relation
\be 
\CN_{m_1 \cdots m_k} \approx \fr{c}{12} m_1 (m_1^2 - 1) \left( 1 + \displaystyle{ \sum_{i=2}^k } \delta_{m_1 m_i} \right) \CN_{m_2 \cdots m_k} \textrm{.}
\ee

Using this recursion relation, we can then obtain an approximate expression for every normalization factor in the semi-classical limit,
\be 
\CN_{\{m_i,k_i\}} = \< L_{m_n}^{k_n} \cdots L_{m_1}^{k_1} L_{-m_1}^{k_1} \cdots L_{-m_n}^{k_n} \> \approx \left( \fr{c}{12} \right)^k \displaystyle{ \prod_{i=1}^n } \left( k_i! m_i^{k_i}(m_i^2 - 1)^{k_i} \right) \textrm{,}
\ee
where again $k = \sum_i k_i$.

In general, we cannot actually use these $k$-graviton states to construct our projection operators, because \emph{this basis is not orthogonal}. For example, consider the inner product
\be
\fr{ \< L_p L_{-m} L_{-n} \> }{\sqrt{ \CN_p \, \CN_{m,n}} } = \fr{n(n^2-1)(2m+n)}{\sqrt{p(p^2-1)n(n^2-1)(\fr{c}{12}m(m^2-1)(1+\delta_{mn})+2mn)}} \delta_{p,m+n} \textrm{.}
\ee
Though these are two distinct states, their inner product is clearly nonzero for $p=m+n$. However, this expression vanishes to leading order in the semi-classical limit $c \ra \oo$,
\be
\fr{ \< L_{m+n} L_{-m} L_{-n} \> }{\sqrt{ \CN_{m+n} \CN_{m,n}} } \approx \fr{n(n^2-1)(2m+n)}{\sqrt{\fr{c}{12} mn(m+n)(m^2-1)(n^2-1)((m+n)^2-1)(1+\delta_{mn})}} \sim \fr{1}{\sqrt{c}} \textrm{,} \nn \\
\ee
such that these two states become approximately orthogonal. This behavior is in fact quite general, and applies to all inner products of distinct graviton states. At some level, this is rather unsurprising, as the limit $c \ra \oo$ in a CFT is equivalent to the limit $G_N \ra 0$ in AdS, such that interactions between gravitons are greatly suppressed. Our basis is therefore approximately orthogonal in the large-$c$ limit, and we can construct the approximate projection operator
\be
\label{eq:VirasoroIdentityProjector}
\CP_{0} \approx \sum_{\{m_i,k_i\}} \fr{L_{-m_1}^{k_1} \cdots L_{-m_n}^{k_n} |0\>\<0| L_{m_n}^{k_n} \cdots L_{m_1}^{k_1}}{\CN_{\{m_i,k_i\}}} \textrm{.}
\ee

\subsection{$T_{\mu \nu}$ correlators and the identity Virasoro block}

We can now use our approximate projector to determine the holomorphic identity block through the relation
\be
\CV_0(z) = \fr{\< \phi_1 (\oo) \phi_1(1) \CP_0 \phi_2(z) \phi_2(0) \>}{\< \phi_1 (\infty) \phi_1(1) \> \< \phi_2(z) \phi_2(0) \>} \textrm{.}
\ee
Since we are working in the graviton basis, we need to calculate correlation functions of the form
\be
\< \phi_1 (\oo) \phi_1(1) L_{-m_1}^{k_1} \cdots L_{-m_n}^{k_n}\> \, , \, \<L_{m_n}^{k_n} \cdots L_{m_1}^{k_1} \phi_2(z) \phi_2(0) \> \textrm{.}
\ee
Our approach will be quite similar to the normalization factor calculations in the previous section. We can simply commute the Virasoro generators through the various scalar operators $\phi_i$, using the commutation relation
\be
\comm{L_{-m}}{\phi_i(w)} = h_i(1-m) w^{-m} \phi_i + w^{1-m} \partial \phi_i \textrm{,}
\ee
where $h_i$ is the holomorphic scaling dimension of $\phi_i$ and $w = x^0 + i x^1$.  For a review of this and various related techniques for computing these correlators see e.g. \cite{Ginsparg}.

As a simple example of this process, let us consider a general one-graviton correlation function. Using this commutation relation, we can obtain the expression
\bq
\begin{split}
\< \phi_i (w_1) \phi_i (w_2) L_{-m} \> &= - \< \comm{L_{-m}}{\phi_i(w_1)} \phi_i(w_2) \> - \< \phi_i(w_1) \comm{L_{-m}}{\phi_i(w_2)} \> \\
&= \left( h_i(m-1) (w_1^{-m} + w_2^{-m}) - w_1^{1-m} \partial_1 - w_2^{1-m} \partial_2 \right) \< \phi_i(w_1) \phi_i(w_2) \> \textrm{.}
\end{split}
\eq
If we use the known two-point correlation function
\be
\< \phi_i(w_1) \phi_i(w_2) \> = \fr{1}{|w_{12}|^{4h_i}} \textrm{,}
\ee
we can calculate the exact one-graviton correlator,
\be
\< \phi_i (w_1) \phi_i (w_2) L_{-m} \> = h_i \left( (m-1) (w_1^{-m} + w_2^{-m}) + \fr{2}{w_{12}} ( w_1^{1-m} - w_2^{1-m} ) \right) \< \phi_i(w_1) \phi_i(w_2) \> \textrm{.} \nn \\
\ee
Similarly, we can obtain the other correlation function
\be
\< L_m \phi_i (w_1) \phi_i (w_2) \> = h_i \left( (m+1) (w_1^m + w_2^m) - \fr{2}{w_{12}} ( w_1^{1+m} - w_2^{1+m} ) \right) \< \phi_i(w_1) \phi_i(w_2) \> \textrm{.} \nn \\
\ee
Combining all of these results, we find the full one-graviton contribution to the identity block
\bq
\begin{split}
\CV_0^{(k=1)}(z) &= \sum_{m=2}^{\oo} \fr{ \<\phi_1(\oo) \phi_1(1) L_{-m}\>\<L_m \phi_2(z) \phi_2(0) \>}{\CN_m \<\phi_1(\oo) \phi_1(1)\> \<\phi_2(z) \phi_2(0)\>} \\
&= 12 \fr{h_1 h_2}{c} \sum_{m=2}^{\oo} \fr{(m-1)^2}{m(m^2-1)} z^m = 2 \fr{h_1 h_2}{c} z^2 \, \phantom{}_2 F_1 (2,2;4;z) \textrm{,}
\end{split}
\eq
which is the precise form of the global conformal block of the one-graviton global conformal primary $L_{-2}$. This result is unsurprising, because the other one-graviton operators $L_{-m}$ are all global conformal descendants of $L_{-2}$.

Let us now consider the more general $k$-graviton correlator,
\be
\< \phi_i(w_1) \phi_i(w_2) L_{-m_1}^{k_1} \cdots L_{-m_n}^{k_n} \> \textrm{.}
\ee
Just as before, we can commute the various $L_{-m_i}$ operator through the two scalar operators to obtain the general expression
\be
\prod_{j=1}^n \Big( h_i(m_j-1)(w_1^{-m_j} + w_2^{-m_j}) - w_1^{1-m_j}\partial_1 - w_2^{1-m_j}\partial_2 \Big)^{k_j} \< \phi_i(w_1) \phi_i(w_2) \> \textrm{.}
\ee
These differential operators clearly do not commute, and computing the resulting expression will generally become intractable. However, if we consider the limit $c \ra \oo$ at fixed $\fr{h_i}{\sqrt{c}}$, we only need to consider terms with leading powers of $h_i$. The result then simplifies to the approximate form
\be
h_i^k \prod_{j=1}^n \left( (m-1) (w_1^{-m} + w_2^{-m}) + \fr{2}{w_{12}} ( w_1^{1-m} - w_2^{1-m} ) \right)^{k_j} \< \phi_i(w_1) \phi_i(w_2) \> \textrm{.}
\ee
\emph{We emphasize that the rest of this section will be studying the limit $h_1, h_2, c \to \infty$ with $h_1 / c \to 0$ and  $h_2 / c \to 0$ but $h_1 h_2 /c$ fixed and finite.}

We can now determine the general $k$-graviton contribution to the identity Virasoro block, which is associated with the approximate projection operator
\be
\CP_0^{(k)} \approx \sum_{\{m_i,k_i\}} \fr{L_{-m_1}^{k_1} \cdots L_{-m_n}^{k_n} |0\>\<0| L_{m_n}^{k_n} \cdots L_{m_1}^{k_1}}{\CN_{\{m_i,k_i\}}} \textrm{.}
\ee
Inserting this projection operator into the four-point correlator, we obtain
\be
\CV_0^{(k)}(z) = \fr{\< \phi_1 (\oo) \phi_1(1) \CP_0^{(k)} \phi_2(z) \phi_2(0) \>}{\< \phi_1 (\infty) \phi_1(1) \> \< \phi_2(z) \phi_2(0) \>} \approx \left( \fr{12h_1h_2}{c} \right)^k \sum_{\{m_i,k_i\}} \prod_{i=1}^n \fr{(m_i-1)^{2k_i}}{k_i!m_i^{k_i}(m_i^2-1)^{k_i}} z^{k_i m_i} \textrm{.} \nn \\
\ee
Now the crucial step is to note that the contribution of each of the $k$-gravitons commutes with the others, so we can write the entire $k$-graviton piece of the conformal block in the limit of interest as
\be
\CV_0^{(k)}(z) \approx \fr{1}{k!} \left( \fr{12h_1h_2}{c} \sum_{m=2}^\infty \fr{(m-1)^2}{m(m^2-1)} z^m \right)^k \textrm{.}
\ee
The expression in parentheses is precisely the one-graviton contribution we found earlier.  Now when we sum over $k$, we find that the result exponentiates!  Thus we have determined the full expression for the identity holomorphic block in our restricted semi-classical limit
\be
\CV_0(z) = \sum_{k=0}^{\oo} \CV_0^{(k)}(z) \approx \sum_{k=0}^\infty \fr{1}{k!} \left( 2 \fr{h_1h_2}{c} z^2 \, \phantom{}_2 F_1 (2,2;4;z)  \right)^k = \exp \left[ 2 \fr{h_1h_2}{c} z^2 \, \phantom{}_2 F_1 (2,2;4;z) \right] \textrm{,} \nn \\
\ee
with a similar result for the antiholomorphic block $\bar{\CV}_0(\bar{z})$.  In the limit we are considering, with $c \to \infty$ with $h_1 h_2 /c$ fixed, this result for the identity Virasoro conformal block should hold for all values of $z$.

\section{Review of monodromy method for the Virasoro blocks}
\label{app:MonodromyMethod}

In this appendix we provide a self-contained review of what we refer to as the `monodromy method' for computing Virasoro conformal partial waves in the semi-classical limit.  Although the method may be well known to experts,  we have included this appendix for the sake of completeness.  Our discussion closely follows \cite{HarlowLiouville,HartmanLargeC}.  We will now give a brief sketch of the main ideas behind the monodromy method, and then we will discuss each step in detail in the subsections that follow.

The semi-classical limit is defined as the large central charge limit $c \to \infty$ with the ratios $h /c$ of conformal dimensions to the central charge kept finite.  It is believed that in this limit, the Virasoro conformal partial waves take the form
\be
\< \CO_1(x_1) \CO_2(x_2) |\alpha \> \< \alpha | \CO_3(x_3) \CO_4(x_4) \> =  {\cal F}_\alpha(x_i) \approx e^{- \frac{c}{6} f(x_i)},
\label{eq:blocksaddle}
\ee
where $f(x_i)$ approaches some fixed function of $x_i$ and the various ratios $h/c$ in the semi-classical limit.  The $\approx$ sign indicates that we are dropping subleading corrections in our $c \to \infty$ limit.  As far as we know this statement has not been rigorously proven, but we will see very good evidence for it below by making use of Liouville theory.  In the much more restrictive limit of appendix \ref{app:Direct} we essentially gave a proof by computing an explicit sum over states.  In what follows we will simply assume this semi-classical scaling behavior.

The next step is to insert into the correlator a `light' operator $\hat{\psi}(z)$ whose dimension  is fixed as $c \to \infty$. We will argue that the leading semi-classical behavior is unchanged, but the conformal block is multiplied by a wavefunction $\psi(z)$:
\be
\Psi(x_i, z) &\equiv & \< \CO_1 \CO_2  | \alpha \> \< \alpha |\hat{\psi}(z) \CO_3 \CO_4 \> = \psi(z,x_i) {\cal F}_\alpha(x_i).
\label{eq:psidef}
\ee
Note that $\psi(z,x_i)$ is just a function, whereas $\hat{\psi}$ is an operator. This formula defines $\psi(z,x_i)$; the content of the equation is that $\psi$ and its derivatives are $\CO(e^{c^0})$. This is extremely powerful, because we can take $\hat{\psi}$ to be any light operator we like, including one of the degenerate operators in the theory.   In particular, we can choose an operator that obeys the shortening condition
\be
\left( L_{-2} -\frac{3}{2(2h_\psi+1)} L_{-1}^2 \right) | \psi\> =0.
\label{eq:shortening}
\ee
Acting with $\left( L_{-2} -\frac{3}{2(2h_\psi+1)} L_{-1}^2 \right)$ on $\hat{\psi}$ inside $\Psi(z_i, z)$ then implies the  differential equation in the $z$ variable
\be 
\psi''(z) + T(z) \psi(z) = 0,
\ee 
where $T(z)$ is given by
  \be
  \frac{c}{6}  T(z) =  \frac{h_1}{z^2} + \frac{h_2}{(z-x)^2 } + \frac{h_3}{(1-z)^2} + \frac{h_1+h_2+h_3-h_4}{z (1-z)} - \frac{c}{6} c_2 (x) \frac{x(1-x)}{z(z-x)(1-z)} 
  \label{eq:source}
  \ee
 after setting $x_1=0, x_2=x, x_3=1, x_4 = \infty$, with $c_2 = \frac{\partial}{\partial x_2} f(x_i)$.

As a final step, it turns out that $\psi(z)$ must have a specific monodromy, again because the degeneracy of $\hat{\psi}$ is very constraining.  In particular, if we study the OPE 
  \be
   \CO_3(0) \CO_4(x) = \sum_\beta c_{34\beta}(x) \CO_\beta(0)
   \ee
inside $\< \alpha | \hat{\psi}(z) \CO_3 \CO_4 \>$ in (\ref{eq:psidef}), the shortening condition (\ref{eq:shortening}) implies that only operators $\CO_\beta$ with one of two different possible weights $h_\beta$ can contribute.  Thus, moving $\psi(z)$ around a cycle that encloses $x_1$ and $x_2$ must have monodromy consistent with these two weights.  This is sufficient to determine $c_2(x)$, and therefore $f(x)$.
    
Now we will  go through each of these points in more detail.

\subsection{Scaling of the semi-classical action}
\label{sec:scaling}

The first key point is that conformal blocks at large central charge are believed to behave like $\sim e^{-\frac{c}{6} f}$, {\it i.e.}
\be
\lim_{c\rightarrow \infty} \frac{1}{c} \log {\cal F} &=& -\frac{1}{6} f(x_i) < \infty.
\label{eq:semiclassical}
\ee
One piece of evidence for this result, and the origin of the term `semi-classical' limit, comes from Liouville theory.  This is a theory with action
\be
S &=& \frac{1}{4b^2} \int d^2 x \sqrt{g}\left(  g^{\alpha \beta}  \partial_\alpha \phi_c \partial_\beta \phi_c + 2(1+b^2) R \phi_c + 16 \lambda  e^{\phi_c}\right),
\ee
where $R$ is the Ricci scalar and $b$ is a parameter related to the central charge $c$ by
\be
c = 1+ 6 \left( b+ \frac{1}{b} \right)^2 \stackrel{b\ll 1}{\sim} 6 b^{-2}. 
\ee
The Liouville theory has a continuous spectrum, with correlators that receive contributions from conformal blocks of arbitrary dimension and spin, so it is a useful laboratory for studying conformal blocks.  Roughly speaking, we can obtain semi-classical conformal blocks by projecting them out of Liouville correlators.

At small $b$ and fixed $\lambda$, the equation of motion for $\phi_c$ is
\be
\partial \bar{\partial} \phi_c = 2 \lambda e^{\phi_c},
\ee
with boundary condition $\phi_c  \sim -2 \log(z \bar{z}) + \CO(1)$ at $z \rightarrow \infty$, so $\<\phi_c\> \sim \CO(c^0)$. Thus, at small $b$, the action should have a semi-classical limit 
\be
S_{cl} &\stackrel{b \ll 1}{=}& \frac{3 c}{2} \int d^2 x \sqrt{g}\left(  g^{\alpha \beta}  \partial_\alpha \phi_c \partial_\beta \phi_c + 2 R \phi_c + 16 \lambda  e^{\phi_c}\right),
\ee
which implies the scaling in (\ref{eq:semiclassical}).  

Primary operators in Liouville theory can be constructed by taking exponentials, {\it i.e.}
\be
V_\alpha \equiv e^{ \frac{\alpha}{b} \phi_c}.
\ee
The weight of such an operator is $h_V = \alpha( b+ \frac{1}{b} - \alpha) \stackrel{b\ll1}{\sim} \alpha \sqrt{\frac{c}{6}} - \alpha^2$.  Thus, in order to take $c \rightarrow \infty$ with $h_V/c$ fixed, we take $\alpha \sim \CO(\sqrt{c})$. Taking $\alpha = \frac{a}{b}$, we can solve for $a$ in terms of $h_V$, finding $a = \frac{1}{2} ( 1 \pm \sqrt{1- 24 h_V/c})$.  So these  ``heavy'' operators can be written as
\be
V = e^{ \frac{a}{b^2} \phi_c} = \exp \left( c\frac{( 1 \pm \sqrt{1- 24 h_V/c})}{12 } \phi_c \right).
\ee
When we insert one of these operators in the path integral, it has the effect of shifting $S_{cl}$ of the Liouville theory by $\CO(c)$, and of shifting the equations of motion for $\phi_c$ by $\CO(c^0)$.  This argument falls short of a proof  of the scaling of $f(x_i)$ because we have only estimated the scaling of the correlators.  We need to project the correlators  onto conformal blocks to determine the scaling of $\log \mathcal{F}$, and so we have not proven that the individual blocks themselves scale as desired.

If we want to construct a light operator, with dimension that scales like $c^0$, then we should take $\alpha \sim \frac{1}{\sqrt{c}} \sim b$.  Such operators are of the form $V = e^{\CO(c^0) \phi_c}$, and their insertions only shift the semi-classical Liouville action by $\CO(c^0)$. 

\subsection{Insertion of the degenerate operator}

The claim that correlators behave like $e^{-\frac{c}{6} f}$ in the semi-classical $c\rightarrow \infty$ has far-reaching consequences once we ask what happens when we insert additional light operators $\hat{\psi}$, {\it i.e.} operators with dimensions $\sim \CO(1)$, in correlators.  The effect of adding such an operator is to multiply the correlator by a wavefunction $\psi(z,x_i)$ for the position of the insertion of $\hat{\psi}$:
\be
\sum_k \< \CO_1 \CO_2 | \alpha;k  \> \< \alpha;k | \hat{\psi}(z) \CO_3 \CO_4 \> = \psi(z,x_i) \sum_k  \< \CO_1 \CO_2 | \alpha; k \> \< \alpha ; k| \CO_3 \CO_4 \>,
\label{eq:lightfactorization}
\ee
where we have made the sum over descendant states explicit via the $k$ label. In the above equation, as in all sums over states of the form $\sum_i | i \> \< i |$, there is implicit position dependence in the sum, because the states must be inserted on a ball that separates the fields on the left from the fields on the right; equivalently, one can write the OPE in terms of sums over operators.  One can take the above equation as a definition of $\psi(z,x_i)$; as stated above, the content of the equation is that $\psi(z,x_i) \sim \CO(e^{c^0})$.  
  We can investigate this assumption by using the definition of the conformal blocks as a sum over states. Define 
\be
\psi_k(z, x_i) \equiv \frac{\< \alpha; k | \hat{ \psi}(z) \CO_3(x_3) \CO_4(x_4) \>}{ \< \alpha ; k | \CO_3(x_3) \CO_4(x_4) \> },
\label{eq:psikdef}
\ee  
so that
\be
\< \CO_1(x_1) \CO_2(x_2) | \alpha; k  \> \<  \alpha; k | \hat{\psi}(z) \CO_3(x_3) \CO_4(x_4) \> = \psi_k(z,x_i) \< \CO_1(x_1) \CO_2(x_2) | \alpha; k \> \< \alpha; k |  \CO_3(x_3) \CO_4(x_4) \> .\nn\\
\ee
Let $k_0$ be the lowest level  so that $\psi_{k_0}$ in (\ref{eq:psikdef}) does not vanish.  Then, equation (\ref{eq:lightfactorization}) follows  if $\frac{\psi_k(z)}{\psi_{k_0}(z)}$ is $\CO(e^{c^0})$ at  $c\rightarrow \infty$ for a light operator $\hat{\psi}$.  To understand why this should be true, we will first assume that $\psi_{k_0}$ is of order $\CO(e^{c^0})$, due to $\hat{\psi}$ being a light operator.  Then, we can look at how $\psi_k$ for general $k$ is related to $\psi_{k_0}$ by examining the action of the Virasoro operator $L_{m}$ inside the correlator:
\be
\< \alpha; k_0 | L_m \hat{\psi}(z) \CO_3(x_3) \CO_4(x_4) \> &=& \sum_{i=3,4,z} \left( \frac{(m-1) h_i}{x_i^m} - \frac{1}{x_i^{m-1}} \partial_i \right)\< \alpha; k_0 | \hat{\psi}(z) \CO_3(x_3) \CO_4(x_4) \> \nn\\
 &=&  \sum_{i=3,4,z} \left( \frac{(m-1) h_i}{x_i^m} - \frac{1}{x_i^{m-1}} \partial_i \right) \psi_{k_0}(z, x_3, x_4) \< \alpha; k_0 |  \CO_3(x_3) \CO_4(x_4)  \> \nn\\
  &\cong & \psi_{k_0}(z, x_i) \sum_{i=3,4} \left( \frac{(m-1) h_i}{x_i^m} - \frac{1}{x_i^{m-1}} \partial_i \right)  \< \alpha; k_0 | \CO_3(x_3) \CO_4(x_4)  \> \nn\\ 
  &=& \psi_{k_0}(z, x_i) \< \alpha; k_0 | L_m\CO_3(x_3) \CO_4(x_4)  \>.
\ee
 The key step in in the third line, where ``$a \cong b$'' means $\frac{a}{b} = \CO(e^{c^0})$.  This step is justified because we can take $h_z$ and $\partial_z$ as $\CO(e^{c^0})$ since $\hat{\psi}$ is a light operator and $\psi_{k_0}$ is $\CO(e^{c^0})$, whereas $h_3, h_4 $ and $\partial_3, \partial_4 \sim \CO(c)$.  Dividing both sides of this equation by $\< \alpha; k_0 | L_m\CO_3(x_3) \CO_4(x_4)  \> = \< \alpha; m+k_0 | \CO_3(x_3) \CO_4) \>$ and being a bit schematic with the indices labeling the level of the descendants, we obtain
\be
\psi_{k_0+m}(z,x_i) &=& \psi_{k_0}(z,x_i),
\ee
whose consequence is (\ref{eq:lightfactorization}). 

\subsection{Differential equation from the degeneracy condition}

Next, we want to explore the consequences of the shortening condition (\ref{eq:shortening}) for correlators of $\hat{\psi}$ with four heavy operators.  The idea is that (\ref{eq:shortening}) becomes a differential equation for the correlator (see e.g. \cite{Ginsparg})
\be
0 &=& \left( \frac{3}{2(2h_\psi+1)} \partial_z^2 + \sum_{i=1}^4 \left( \frac{h_i}{(z-x_i)^2} + \frac{1}{z-x_i}\partial_i \right) \right) \< \CO_1 \CO_2 \hat{\psi} \CO_3 \CO_4 \>  \nn\\
 &\stackrel{b \ll 1}{=} &  \left( \frac{c}{6} \partial_z^2 + \sum_{i=1}^4 \left( \frac{h_i}{(z-x_i)^2} + \frac{1}{z-x_i}\partial_i \right) \right) \< \CO_1 \CO_2 \hat{\psi} \CO_3 \CO_4 \> ,
\ee
where in the second line we have used the weight of the degenerate operator
\be
h_\psi = -\frac{1}{2}  -  \frac{3b^2}{4},
\ee
and $c \approx \frac{6}{b^2}$ at $b \ll 1$.  We would like  to argue that this equation is satisfied not only for the correlator, but for each of its constituent conformal blocks.  The justification for this is that each conformal block has a different monodromy in $z$, determined by the weight of the block itself.  So we have
\be
0 &=& \left( \frac{c}{6} \partial_z^2 + \sum_{i=1}^4 \left( \frac{h_i}{(z-x_i)^2} + \frac{1}{z-x_i}\partial_i \right) \right) \psi(z,x_i) e^{-\frac{c}{6} f(x_i)} \nn\\
 &=& \frac{c}{6} \left( \partial_z^2 \psi(z,x_i) + T(z,x_i) \psi(z,x_i) \right),
 \label{eq:monodromydiffeq}
\ee
where
\be
T(z,x_i) &=& \sum_{i=1}^4 \frac{\epsilon_i}{(z-x_i)^2} - \frac{c_i}{z-x_i}, \qquad 
c_i \equiv \frac{\partial}{\partial x_i} f, \qquad \epsilon_i \equiv \frac{ 6 h_i}{c},
\ee
and we have again used the fact that $\psi \sim \CO(e^{c^0})$, so we can neglect $\partial_i$ derivatives acting on it.  Finally, $T(z,x_i)$ itself is further constrained by a conformal Ward identity, as it is exactly the wavefunction that arises when we compute the $\< \hat T(z) \CO_1 \CO_2 \CO_3 \CO_4\>$ five-point function, where the energy-momentum tensor $\hat T(z)$ should not be confused with its wavefunction $T(z,x_i)$:
\be
\< \hat T(z) \CO_1 \CO_2 \CO_3 \CO_4\> &=&  \sum_{i=1}^4 \left( \frac{h_i}{(z-x_i)^2} + \frac{1}{z-x_i}\partial_i \right)\< \CO_1 \CO_2 \CO_3 \CO_4\>  \nn\\
 &=&- \frac{c}{6} T(z,x_i) \< \CO_1 \CO_2 \CO_3 \CO_4\>.
\ee
Therefore $T(z,x_i)$ must decay\footnote{This is easy to see by taking the $\CO_1\CO_2\CO_3\CO_4 \supset c_{1234T}(x_i) T(0)$ OPE.} like $z^{-4}$ as $z \rightarrow \infty$ , which implies three constraints:
\be
\sum_i c_i = 0 , \qquad 
\sum_i \left(c_i x_i - \frac{6 h_i}{c}\right) = 0, \qquad 
\sum_i \left(c_i x_i^2 -  \frac{12 h_i}{c} x_i\right) = 0.
\ee
Taking $x_1 =0, x_2=x, x_3 =1, x_4 = \infty$ then leads us to equation (\ref{eq:source}).

\subsection{Constraint on $h_\beta$ and monodromy}

Finally, we need to constrain the monodromy of $\psi(z)$ to determine the function $f(x_i)$ which defines the semi-classical conformal block.  First, let us consider the constraint of the shortening condition for $\hat{\psi}$ on three-point functions 
\be
V_{\alpha \beta \psi} = \< \CO_\alpha(x_1) \CO_\beta(x_2)  \hat{\psi}(x_3) \> = \frac{C_{\alpha \beta \psi}}{x_{12}^{(h_\alpha+h_\beta-h_\psi)}x_{13}^{(h_\alpha+h_\psi-h_\beta)}x_{23}^{(h_\psi+h_\beta-h_\alpha)}}.
\ee
It is straightforward to act on this with the appropriate shortening operator for $\hat \psi$ to see
\be
0&=&\left( - \frac{3}{2(2h_\psi+1)} \partial_3^2 + \sum_{i=1,2} \left( \frac{h_i}{(x_3-x_i)^2} + \frac{1}{x_3-x_i}\partial_i \right) \right)  V_{\alpha \beta \psi} \nn\\
& = & \left(\frac{2 h_{\psi } \left(h_{\alpha }+h_{\beta }\right)+6 h_{\alpha } h_{\beta }-3 h_{\alpha }^2+h_{\alpha }-3 h_{\beta }^2+h_{\beta }+h_{\psi
   }^2-h_{\psi }}{4 h_{\psi }+2}\right) V_{\alpha \beta \psi} \frac{x_{12}^2}{x_{13}^2 x_{23}^2}
\ee
One can solve this algebraic equation for $h_\beta$ as a function of $h_\alpha$ and set $h_\psi = -\frac{1}{2} - \frac{3 b^2}{4}$.  In the limit $b\ll 1$ with $h_\alpha b^2$ fixed, one finds
\be
h_\beta- h_\alpha - h_\psi = \frac{1}{2} \left( 1 \pm \sqrt{1- 4 b^2 h_\alpha} \right).
\ee
We want to know the monodromy of $\psi(z)$ as $\hat{\psi}$ encircles $x_1$ and $x_2$ in the four-point function $\< \CO_1 \CO_2 | \alpha \> \< \alpha | \hat{\psi} \CO_3 \CO_4 \>$.  To relate this to the argument above, we take the OPE of $\CO_3 \CO_4 = \sum_\beta c_{34 \beta} \CO_\beta$.  
Our analysis of the 3-pt function shows that $\sum_\beta c_{34\beta} \<\alpha | \hat{\psi} \CO_\beta\>$  gets contributions only from $\CO_\beta$ with $h_\beta$ such that $\< \CO_\alpha(y) \hat{\psi}(z) \CO_\beta(\frac{x_3+x_4}{2}) \> \sim (z-y)^{-(h_\psi + h_\alpha - h_\beta)} = (z-y)^{\frac{1 \pm \sqrt{1- 4 b^2 h_\alpha}}{2}}$ as $z$ encircles $y$. Since the sum over states $|\alpha\>$ arises from the $\CO_1 \CO_2$ OPE, this cycle must enclose both $x_1 $ and $x_2$ when we apply it to $\psi(z)$.  Thus, under a cycle encircling $x_1$ and $x_2$ but not $x_3$ and $x_4$, the solutions to the differential equation (\ref{eq:monodromydiffeq}) must have monodromy
\begin{equation}
M = \left( \begin{array}{cc} e^{ i \pi (1 + \sqrt{1- 24 h_\alpha/c)}} & 0 \\
0 & e^{ i \pi (1 - \sqrt{1- 24 h_\alpha/c})} \end{array} \right) = - \left( \begin{array}{cc} e^{i \pi \Lambda_\alpha } & 0 \\ 0 & e^{-i \pi \Lambda_\alpha} \end{array} \right), \qquad \Lambda_\alpha = \sqrt{1- 24 h_\alpha/c}, 
\label{eq:monodromymatrix}
\end{equation}
in a basis that diagonalizes $M$.   This fact combined with the results of the previous subsection allows us to determine the semi-classical conformal block using the monodromy method.  Note that for the identity or vacuum conformal block this means that $M$ must be the $2 \times 2$ identity matrix, which is identical in all bases.  This leads to further simplifications for the monodromy method when applied to the identity conformal block.

\section{Computing Virasoro blocks via the monodromy method}
\label{app:ComputingVirasoroViaMonodromy}

We will now use the monodromy method reviewed in appendix \ref{app:MonodromyMethod} to compute the Virasoro conformal blocks in a semi-classical limit more general than that which was considered in appendix \ref{app:Direct}.  Specifically, we will be able to determine the conformal block for a primary of weight $h_p$ in a correlator of the form
\be
\langle \phi_1(0) \phi_1(x) \phi_2(1) \phi_2(\infty) \rangle
\ee
in the limit that
\be
c \to \infty, \ \textrm{ and  } \ \frac{h_i}{c} \ \ \mathrm{fixed},
\ee
followed by a perturbative expansion to linear order in $h_1/c$ and $h_p/c$, but working non-perturbatively in $h_2/c$.   Note that working to linear order in $h_1 / c$ in the computation of $f$ for the Virasoro block $\mathcal{F} = e^{-\frac{c}{6} f}$ means that we are neglecting terms of order $h_1^2 / c$ in the exponent of $\mathcal{F}$.  To use the monodromy method we are already neglecting order one terms in the exponent of $\mathcal{F}$, so strictly speaking, we need to take $h_1^2 / c \lesssim 1$ for a self-consistent approximation.  This makes it possible to use the CFT bootstrap to study AdS$_3$ setups where a probe object orbits a finite mass deficit angle or a BTZ black hole. For the reader just looking to find the results, the formulas we compute for the conformal blocks are equations (\ref{eq:BTZschannel}) and (\ref{eq:testmassschannel}). 

\subsection{S-channel Virasoro blocks}

As discussed in appendix \ref{app:MonodromyMethod}, we would like to solve the differential equation 
\be 
\psi''(z) + T(z) \psi(z) = 0
\ee 
where $T(z)$ is given by equation (\ref{eq:source}).
Then we must impose that the pair of solutions for $\psi$ (there are two, since the differential equation is second order) have monodromy according to (\ref{eq:monodromymatrix}) when we take $z$ around $0$ and $x$; this determines the function $c_2(x)$.  Once $c_2$ is fixed we can use the relation $c_2 = \frac{\partial}{\partial x_2} f(x_i)$ to determine the semi-classical conformal block
\be
{\cal F}(x_i) \approx e^{- \frac{c}{6} f(x_i)}
\ee

For our particular semi-classical limit let us define $\epsilon_i \equiv \frac{6 h_i}{c}$.  
We write the solutions for $\psi$ as 
\be
\psi= \psi^{(0)}+\epsilon_1 \psi^{(1)}+\epsilon_1^2 \psi^{(2)} + \dots.
\ee
Then we can write
\be
T(z) &=& \epsilon_2 \frac{1}{(1-z)^2} + \epsilon_1 \left( \frac{ 1}{z^2} + \frac{1}{(z-x)^2} + \frac{2}{z(1-z)} - \frac{c_2}{\epsilon_1} \frac{x(1-x)}{z(z-x)(1-z)} \right)
\ee
We can immediately solve the differential equation for $\psi^{(0)}$ to find the two solutions
\be
\psi_{1,2}^{(0)}(z) &=& (1-z)^{\frac{1\pm \sqrt{1-4 \epsilon_2}}{2}}
\ee
Notice that  the exponent transitions from real to complex exactly when the large mass $h_2$ develops a horizon in AdS$_3$.  To see this, recall that $c = \frac{3}{2G}$ so we have
\be
m_2  = 2 h_2 = \frac{\epsilon_2 c}{3} =  \frac{  \epsilon_2}{2G}
\ee
Thus, exactly at $\epsilon_2 = \frac{1}{4}$, the mass reaches the critical mass $\frac{1}{8 G}$ to make a BTZ black hole.

To solve for $\psi$ at higher orders in $\epsilon_1$, it is useful to use our zeroth order solutions in order to reduce the second order differential equation to a first order differential equation using the method of variation of parameters.  In this method, given an inhomogeneous ODE of the form
\be
y''(z)+a(z) y(z) = b(z)
\label{eq:generalODE}
\ee
and two solutions $y_i(z)$ to the homogeneous ODE $y''(z)+a(z)y(z) =0$, we can find a solution of the form
\be
y_p(z) = f_1(z) y_1(z) + f_2(z) y_2(z)
\ee
through
\be
f_1'(z) = - \frac{y_2(z) b(z)}{W(z)}, \qquad f_2'(z) = \frac{y_1(z) b(z)}{W(z)}
\ee
where 
\be
W(z) \equiv y_1(z) y'_2(z) - y'_1(z)y_2(z).
\ee
is the Wronskian determinant.

To bring our problem into this form, we divide up $T$ into a zero-th order piece $T^{(0)}$ and a correction $T^{(1)}$:
\be
T &=& T^{(0)} + \epsilon_1 T^{(1)} + \epsilon_1^2 T^{(2)} + \dots \nn\\
T^{(0)} &=& \epsilon_2 \frac{1}{(1-z)^2} \nn\\
T^{(1)} &=& \left( \frac{ 1}{z^2} + \frac{1}{(z-x)^2} + \frac{2}{z(1-z)} - \frac{c_2^{(1)}}{\epsilon_1} \frac{x(1-x)}{z(z-x)(1-z)} \right). 
\ee
At linear order in $\epsilon_1$, our differential equation takes the form
\be
(\psi_i^{(1)})'' + T^{(0)}  \psi_i^{(1)} &=& - T^{(1)} \psi_i^{(0)}.
\ee
Now we can determine $\psi_i^{(1)}$. We simply need to integrate
\be
\psi^{(1)}_i &=& \psi_1^{(0)} \int dz \frac{ - \psi_2^{(0)} (- T^{(1)} \psi_i^{(0)})}{W} + \psi_2^{(0)} \int dz \frac{\psi_1^{(0)} (-T^{(1)} \psi_i^{(0)})}{W} .
\label{eq:variationofparameters}
\ee
These integrals can be performed in closed form in terms of logarithms and hypergeometric functions, which allows one to read off their monodromy properties.  

We want to demand that the solutions $\psi_i^{(1)}$ transform with eigenvalues given by (\ref{eq:monodromymatrix}) as $z$ encircles $0$ and $x$ in order to determine the function $c_2(x)$.  The method of variation of parameters automatically gives $\psi^{(1)}$ in a form that is decomposed into a basis of the zero-th order solutions multiplied by coefficients that are functions of $z$. 
Let us analyze the coefficient of $\psi_1^{(0)}$ first, since it is simpler:
\be
\int dz \frac{ - \psi_2^{(0)} (- T^{(1)} \psi^{(0)}_1)}{W} &=& \frac{\left(\frac{c_2}{\epsilon_1} (1-x)+1\right) \log (\frac{z}{z-x}
)+\frac{(x-2) z+x}{z (z-x)}}{\sqrt{1-4 \epsilon _2}}
\ee
It is easy to see that this returns to itself after a rotation of $z =r e^{i \phi}$ with  $\phi$ from $0$ to $2\pi$ if $r>x$, since we never cross the branch cut of the logarithm.  This can also be seen by noting that the two poles of the integrand at $z=0$ and $z=x$ have opposite residues.  This means that this term does not contribute to the monodromy of $\psi^{(1)}$. 
Now consider the second term:
\be
\int dz \frac{\psi_1^{(0)} (-T^{(1)} \psi_1^{(0)})}{W} = \int dz \frac{(1-z)^{\sqrt{1-4 \epsilon _2}} \left(\frac{c_2 (x-1) x z (x-z)}{\epsilon _1}-x^2 (z+1)+2 x z (z+1)-2 z^2\right)}{z^2 \sqrt{1-4 \epsilon_2} (x-z)^2}  \nn \\
\ee
After either a direct evaluation, or an examination of the residues of the poles at $z=0$ and $z=x$, we find that under a $2\pi$ phase rotation, the integral shifts by a monodromy $(\delta M_{0x})_{12}$ given by
\begin{equation}
(\delta M_{0x})_{12} = \frac{2 \pi i}{\alpha_2 } \left(  \left(\alpha _2 -1\right)- \left(\frac{c_2(x)}{\epsilon_1} (x-1)-1\right) (1-x)^{\alpha _2}+   \frac{c_2(x)}{\epsilon_1} (x-1)+\alpha _2 (1-x)^{\alpha _2}  \right) , \\
\label{eq:BTZmonodromy}
\end{equation}
where $\alpha_2 \equiv \sqrt{1-4\epsilon_2}$. The calculation for $\psi_2^{(1)}$ follows from the same calculation but with $\alpha_2 \rightarrow - \alpha_2$.  At this order, we have therefore found the monodromy matrix is
\be
\delta M_{0x} = \left( \begin{array}{cc} 0 & (\delta M_{0x})_{12} \\ (\delta M_{0x})_{21} & 0 \end{array} \right),
\ee
where $(\delta M_{0x})_{21} [ \alpha_2] = -(\delta M_{0x})_{12} [- \alpha_2]$. 
The eigenvalues of $M_{0x}$ at this order are therefore
$1 \pm \left[ (\delta M_{0x})_{12}(\delta M_{0x})_{21}\right]^{1/2}$. 
By inspection of (\ref{eq:monodromymatrix}) expanded to linear order in $h_p$, we can therefore identify $\sqrt{(\delta M_{0x})_{12}(\delta M_{0x})_{21}}$ as $2 i \pi \epsilon_p$, or equivalently
\be
(\delta M_{0x})_{12}(\delta M_{0x})_{21} =- 4 \pi^2 \epsilon_p^2.
\ee
This equation can easily be solved for $c_2$:
\be
c_2 = \frac{ \epsilon_1 \left( -1+\alpha_2 + (1-x)^{\alpha_2}(1+\alpha_2) \right) \pm  \alpha_2 (1-x)^{\frac{\alpha_2}{2}} \epsilon_p}{(1-x)(1-(1-x)^{\alpha_2})}.
\ee  
 Finally, this can be integrated to get the conformal block at $\CO(\epsilon_1, \epsilon_p)$ and any $\epsilon_2$. We choose the integration constant and the sign of $\pm$ in the above equation so that $f(z) \sim 2 (\epsilon_1 -\epsilon_p )\log(z)$ at $z\sim 0$, to obtain
\begin{equation}
f(z) 
 = (2 \epsilon_1 - \epsilon_p) \log \left(\frac{1-(1-z)^{\alpha _2}}{\alpha _2}\right)+\epsilon_1 \left(1-\alpha _2\right) \log (1-z) + 2\epsilon_p \log \left( \frac{1+(1-z)^{\frac{\alpha_2}{2}}}{2} \right).
 \label{eq:BTZschannel}
\end{equation}
This gives us the conformal block in the limit we desired, where one operator $h_1$ is a `test mass' and the other operator of dimension $h_2 \propto c$ would create a finite deficit angle or a BTZ black hole in AdS.  

Let us pause to note the approximations we have made.  Aside from the limit $c \to \infty$ with $h_i / c$ fixed, we have also expanded the function $f$ in the conformal block $\mathcal{F} \approx e^{-\frac{c}{6} f}$ in $h_1 / c$.  Since we have only computed $f$ to first order in $h_1 / c$, we are dropping terms of order $h_1^2 / c^2$, which means that we have ignored effects of order $h_1^2 / c$ in the exponent.  By pushing the monodromy method further and working to higher order in $h_1/c$, we could control these neglected terms.  However, the monodromy method always neglects terms of order $1 \ll c$ in the exponent of $\mathcal{F}$.

As a check,  we can look at the identity block $\epsilon_p =0$ and compare to our results from the direct approach.   Replacing $\alpha_2 = \sqrt{1-4\epsilon_2}$ and expanding to $\CO(\epsilon_2)$ to compare with the result of appendix \ref{app:Direct}, one finds
\be
\frac{1}{\epsilon_1} f(z) &=& 2 \log(z) -\frac{\epsilon_2}{3} z^2 {}_2F_1(2,2,4,z) \nn\\
 && + \epsilon_2^2 \left(\frac{4 (z-1) \log ^2(1-z)}{z^2}+\left(\frac{4}{z}-2\right) \log (1-z)+8\right) + \CO(\epsilon_2^3)
 \ee
We see that the second term matches, as expected.  We have also checked that  equation (\ref{eq:BTZschannel}) agrees with the recursion relation method \cite{ZamolodchikovRecursion,Zamolodchikov:1995aa, HartmanLargeC} when we expand in small $z$.

\subsection{S-channel Virasoro blocks at quadratic order}

We can also obtain the conformal blocks at order  $\CO(\epsilon_1^2,  \epsilon_p^2)$ if we set $\epsilon_2 = \epsilon_1  \equiv \epsilon$.  To do this, we take our first order solutions in the limit of small $\epsilon_2$ and substitute them back into (\ref{eq:variationofparameters}). The resulting expression for $\psi_{1,2}^{(2)}$ simply contains logarithms and dilogarithms, and thus the monodromy can straightforwardly be matched to (\ref{eq:monodromymatrix}) at second order in $h_p/c$.  We find the result:
\be
f^{(1)} (z) &=& (2 \epsilon-\epsilon_p) \log(z) + 2\epsilon_p \log \left( \frac{ 1+\sqrt{1-z}}{2}\right) , \nn\\
f^{(2)}(z) &=& 2 \left(2 \epsilon ^2-\epsilon _p^2\right) \log (1-z)+4 \epsilon _p^2 \log \left(\frac{1}{2}
   \left(\sqrt{1-z}+1\right)\right)\nn\\
    && +\frac{2 \left(z \left(\epsilon _p-2 \epsilon \right){}^2+\log (1-z) \left(\epsilon _p-2
   \sqrt{1-z} \epsilon \right){}^2\right)}{z}.
   \label{eq:testmassschannel}
   \ee
A feature of this result is that $\epsilon_p$ terms contain no divergences at $z\rightarrow 1$ at this order:
\be
f(1-y) \stackrel{y \ll 1}{\sim} 4 \epsilon^2 \log(y) + \CO(y^0, \epsilon^3, \epsilon_p^3).
\ee

\section{T-channel Virasoro blocks}
\label{app:TChannelBlocks}

In this appendix we will study the Virasoro blocks in the t-channel, based on the primary exchange
\be
\langle \phi_1 \phi_2 | \CO_p \rangle \langle \CO_p | \phi_1 \phi_2 \rangle .
\ee
We will analyze the particular semiclassical heavy/light or probe limit \cite{Behan:2014dxa}, where $h_1^2 \ll h_2, h_p, c$.  We will study two further limits which, when combined, are sufficient for discussions of the t-channel blocks on the right-hand side of the bootstrap equation (\ref{eq:ViraBootstrap}) in section \ref{sec:VirasoroBootstrap}.  For the first limit, we define $\delta h \equiv h_p - h_2$ and then we assume $\delta h^2 \ll h_2, c$.    This is the limit that is relevant for the anti-holomorphic part of the Virasoro blocks.  In the second limit we take $h_p \gg h_2, c$ in order to obtain 2d Virasoro blocks with large spin and fixed twist.  This is discussed at the end of this appendix.

The Virasoro blocks greatly simplify in the first limit, so that they are dominated solely by the exchange of the primary $\CO_p$.   To see this, note that the three-point function is
\be
\< \phi_1(y_1) \phi_2(y_2) \CO_p(y_3) \> = \frac{1}{y_{12}^{h_1 - \delta h} y_{13}^{2h_2 +\delta h - h_1} y_{23}^{h_1 + \delta h}} .
\ee
Now, when we act on $\CO_p$ with $L_{-n}$ and take $y_3 \rightarrow 0$, we find
\be
\frac{\< \phi_2(\infty) \phi_1(1) L_{-n} | \CO_p\>}{\< \phi_2(\infty) \phi_1(1)| \CO_p\>} = n h_1 + \delta h .
\ee
Similarly, the conjugate gives
\be
\frac{\< \CO_p | L_n \phi_1(z) \phi_2(0) \>}{\< \CO_p | \phi_1(z) \phi_2(0)\>} = z^n (n h_1 + \delta h).
\ee
The point is that both of these ratios of 3-pt functions are proportional to $h_1$ and $\delta h$, but they never involve $h_2$ or $c$.  This persists if we study more general descendant states.

These computations are relevant for the t-channel blocks if we study a modified version of the `graviton basis' of equation (\ref{eq:GravitonBasis}), where we also include the $L_{-1}^k$ operators.  This is necessary because the state $\CO_p |0 \rangle = | \CO_p \rangle$ will not be annihilated by these global conformal generators.  So we have a modified version of the projector in equation (\ref{eq:VirasoroIdentityProjector}) 
\be
\CP_{\CO_p} \approx \sum_{\{m_i,k_i\}} \fr{L_{-m_1}^{k_1} \cdots L_{-m_n}^{k_n}  L_{-1}^{k_0} | \CO_p \>\<\CO_p | L_{1}^{k_0}  L_{m_n}^{k_n} \cdots L_{m_1}^{k_1}}{\CN^{\CO_p}_{\{m_i,k_i\}}} ,
\ee
which we might use to compute the Virasoro block. The modified normalization 
\be
\CN^{\CO_p}_{\{m_i,k_i\}} = \<\CO_p |  L_{1}^{k_0}  L_{m_n}^{k_n} \cdots L_{m_1}^{k_1}   L_{-m_1}^{k_1} \cdots L_{-m_n}^{k_n}  L_{-1}^{k_0}  | \CO_p \>
\ee
has a single important feature -- namely that in this particular semiclassical limit, we obtain an extra factor of either $c$ or $h_p \approx h_2$ from each additional $L_m$.  Thus the contribution of descendants to this Virasoro block is always suppressed as a power of one of the ratios
\be
\frac{h_1^2}{h_2} , \ \frac{\delta h^2}{h_2}, \ \frac{h_1^2}{c} , \ \frac{\delta h^2}{c} \ll 1,
\ee  
which are small in the probe limit.  So in the t-channel, in this heavy / light probe semiclassical limit, not only is it sufficient to use the global blocks for the 2d bootstrap; in fact, it is sufficient to simply use the OPE limit, or the result of primary exchange!

To use the Virasoro blocks at high spin, as is necessary in section \ref{sec:VirasoroBootstrap}, we also need to study a very different limit where 
\be
h_p \gg h_1, h_2 .
\ee
Combining an anti-holomorphic Virasoro block with $\bar h_p \approx \bar h_2$ and a holomorphic block with $h_p \gg h_1, h_2$ allows us to construct a block with twist $\tau \approx \Delta_2$ but with large $\ell = h_p - \bar h_p$.  Fortunately this large $h_p$ limit has already been studied \cite{Zamolodchikov:1995aa}, see appendix D of \cite{HarlowLiouville} for a thorough discussion using the monodromy method.  The result is that
\be
{\cal F}(z) \sim (16 q)^{h_p - \frac{c}{24}} \theta_3(q)^{\frac{c}{2} - 8 h_1 - 8 h_2} z^{\frac{c}{24} - h_1 -h_2} (1-z)^{\frac{c}{24} - h_1 -h_2} ,
\ee
where 
\be
q = e^{-\pi K(1-z)/K(z)}, \qquad \theta_3(q) = \sum_{n=-\infty}^\infty q^{n^2} = \sqrt{ \frac{2}{\pi} K(z)},
\ee
and $K$ is the elliptic function
\be
K(z) = \frac{1}{2} \int_0^1 \frac{dt}{\sqrt{t(1-t) (1-z t)}} .
\ee
This is the result for the operators inserted at $x_1=0,x_2=z,x_3=1,x_4=\infty$.  To apply this to the t-channel of the bootstrap equation, we need to map to $x_1=0, x_2=\infty, x_3=z, x_4=1$, which corresponds to
\be
{\cal F}(z) \rightarrow \frac{1}{z^{2h_1}} {\cal F} \left(1-\frac{1}{z} \right).
\ee
Expanding near $z \sim 0$, we find
\be
{\cal F}(z) \sim z^{\frac{c}{24} - 2 h_1},
\ee
which should be compared with the singularity $z^{-2h_1}$ of the identity block in the s-channel.  Clearly there is a mismatch in the power of the singularity, and since $c \gg 1$, the singularity of the Virasoro blocks is much weaker in the limit $h_p \gg h_1, h_2,c$ at small $z$.

\section{Calculation of deficit angle spectrum}

In this appendix, we present a more detailed calculation of the results discussed in section \ref{sec:deficitanglespectrum}. Specifically, we will use the 2d bootstrap equation to place bounds on the coefficients of $t$-channel global conformal blocks. These bounds provide rigorous evidence that the large $\ell$ spectrum of 2d CFTs with large central charge matches that of deficit angles in AdS$_3$.

\subsection{Bootstrap equation in the lightcone OPE limit}

In the limit $u \ll v \ll 1$, the bootstrap equation takes the approximate form
\be
1 \approx \alpha^{-\Delta_1} z^{\half(\Delta_1 + \Delta_2 - \Delta_{12})} v^{-\half (\alpha \Delta_1 + \Delta_2)} \sum_{\tau,\ell} P_{\tau,\ell} \, k'_{2\ell} (1-z) \, 2^\tau v^{\fr{\tau}{2}} k'_{\tau}(v) \textrm{.}
\ee
where $k'_{2\beta}(x) = x^\beta \phantom{}_2 F_1 (\beta-\half \Delta_{12},\beta-\half \Delta_{12};2\beta;x)$.
The left side of this expression is clearly constant and finite, so the $u,v$-dependence of the right side must also vanish. The small $v$ behavior of each term in this series is approximately $v^{\half(\tau - \alpha \Delta_1 - \Delta_2)}$, which greatly constrains the possible twists $\tau$ that can dominate at large $\ell$.

In particular, there \textit{must} exist operators with $\tau \approx \alpha \Delta_1 + \Delta_2$ in order to produce a constant result in the limit $v \ra 0$. For the right side to also be independent of $u \approx z$, there must actually be an infinite tower of conformal blocks with twist accumulating at $\alpha \Delta_1 + \Delta_2$ as $\ell \to \infty$, such that the full sum introduces a power-law singularity in $z$ not possessed by any individual term. In the small $v$ limit, where these conformal blocks provide the dominant contribution, we can approximate the bootstrap equation as
\be
1 \approx 2^{\tau_0} \alpha^{-\Delta_1} z^{\half(\Delta_1 + \Delta_2 - \Delta_{12})} \sum_{\ell} P_{\tau_0,\ell} \, k'_{2\ell} (1-z) \textrm{,}
\ee
where $P_{\tau_0,\ell}$ can be formally thought of as the sum of all conformal block coefficients with twist within some small range centered about $\tau_0 \equiv \alpha \Delta_1 + \Delta_2$.

Following the work of \cite{Fitzpatrick:2012yx}, the sum over $\ell$ can be written as an integral over a conformal block coefficient density $f_0 (\ell)$,
\be
\sum_{\ell} P_{\tau_0,\ell} \, k'_{2\ell} (1-z) = \int_0^\oo d\ell \, f_0 (\ell) \, k'_{2\ell} (1-z) \textrm{,}
\ee
where $f_0(\ell)$ is defined as
\be
f_0 (\ell) \equiv \sum_{\ell'} P_{\tau_0,\ell'} \, \delta(\ell - \ell') \textrm{.}
\ee
In the following section we will derive bounds on the structure of $f_0(\ell)$ which indicate that it is of the form
\be
f_0(\ell) = A_0 \fr{\G^2(\ell)}{\G(2\ell)} \ell^{\Delta_1 + \Delta_2 - 1} \textrm{.}
\ee
Assuming this form, we can rewrite the bootstrap equation as
\be
1 \approx 2^{\tau_0+1} \alpha^{-\Delta_1} z^{\half(\Delta_1 + \Delta_2)} A_0 \int_0^\oo d\ell \, \ell^{\Delta_1 + \Delta_2 - 1} K_{\Delta_{12}}(2\ell \sqrt{z}) \textrm{.}
\ee
This expression can be used to fix the value of $A_0$, which in turn provides the result
\be
P_{\tau_0,\ell} \approx \fr{4 \sqrt{\pi} \alpha^{\Delta_1}}{2^{\tau_0+2\ell}\G(\De_1) \G(\De_2)} \ell^{\Delta_1+\Delta_2- \fr{3}{2}} \approx 2^{\De_1(1-\alpha)} \alpha^{\De_1} P^{\textrm{GFT}}_{\Delta_1 + \Delta_2,\ell} \qquad (\ell \gg 1) \textrm{.}
\ee
At large $\ell$, the approximate conformal block coefficients have been related to those of GFT, with an $\alpha$-dependent coefficient.  This is not strictly obligatory, since we are only constraining the accumulation at large $\ell$, and not the contribution of each individual block, but it provides a plausible expectation.  It should be noted that we are using the Virasoro blocks in the semi-classical limit, so this result will be corrected by $1/c$ effects.

We can extend this argument to higher twists by considering the bootstrap equation to all orders in $v$,
\be
\alpha^{\Delta_1} v^{-\half \Delta_1 (1-\alpha)} \left( \fr{1-v}{1-v^\alpha} \right)^{\Delta_1} \approx \left( \fr{u}{v} \right)^{\half(\Delta_1 + \Delta_2)} u^{-\half \Delta_{12}} \sum_{\tau,\ell} P_{\tau,\ell} \, g_{\tau,\ell}(v,u) \textrm{,}
\ee
which can be rewritten in the more useful form
\be
(1-v^\alpha)^{-\Delta_1} \approx \alpha^{-\Delta_1} z^{\half(\Delta_1 + \Delta_2 - \Delta_{12})} v^{-\fr{\tau_0}{2}} (1-v)^{-\De_{12}} \sum_{\tau,\ell} P_{\tau,\ell} \, g_{\tau,\ell}(v,u) \textrm{,}
\ee
where we have used the relation $u \approx z(1-v)$. We can now subtract the $\tau_0$ contributions from both sides of this expression. Since we are specifically working in the limit $\De_1 \ll \De_2$, such that $\tau_0 \approx \De_2$, we can use the approximate global conformal blocks derived in appendix \ref{app:ConformalBlocks} to calculate the approximate $\tau_0$ contribution,
\be
\sum_{\ell} P_{\tau_0,\ell} \, g_{\tau_0,\ell} (v,u) \approx \alpha^{\De_1} z^{-\half(\De_1+\De_2-\De_{12})} v^{\fr{\tau_0}{2}} (1-v)^{\Delta_{12}} \textrm{.}
\ee
Notice that this expression is of precisely the right form to cancel the overall prefactor, such that the $\tau_0$ contribution is simply 1, with no subleading corrections in $v$. Our modified bootstrap equation then becomes
\be
(1-v^\alpha)^{-\Delta_1} - 1 \approx \alpha^{-\Delta_1} z^{\half(\Delta_1 + \Delta_2 - \Delta_{12})} v^{-\fr{\tau_0}{2}} (1-v)^{-\De_{12}} \sum_{\tau>\tau_0,\ell} P_{\tau,\ell} \, g_{\tau,\ell}(v,u) \textrm{.}
\ee

We can now repeat our earlier procedure with this modified bootstrap equation. Expanding the left side as a power series in $v^\alpha$ and taking the small $v$ limit, we obtain the relation
\be
\De_1 v^{\alpha} \approx \alpha^{-\Delta_1} z^{\half(\Delta_1 + \Delta_2 - \Delta_{12})} v^{-\fr{\tau_0}{2}} \sum_{\tau>\tau_0,\ell} P_{\tau,\ell} \, k'_{2\ell} (1-z) \, 2^\tau v^{\fr{\tau}{2}} k'_{\tau}(v) \textrm{.}
\ee
For this expression to be satisfied, there must be an infinite tower of conformal blocks with twist $\tau \approx \alpha (\De_1 + 2) + \De_2$. To find the corresponding conformal block coefficients, we can again consider the limit $v \ra 0$, where these operators are the dominant contribution,
\be
\De_1 \approx 2^{\tau_1} \alpha^{-\Delta_1} z^{\half(\Delta_1 + \Delta_2 - \Delta_{12})} \sum_{\ell} P_{\tau_1,\ell} \, k'_{2\ell} (1-z) \textrm{,}
\ee
where we have introduced the generalized notation $\tau_n \equiv \alpha(\De_1 + 2n) + \De_2$. We can also define a generalized conformal block coefficient density $f_n(\ell)$, such that
\be
\sum_{\ell} P_{\tau_n,\ell} \, k'_{2\ell} (1-z) = \int_0^\oo d\ell \, f_n (\ell) \, k'_{2\ell} (1-z) \textrm{.}
\ee
The bounds we will derive in the following section indicate that this more general density is also of the form
\be
f_n(\ell) = A_n \fr{\G^2(\ell)}{\G(2\ell)} \ell^{\Delta_1 + \Delta_2 - 1} \textrm{.}
\ee
Assuming this form for our case of $n=1$, the modified bootstrap equation becomes
\be
\De_1 \approx 2^{\tau_1+1} \alpha^{-\Delta_1} z^{\half(\Delta_1 + \Delta_2)} A_1 \int_0^\oo d\ell \, \ell^{\Delta_1 + \Delta_2 - 1} K_{\Delta_{12}}(2\ell \sqrt{z}) \textrm{.}
\ee
Solving this expression for $A_1$, we then find the conformal block coefficients
\be
P_{\tau_1,\ell} \approx \fr{4 \sqrt{\pi} \De_1 \alpha^{\Delta_1}}{2^{\tau_1+2\ell}\G(\De_1) \G(\De_2)} \ell^{\Delta_1+\Delta_2- \fr{3}{2}} \qquad (\ell \gg 1) \textrm{.}
\ee

We therefore find coefficients of a very similar form to those for $n=0$. Inspired by those previous results, let's compare this expression to the coefficients of GFT \cite{Unitarity},
\be
P^{\textrm{GFT}}_{\De_1 + \De_2 + 2n,\ell} \approx \fr{(\De_1)_n}{n!2^{2n}} P^{\textrm{GFT}}_{\De_1 + \De_2,\ell} \textrm{,}
\ee
where we have specifically taken the limit $\De_1,n \ll \De_2 \ll \ell$. We therefore have the relation
\be
P_{\tau_1,\ell} \approx 2^{(\De_1+2)(1-\alpha)} \alpha^{\De_1} P^{\textrm{GFT}}_{\De_1 + \De_2 + 2,\ell} \qquad (\ell \gg 1) \textrm{.}
\ee
with the same caveat as above, namely that we can really only constrain the large $\ell$ accumulation, and not the contribution of each individual term.

We can continue to repeat this procedure to find the coefficients for increasing values of $n$. To see this most clearly, we expand the left side of the bootstrap equation as a series in $v^\alpha$,
\be
\sum_{n=0}^\oo \fr{(\De_1)_n}{n!} v^{n\alpha} \approx \alpha^{-\Delta_1} z^{\half(\Delta_1 + \Delta_2 - \Delta_{12})} v^{-\fr{\tau_0}{2}} (1-v)^{-\De_{12}} \sum_{\tau,\ell} P_{\tau,\ell} \, g_{\tau,\ell}(v,u) \textrm{.}
\ee
For $n \ll \De_2$, each individual $v^{n\alpha}$ term in the series on the left corresponds to the full contribution of the $\tau_n$ tower of conformal blocks on the right side. Our procedure can be iterated to find the corresponding coefficients $P_{\tau_n,\ell}$, but we can already see the full answer from this expression. The factor of $\fr{(\De_1)_n}{n!}$ in the power series is precisely the factor needed to reproduce the appropriate GFT coefficients, such that we obtain the general relation
\be
P_{\tau_n,\ell} \approx 2^{(\De_1+2n)(1-\alpha)} \alpha^{\De_1} P^{\textrm{GFT}}_{\De_1 + \De_2 + 2n,\ell} \qquad (\ell \gg 1) \textrm{.}
\ee
We therefore see that in the limit $\alpha \ra 1$, with vanishing deficit angle, the large $\ell$ spectrum of operators and conformal block coefficients for any CFT with large central charge perfectly reproduces that of a generalized free theory.  This is precisely what we would expect, as it corresponds to the $c \to \infty$ limit with fixed $\Delta_1$ and $\Delta_2$.

\subsection{Bounds on coefficient density}

We will now place bounds on the asymptotic behavior of the conformal block coefficient density $f_n(\ell)$. More specifically, we will prove that given a function $\CL_n(z)$, defined as
\be
\CL_n(z) \equiv \int_0^\oo d\ell \, f_n(\ell) k'_{2\ell}(1-z) \textrm{,}
\label{eq:spectraldecomp}
\ee
which behaves like $z^{\half(\De_{12}-a)}$ at small $z$, then the integrated density 
\be
F_n(L) \equiv \int_0^L d\ell \, \fr{\G(2\ell)}{\G^2(\ell)} f_n(\ell) \textrm{,}
\ee
behaves at large $L$ like
\be
\lim_{L \rightarrow \infty} F_n(L) = \frac{A_n}{\frac{a}{2} \Gamma^2(\frac{a}{2})} \textrm{.}
\ee

First, we establish an upper bound on $F_n(L)$. This discussion will be almost identical to a similar proof in \cite{Fitzpatrick:2012yx}, which interested readers may consult for more details.  For simplicity, we define the function
\be
h(\ell,z) \equiv \fr{\G^2(\ell)}{\G(2\ell)} k'_{2\ell}(1-z) \textrm{,}
\ee
which is a positive, decreasing function of $\ell$ at any fixed $z$. Since the integrand of $\CL_n(z)$ is non-negative, we can place the bound
\be
\CL_n(z) \geq h(L,z) F_n(L) \textrm{,}
\ee
for any value of $L$. As shown in appendix \ref{app:ConformalBlocks}, at large $L$ and fixed $\lambda \equiv L\sqrt{z}$ the function $h(L,z)$ takes the approximate form
\be
\lim_{L\ra\oo} h(L,z) = 2 z^{\half \Delta_{12}} K_{\De_{12}}(2\lambda) \textrm{.}
\ee
Combining these two results, we obtain the upper bound
\be
F_n(L) \leq \fr{L^a}{2 \lambda^a K_{\De_{12}}(2\lambda)} \qquad (L \gg 1) \textrm{.}
\label{eq:FnLupperbound}
\ee
The parameter $\lambda$ is arbitrary and $L$-independent, such that we can identify this upper bound as $A_U L^a$. 

We now turn to establishing that the behavior of $F_n(L)$ is exactly power-law, and that the coefficient can be determined. We would like to be able to use the limit (\ref{eq:besselappx}) to work with the simpler function $K(2 \ell \sqrt{z})\equiv z^{\frac{1}{2} \Delta_{12}} K_{\Delta_{12}}(2 \ell \sqrt{z})$ instead of $k'_{2\ell}(1-z)$ in (\ref{eq:spectraldecomp}).  To do this, we can split up the representation of $\CL(x)$ into two parts:
\be
\CL_n(z) &=& z^{\frac{1}{2} \Delta_{12}} \int_0^{\frac{y_*}{\sqrt{z}}} \frac{\Gamma(2\ell)}{2\Gamma^2(\ell)} d \ell f(\ell) K(2\ell \sqrt{z}) + \int_{\frac{y_*}{\sqrt{z}}}^\infty d\ell f(\ell) k'_{2\ell}(1-z)
\label{eq:twoparts}
\ee
The advantage is that the second integral gives a negligible contribution to the small $z$ limit of $\CL(z)$ when $y_*$ is large, in the sense that 
\be
\lim_{y_* \rightarrow \infty} \lim_{z\rightarrow 0} z^{\frac{1}{2}(a-\Delta_{12})} \int_{\frac{y_*}{\sqrt{x}}}^\infty d\ell f(\ell) k'_{2\ell}(1- z) =0
\ee
This  follows from the fact that the contribution from $k'_{2\ell}(1-z)$ shuts down exponentially at large $\ell$, so $f(\ell)$ would have to grow exponentially in order to avoid the above conclusion, which would violate the upper bound (\ref{eq:FnLupperbound}). 
Thus we can work with $K(2 \ell \sqrt{z})$.  
We can simplify our formulas a bit by defining $x = 2 \sqrt{z}$, as well as $G(x) = 2^{-1} (x/2)^{-\Delta_{12}} {\cal L}(x)$ and $\tilde{f}(\ell) = \frac{\Gamma(2\ell)}{\Gamma^2(\ell)}f_n(\ell)$.   Now, we can take $y_* \rightarrow \infty$ and consider
\be
G(x) = \int_0^\infty d\ell \tilde{f}(\ell) K(\ell x) \sim 2^{-a-1}x^{-a}
\ee
at $x\sim 0$.  
We want the limiting behavior of $F(L)$:
\be
F(L) &=& \int_0^L d\ell \tilde{f}(\ell)
\ee
The rest of the proof will be a straightforward generalization of a proof of the Hardy-Littlewood theorem due to Karamata \cite{titchmarsh}.  To do this, we define the linear functional $L[g](x)$:
\be
L[g](x) &\equiv& \int_0^\infty d\ell \bar{f}(\ell) K(\ell x) g(K(\ell x)) .
\ee
We want to prove that at $x\sim 0$, this linear functional behaves like
\be
L[g](x) \sim 2^{-a-1} x^{-a} \frac{\int_0^\infty d\ell \ell^{a-1} K(\ell) g(K(\ell))}{\int_0^\infty d\ell \ell^{a-1} K(\ell)}.
\label{eq:linearfunc}
\ee
To do this, we will show that it behaves this way on a dense set of functions:
\be
g_n(y) \equiv \frac{1}{y} K((n+1)K^{-1}(y)),
\ee
where $K^{-1}$ indicates the inverse function of $K$ (which exists by the monotonicity of $K$).\footnote{  
The fact that $g_n(y)$ are a dense set of functions in the space of piece-wise continuous functions on  $(0, \lim_{y\rightarrow 0}K(y))$, which is equivalent to the condition that $\{ K(n x)\}_{n \in \mathbb{N}}$ are dense,
follows from the fact that one can turn $K_\nu(n x)$ into $e^{-n x}$ by an invertible integral transform.  Specifically, take $f(t)= t^{\nu-1} e^{-n t}$, perform a Hankel transform, to get $H[f](s) \propto \frac{s^\nu}{(s^2 + n^2)^{\nu+1/2}}$, divide by $s^\nu$, and perform a Fourier transform to get $\propto z^\nu K_\nu(n z), (z>0)$.  For $\nu=0$, the combination of these two transforms is an Abel transform.  }
On this set of functions, the relation (\ref{eq:linearfunc}) follows straightforwardly. First,
\be
L[g_n](x) = \int_0^\infty d\ell \tilde{f}(\ell) K(\ell x) g_n(K(\ell x)) = \int_0^\infty d\ell \tilde{f}(\ell) K(\ell x (n+1)) \sim \frac{2^{-a-1}}{(x(n+1))^a} . \nn \\
\label{eq:firstlinfunc}
\ee
Second,
\be
\int_0^\infty d\ell \ell^{a-1} K(\ell) g_n(K(\ell)) = \int_0^\infty d\ell \ell^{a-1} K((n+1)\ell) = \frac{1}{(n+1)^a} \int_0^\infty d\ell \ell^{a-1} K(\ell).
\label{eq:secondlinfunc}
\ee
By comparison of (\ref{eq:firstlinfunc}) with (\ref{eq:secondlinfunc}), we see that equation (\ref{eq:linearfunc}) holds for the dense set of functions $g_n(x)$.  

The last step is to define a function $\bar{g}(x)$:
\be
\bar{g}(x) &\equiv & \left\{ \begin{array}{ll} 0 & x> K(\lambda) \\ \frac{1}{x} & x < K(\lambda) \end{array} \right.
\ee
Here, $\lambda$ is a fixed real number in $(0,\infty)$; its specific value is not important.  Now, we take $x=\lambda/L$ and evaluate
\be
L[\bar{g}](\lambda/L) = \int_0^\infty d\ell \tilde{f}(\ell) K(\frac{\ell \lambda}{L}) g(K(\frac{\ell \lambda}{L})) = \int_0^L d\ell \tilde{f}(\ell) = F(L) .
\ee
On the other hand, by equation (\ref{eq:linearfunc}), in the limit of large $L$, we have 
\be
L[\bar{g}](\lambda/L) &\sim& 2^{-a-1} \frac{L^a}{\lambda^a}\frac{\int_0^{\frac{L}{\lambda}} d\ell \ell^{a-1} }{\int_0^\infty d\ell \ell^{a-1} K(\ell)} = 2^{-a-1} L^a \frac{1}{a \int_0^\infty d\ell \ell^{a-1} K(\ell)} .
\ee
Thus, we have shown that
\be
F(L) \sim  \frac{(L/2)^a}{2a \int_0^\infty d\ell \ell^{a-1} K(\ell)}
\ee
in the limit of large $L$.

\bibliographystyle{JHEP}
\bibliography{DeficitAnglesBib}

\providecommand{\href}[2]{#2}\begingroup\raggedright\begin{thebibliography}{10}

\bibitem{FerraraOriginalBootstrap1}
S.~Ferrara, A.~Grillo, and R.~Gatto, {\it {Tensor representations of conformal
  algebra and conformally covariant operator product expansion}},  {\em Annals
  Phys.} {\bf 76} (1973) 161--188.

\bibitem{PolyakovOriginalBootstrap2}
A.~M. Polyakov, {\it {Non-Hamiltonian Approach to the Quantum Field Theory at
  Small Distances}},  {\em Zh.Eksp.Teor.Fiz.} (1973).

\bibitem{Rattazzi:2008pe}
R.~Rattazzi, V.~S. Rychkov, E.~Tonni, and A.~Vichi, {\it {Bounding scalar
  operator dimensions in 4D CFT}},  {\em JHEP} {\bf 12} (2008) 031,
  [\href{http://xxx.lanl.gov/abs/0807.0004}{{\tt arXiv:0807.0004}}].

\bibitem{Callan:1973pu}
J.~Callan, Curtis~G. and D.~J. Gross, {\it {Bjorken scaling in quantum field
  theory}},  {\em Phys.Rev.} {\bf D8} (1973) 4383--4394.

\bibitem{AldayMaldacena}
L.~F. Alday and J.~M. Maldacena, {\it {Comments on operators with large spin}},
   {\em JHEP} {\bf 0711} (2007) 019,
  [\href{http://xxx.lanl.gov/abs/0708.0672}{{\tt arXiv:0708.0672}}].

\bibitem{KomargodskiZhiboedov}
Z.~Komargodski and A.~Zhiboedov, {\it {Convexity and Liberation at Large
  Spin}},  {\em JHEP} {\bf 1311} (2013) 140,
  [\href{http://xxx.lanl.gov/abs/1212.4103}{{\tt arXiv:1212.4103}}].

\bibitem{Fitzpatrick:2012yx}
A.~L. Fitzpatrick, J.~Kaplan, D.~Poland, and D.~Simmons-Duffin, {\it {The
  Analytic Bootstrap and AdS Superhorizon Locality}},  {\em JHEP} {\bf 1312}
  (2013) 004, [\href{http://xxx.lanl.gov/abs/1212.3616}{{\tt
  arXiv:1212.3616}}].

\bibitem{BTZ}
M.~Banados, C.~Teitelboim, and J.~Zanelli, {\it {The Black hole in
  three-dimensional space-time}},  {\em Phys.Rev.Lett.} {\bf 69} (1992)
  1849--1851, [\href{http://xxx.lanl.gov/abs/hep-th/9204099}{{\tt
  hep-th/9204099}}].

\bibitem{JP}
I.~Heemskerk, J.~Penedones, J.~Polchinski, and J.~Sully, {\it {Holography from
  Conformal Field Theory}},  {\em JHEP} {\bf 10} (2009) 079,
  [\href{http://xxx.lanl.gov/abs/0907.0151}{{\tt arXiv:0907.0151}}].

\bibitem{Hamilton:2005ju}
A.~Hamilton, D.~N. Kabat, G.~Lifschytz, and D.~A. Lowe, {\it {Local bulk
  operators in AdS/CFT: A Boundary view of horizons and locality}},  {\em
  Phys.Rev.} {\bf D73} (2006) 086003,
  [\href{http://xxx.lanl.gov/abs/hep-th/0506118}{{\tt hep-th/0506118}}].

\bibitem{Katz}
A.~L. Fitzpatrick, E.~Katz, D.~Poland, and D.~Simmons-Duffin, {\it {Effective
  Conformal Theory and the Flat-Space Limit of AdS}},  {\em JHEP} {\bf 1107}
  (2011) 023, [\href{http://xxx.lanl.gov/abs/1007.2412}{{\tt
  arXiv:1007.2412}}].

\bibitem{ElShowk:2011ag}
S.~El-Showk and K.~Papadodimas, {\it {Emergent Spacetime and Holographic
  CFTs}},  {\em JHEP} {\bf 1210} (2012) 106,
  [\href{http://xxx.lanl.gov/abs/1101.4163}{{\tt arXiv:1101.4163}}].

\bibitem{Papadodimas:2012aq}
K.~Papadodimas and S.~Raju, {\it {An Infalling Observer in AdS/CFT}},  {\em
  JHEP} {\bf 1310} (2013) 212, [\href{http://xxx.lanl.gov/abs/1211.6767}{{\tt
  arXiv:1211.6767}}].

\bibitem{AdSfromCFT}
A.~L. Fitzpatrick and J.~Kaplan, {\it {AdS Field Theory from Conformal Field
  Theory}},  {\em JHEP} {\bf 1302} (2013) 054,
  [\href{http://xxx.lanl.gov/abs/1208.0337}{{\tt arXiv:1208.0337}}].

\bibitem{Belitsky:2003ys}
A.~V. Belitsky, A.~Gorsky, and G.~Korchemsky, {\it {Gauge / string duality for
  QCD conformal operators}},  {\em Nucl.Phys.} {\bf B667} (2003) 3--54,
  [\href{http://xxx.lanl.gov/abs/hep-th/0304028}{{\tt hep-th/0304028}}].

\bibitem{Korchemsky:1992xv}
G.~Korchemsky and G.~Marchesini, {\it {Structure function for large x and
  renormalization of Wilson loop}},  {\em Nucl.Phys.} {\bf B406} (1993)
  225--258, [\href{http://xxx.lanl.gov/abs/hep-ph/9210281}{{\tt
  hep-ph/9210281}}].

\bibitem{Monodromy}
A.~Belavin, A.~M. Polyakov, and A.~Zamolodchikov, {\it {Infinite Conformal
  Symmetry in Two-Dimensional Quantum Field Theory}},  {\em Nucl.Phys.} {\bf
  B241} (1984) 333--380.

\bibitem{Cardy:1984rp}
J.~L. Cardy, {\it {Conformal invariance and universality in finite-size
  scaling}},  {\em J.Phys.} {\bf A17} (1984) L385--L387.

\bibitem{Cardy:1986ie}
J.~L. Cardy, {\it {Operator Content of Two-Dimensional Conformally Invariant
  Theories}},  {\em Nucl.Phys.} {\bf B270} (1986) 186--204.

\bibitem{ETH2}
J.~M. Deutsch, {\it Quantum statistical mechanics in a closed system},  {\em
  Phys. Rev. A} {\bf 43} (Feb, 1991) 2046--2049.

\bibitem{ETH}
M.~Srednicki, {\it Chaos and quantum thermalization},  {\em Phys. Rev. E} {\bf
  50} (Aug, 1994) 888--901.

\bibitem{Unitarity}
A.~L. Fitzpatrick and J.~Kaplan, {\it {Unitarity and the Holographic
  S-Matrix}},  {\em JHEP} {\bf 1210} (2012) 032,
  [\href{http://xxx.lanl.gov/abs/1112.4845}{{\tt arXiv:1112.4845}}].

\bibitem{Maldacena:2011jn}
J.~Maldacena and A.~Zhiboedov, {\it {Constraining Conformal Field Theories with
  A Higher Spin Symmetry}},  {\em J.Phys.} {\bf A46} (2013) 214011,
  [\href{http://xxx.lanl.gov/abs/1112.1016}{{\tt arXiv:1112.1016}}].

\bibitem{CFTquasinormal}
D.~Birmingham, I.~Sachs, and S.~N. Solodukhin, {\it {Conformal field theory
  interpretation of black hole quasinormal modes}},  {\em Phys.Rev.Lett.} {\bf
  88} (2002) 151301, [\href{http://xxx.lanl.gov/abs/hep-th/0112055}{{\tt
  hep-th/0112055}}].

\bibitem{Maldacena:1997re}
J.~M. Maldacena, {\it {The Large N limit of superconformal field theories and
  supergravity}},  {\em Adv.Theor.Math.Phys.} {\bf 2} (1998) 231--252,
  [\href{http://xxx.lanl.gov/abs/hep-th/9711200}{{\tt hep-th/9711200}}].

\bibitem{GKP}
S.~S. Gubser, I.~R. Klebanov, and A.~M. Polyakov, {\it {Gauge theory
  correlators from non-critical string theory}},  {\em Phys. Lett.} {\bf B428}
  (1998) 105--114, [\href{http://xxx.lanl.gov/abs/hep-th/9802109}{{\tt
  hep-th/9802109}}].

\bibitem{Witten}
E.~Witten, {\it {Anti-de Sitter space and holography}},  {\em Adv. Theor. Math.
  Phys.} {\bf 2} (1998) 253--291,
  [\href{http://xxx.lanl.gov/abs/hep-th/9802150}{{\tt hep-th/9802150}}].

\bibitem{JPStringTheory}
J.~Polchinski, {\it {String theory. Vol. 1: An introduction to the bosonic
  string}}, .

\bibitem{BF}
P.~Breitenlohner and D.~Z. Freedman, {\it {Positive Energy in anti-De Sitter
  Backgrounds and Gauged Extended Supergravity}},  {\em Phys.Lett.} {\bf B115}
  (1982) 197.

\bibitem{JoaoMellin}
J.~Penedones, {\it {Writing CFT correlation functions as AdS scattering
  amplitudes}},  {\em JHEP} {\bf 03} (2011) 025,
  [\href{http://xxx.lanl.gov/abs/1011.1485}{{\tt arXiv:1011.1485}}].

\bibitem{CostaEikonal}
L.~Cornalba, M.~S. Costa, and J.~Penedones, {\it {Eikonal Approximation in
  AdS/CFT: Resumming the Gravitational Loop Expansion}},  {\em JHEP} {\bf 09}
  (2007) 037, [\href{http://xxx.lanl.gov/abs/0707.0120}{{\tt
  arXiv:0707.0120}}].

\bibitem{CostaCPW}
L.~Cornalba, M.~S. Costa, J.~Penedones, and R.~Schiappa, {\it {Eikonal
  approximation in AdS/CFT: Conformal partial waves and finite N four-point
  functions}},  {\em Nucl. Phys.} {\bf B767} (2007) 327--351,
  [\href{http://xxx.lanl.gov/abs/hep-th/0611123}{{\tt hep-th/0611123}}].

\bibitem{CostaSW}
L.~Cornalba, M.~S. Costa, J.~Penedones, and R.~Schiappa, {\it {Eikonal
  approximation in AdS/CFT: From shock waves to four-point functions}},  {\em
  JHEP} {\bf 08} (2007) 019,
  [\href{http://xxx.lanl.gov/abs/hep-th/0611122}{{\tt hep-th/0611122}}].

\bibitem{Hawking:1982dh}
S.~Hawking and D.~N. Page, {\it {Thermodynamics of Black Holes in anti-De
  Sitter Space}},  {\em Commun.Math.Phys.} {\bf 87} (1983) 577.

\bibitem{Fitzpatrick:2011hh}
A.~L. Fitzpatrick and D.~Shih, {\it {Anomalous Dimensions of Non-Chiral
  Operators from AdS/CFT}},  {\em JHEP} {\bf 10} (2011) 113,
  [\href{http://xxx.lanl.gov/abs/1104.5013}{{\tt arXiv:1104.5013}}].

\bibitem{Deser:1983tn}
S.~Deser, R.~Jackiw, and G.~'t~Hooft, {\it {Three-Dimensional Einstein Gravity:
  Dynamics of Flat Space}},  {\em Annals Phys.} {\bf 152} (1984) 220.

\bibitem{Deser:1983nh}
S.~Deser and R.~Jackiw, {\it {Three-Dimensional Cosmological Gravity: Dynamics
  of Constant Curvature}},  {\em Annals Phys.} {\bf 153} (1984) 405--416.

\bibitem{Brown:1986nw}
J.~D. Brown and M.~Henneaux, {\it {Central Charges in the Canonical Realization
  of Asymptotic Symmetries: An Example from Three-Dimensional Gravity}},  {\em
  Commun.Math.Phys.} {\bf 104} (1986) 207--226.

\bibitem{Cruz:1994ir}
N.~Cruz, C.~Martinez, and L.~Pena, {\it {Geodesic structure of the (2+1) black
  hole}},  {\em Class.Quant.Grav.} {\bf 11} (1994) 2731--2740,
  [\href{http://xxx.lanl.gov/abs/gr-qc/9401025}{{\tt gr-qc/9401025}}].

\bibitem{Komargodski:2012ek}
Z.~Komargodski and A.~Zhiboedov, {\it {Convexity and Liberation at Large
  Spin}},  {\em JHEP} {\bf 1311} (2013) 140,
  [\href{http://xxx.lanl.gov/abs/1212.4103}{{\tt arXiv:1212.4103}}].

\bibitem{Alday:2013cwa}
L.~F. Alday and A.~Bissi, {\it {Higher-spin correlators}},  {\em JHEP} {\bf
  1310} (2013) 202, [\href{http://xxx.lanl.gov/abs/1305.4604}{{\tt
  arXiv:1305.4604}}].

\bibitem{Alday:2013opa}
L.~F. Alday and A.~Bissi, {\it {The superconformal bootstrap for structure
  constants}},  \href{http://xxx.lanl.gov/abs/1310.3757}{{\tt
  arXiv:1310.3757}}.

\bibitem{Beem:2013sza}
C.~Beem, M.~Lemos, P.~Liendo, W.~Peelaers, L.~Rastelli, et~al., {\it {Infinite
  Chiral Symmetry in Four Dimensions}},
  \href{http://xxx.lanl.gov/abs/1312.5344}{{\tt arXiv:1312.5344}}.

\bibitem{Rychkov:2009ij}
V.~S. Rychkov and A.~Vichi, {\it {Universal Constraints on Conformal Operator
  Dimensions}},  {\em Phys. Rev.} {\bf D80} (2009) 045006,
  [\href{http://xxx.lanl.gov/abs/0905.2211}{{\tt arXiv:0905.2211}}].

\bibitem{Caracciolo:2009bx}
F.~Caracciolo and V.~S. Rychkov, {\it {Rigorous Limits on the Interaction
  Strength in Quantum Field Theory}},  {\em Phys. Rev.} {\bf D81} (2010)
  085037, [\href{http://xxx.lanl.gov/abs/0912.2726}{{\tt arXiv:0912.2726}}].

\bibitem{Poland:2010wg}
D.~Poland and D.~Simmons-Duffin, {\it {Bounds on 4D Conformal and
  Superconformal Field Theories}},  {\em JHEP} {\bf 1105} (2011) 017,
  [\href{http://xxx.lanl.gov/abs/1009.2087}{{\tt arXiv:1009.2087}}].

\bibitem{Rattazzi:2010gj}
R.~Rattazzi, S.~Rychkov, and A.~Vichi, {\it {Central Charge Bounds in 4D
  Conformal Field Theory}},  {\em Phys.Rev.} {\bf D83} (2011) 046011,
  [\href{http://xxx.lanl.gov/abs/1009.2725}{{\tt arXiv:1009.2725}}].

\bibitem{Rattazzi:2010yc}
R.~Rattazzi, S.~Rychkov, and A.~Vichi, {\it {Bounds in 4D Conformal Field
  Theories with Global Symmetry}},  {\em J.Phys.} {\bf A44} (2011) 035402,
  [\href{http://xxx.lanl.gov/abs/1009.5985}{{\tt arXiv:1009.5985}}].

\bibitem{Vichi:2011ux}
A.~Vichi, {\it {Improved bounds for CFT's with global symmetries}},  {\em JHEP}
  {\bf 1201} (2012) 162, [\href{http://xxx.lanl.gov/abs/1106.4037}{{\tt
  arXiv:1106.4037}}].

\bibitem{Poland:2011ey}
D.~Poland, D.~Simmons-Duffin, and A.~Vichi, {\it {Carving Out the Space of 4D
  CFTs}},  {\em JHEP} {\bf 1205} (2012) 110,
  [\href{http://xxx.lanl.gov/abs/1109.5176}{{\tt arXiv:1109.5176}}].

\bibitem{Rychkov:2011et}
S.~Rychkov, {\it {Conformal Bootstrap in Three Dimensions?}},
  \href{http://xxx.lanl.gov/abs/1111.2115}{{\tt arXiv:1111.2115}}.

\bibitem{ElShowk:2012ht}
S.~El-Showk, M.~F. Paulos, D.~Poland, S.~Rychkov, D.~Simmons-Duffin, et~al.,
  {\it {Solving the 3D Ising Model with the Conformal Bootstrap}},  {\em
  Phys.Rev.} {\bf D86} (2012) 025022,
  [\href{http://xxx.lanl.gov/abs/1203.6064}{{\tt arXiv:1203.6064}}].

\bibitem{Liendo:2012hy}
P.~Liendo, L.~Rastelli, and B.~C. van Rees, {\it {The Bootstrap Program for
  Boundary CFT}},  \href{http://xxx.lanl.gov/abs/1210.4258}{{\tt
  arXiv:1210.4258}}.

\bibitem{ElShowk:2012hu}
S.~El-Showk and M.~F. Paulos, {\it {Bootstrapping Conformal Field Theories with
  the Extremal Functional Method}},
  \href{http://xxx.lanl.gov/abs/1211.2810}{{\tt arXiv:1211.2810}}.

\bibitem{Beem:2013qxa}
C.~Beem, L.~Rastelli, and B.~C. van Rees, {\it {The $N=4$ Superconformal
  Bootstrap}},  \href{http://xxx.lanl.gov/abs/1304.1803}{{\tt
  arXiv:1304.1803}}.

\bibitem{Kos:2013tga}
F.~Kos, D.~Poland, and D.~Simmons-Duffin, {\it {Bootstrapping the O(N) Vector
  Models}},  \href{http://xxx.lanl.gov/abs/1307.6856}{{\tt arXiv:1307.6856}}.

\bibitem{Gliozzi:2013ysa}
F.~Gliozzi, {\it {More constraining conformal bootstrap}},  {\em
  Phys.Rev.Lett.} {\bf 111} (2013) 161602,
  [\href{http://xxx.lanl.gov/abs/1307.3111}{{\tt arXiv:1307.3111}}].

\bibitem{El-Showk:2013nia}
S.~El-Showk, M.~Paulos, D.~Poland, S.~Rychkov, D.~Simmons-Duffin, et~al., {\it
  {Conformal Field Theories in Fractional Dimensions}},
  \href{http://xxx.lanl.gov/abs/1309.5089}{{\tt arXiv:1309.5089}}.

\bibitem{Gaiotto:2013nva}
D.~Gaiotto, D.~Mazac, and M.~F. Paulos, {\it {Bootstrapping the 3d Ising twist
  defect}},  \href{http://xxx.lanl.gov/abs/1310.5078}{{\tt arXiv:1310.5078}}.

\bibitem{Bashkirov:2013vya}
D.~Bashkirov, {\it {Bootstrapping the $N=1$ SCFT in three dimensions}},
  \href{http://xxx.lanl.gov/abs/1310.8255}{{\tt arXiv:1310.8255}}.

\bibitem{El-Showk:2014dwa}
S.~El-Showk, M.~F. Paulos, D.~Poland, S.~Rychkov, D.~Simmons-Duffin, et~al.,
  {\it {Solving the 3d Ising Model with the Conformal Bootstrap II.
  c-Minimization and Precise Critical Exponents}},
  \href{http://xxx.lanl.gov/abs/1403.4545}{{\tt arXiv:1403.4545}}.

\bibitem{Maldacena:2012sf}
J.~Maldacena and A.~Zhiboedov, {\it {Constraining conformal field theories with
  a slightly broken higher spin symmetry}},  {\em Class.Quant.Grav.} {\bf 30}
  (2013) 104003, [\href{http://xxx.lanl.gov/abs/1204.3882}{{\tt
  arXiv:1204.3882}}].

\bibitem{Dolan:2011dv}
F.~Dolan and H.~Osborn, {\it {Conformal Partial Waves: Further Mathematical
  Results}},  \href{http://xxx.lanl.gov/abs/1108.6194}{{\tt arXiv:1108.6194}}.

\bibitem{Dolan:2003hv}
F.~A. Dolan and H.~Osborn, {\it {Conformal partial waves and the operator
  product expansion}},  {\em Nucl. Phys.} {\bf B678} (2004) 491--507,
  [\href{http://xxx.lanl.gov/abs/hep-th/0309180}{{\tt hep-th/0309180}}].

\bibitem{RT1}
S.~Ryu and T.~Takayanagi, {\it {Holographic derivation of entanglement entropy
  from AdS/CFT}},  {\em Phys.Rev.Lett.} {\bf 96} (2006) 181602,
  [\href{http://xxx.lanl.gov/abs/hep-th/0603001}{{\tt hep-th/0603001}}].

\bibitem{RT2}
S.~Ryu and T.~Takayanagi, {\it {Aspects of Holographic Entanglement Entropy}},
  {\em JHEP} {\bf 0608} (2006) 045,
  [\href{http://xxx.lanl.gov/abs/hep-th/0605073}{{\tt hep-th/0605073}}].

\bibitem{HartmanLargeC}
T.~Hartman, {\it {Entanglement Entropy at Large Central Charge}},
  \href{http://xxx.lanl.gov/abs/1303.6955}{{\tt arXiv:1303.6955}}.

\bibitem{Bizon}
P.~Bizo\'n and J.~Ja\l~mu\.zna, {\it {Globally regular instability of
  $AdS_3$}},  {\em Phys.Rev.Lett.} {\bf 111} (2013) 041102,
  [\href{http://xxx.lanl.gov/abs/1306.0317}{{\tt arXiv:1306.0317}}].

\bibitem{Zamolodchikov:1985wn}
A.~Zamolodchikov, {\it {Infinite Additional Symmetries in Two-Dimensional
  Conformal Quantum Field Theory}},  {\em Theor.Math.Phys.} {\bf 65} (1985)
  1205--1213.

\bibitem{Bouwknegt:1992wg}
P.~Bouwknegt and K.~Schoutens, {\it {W symmetry in conformal field theory}},
  {\em Phys.Rept.} {\bf 223} (1993) 183--276,
  [\href{http://xxx.lanl.gov/abs/hep-th/9210010}{{\tt hep-th/9210010}}].

\bibitem{Gutperle:2011kf}
M.~Gutperle and P.~Kraus, {\it {Higher Spin Black Holes}},  {\em JHEP} {\bf
  1105} (2011) 022, [\href{http://xxx.lanl.gov/abs/1103.4304}{{\tt
  arXiv:1103.4304}}].

\bibitem{deBoer:2013gz}
J.~de~Boer and J.~I. Jottar, {\it {Thermodynamics of higher spin black holes in
  $AdS_3$}},  {\em JHEP} {\bf 1401} (2014) 023,
  [\href{http://xxx.lanl.gov/abs/1302.0816}{{\tt arXiv:1302.0816}}].

\bibitem{Zamolodchikov:1987}
A.~Zamolodchikov, {\it Conformal symmetry in two-dimensional space: Recursion
  representation of conformal block},  {\em Theoretical and Mathematical
  Physics} {\bf 73} (1987), no.~1 1088--1093.

\bibitem{ZamolodchikovRecursion}
A.~Zamolodchikov, {\it {Conformal Symmetry in Two-Dimensions: An Explicit
  Recurrence Formula for the Conformal Partial Wave Amplitude}},  {\em
  Commun.Math.Phys.} {\bf 96} (1984) 419--422.

\bibitem{Zamolodchikov:1995aa}
A.~B. Zamolodchikov and A.~B. Zamolodchikov, {\it {Structure constants and
  conformal bootstrap in Liouville field theory}},  {\em Nucl.Phys.} {\bf B477}
  (1996) 577--605, [\href{http://xxx.lanl.gov/abs/hep-th/9506136}{{\tt
  hep-th/9506136}}].

\bibitem{Litvinov:2013sxa}
A.~Litvinov, S.~Lukyanov, N.~Nekrasov, and A.~Zamolodchikov, {\it {Classical
  Conformal Blocks and Painleve VI}},
  \href{http://xxx.lanl.gov/abs/1309.4700}{{\tt arXiv:1309.4700}}.

\bibitem{Ginsparg}
P.~H. Ginsparg, {\it {Applied Conformal Field Theory}},
  \href{http://xxx.lanl.gov/abs/hep-th/9108028}{{\tt hep-th/9108028}}.

\bibitem{HarlowLiouville}
D.~Harlow, J.~Maltz, and E.~Witten, {\it {Analytic Continuation of Liouville
  Theory}},  {\em JHEP} {\bf 1112} (2011) 071,
  [\href{http://xxx.lanl.gov/abs/1108.4417}{{\tt arXiv:1108.4417}}].

\bibitem{Behan:2014dxa}
C.~Behan, {\it {Conformal blocks for highly disparate scaling dimensions}},
  \href{http://xxx.lanl.gov/abs/1402.5698}{{\tt arXiv:1402.5698}}.

\bibitem{titchmarsh}
E.~C. Titchmarsh, {\em The theory of functions}, vol.~80.
\newblock London, 1939.

\end{thebibliography}\endgroup

 \end{document}